\documentclass[prd,aps,tightline,twocolumn,
nofootinbib,superscriptaddress]{revtex4}
\usepackage{amsmath}
\usepackage{booktabs}
\usepackage{amssymb}
\usepackage{latexsym}
\usepackage{wrapfig}

\usepackage[dvipdfmx]{graphicx}
\usepackage{fancybox}
\usepackage{color}

\newcommand{\lsim}
{\raise0.3ex\hbox{$\;<$\kern-0.75em\raise-1.1ex\hbox{$\sim\;$}}}
\newcommand{\gsim}
{\raise0.3ex\hbox{$\;>$\kern-0.75em\raise-1.1ex\hbox{$\sim\;$}}}
\begin{document}

\begin{flushleft}
STUPP-14-217, KEK-TH-1712,  KEK-Cosmo-138 \hspace{50mm} 
\today
\end{flushleft}

%\vskip 1.35cm
%\vspace{0.5cm}
\title{ 
Big-bang nucleosynthesis through bound-state effects with a long-lived slepton in the NMSSM
}
%\vspace{1cm}

\author{Kazunori Kohri}
\affiliation{KEK (High Energy Accelerator Research Organization), and Sokendai, 1-1 Oho, Tsukuba 305-0801, Japan}

\author{Masafumi Koike}
\affiliation{Department of Information Science, Utsunomiya University, Yoto, Utsunomiya, Tochigi 321-8585, Japan}

\author{Yasufumi Konishi}
\affiliation{Department of Physics, Saitama University, 
     Shimo-okubo, Sakura-ku, Saitama, 338-8570, Japan}

\author{Shingo Ohta}
\affiliation{Department of Physics, Saitama University, 
     Shimo-okubo, Sakura-ku, Saitama, 338-8570, Japan}
     
\author{Joe~Sato}
\affiliation{Department of Physics, Saitama University, 
     Shimo-okubo, Sakura-ku, Saitama, 338-8570, Japan}
     
\author{Takashi Shimomura}
\affiliation{Department of Physics, Niigata University, Niigata, 950-2181, Japan} 
%\affiliation{Max-Planck-Institut fur Kernphysik, Saupfercheckweg 1, D-69117 Heidelberg, Germany}

\author{Kenichi Sugai}
\affiliation{Department of Physics, Saitama University, 
     Shimo-okubo, Sakura-ku, Saitama, 338-8570, Japan}    

\author{Masato Yamanaka}
\affiliation{Department of Physics, Nagoya University, Nagoya 464-8602, Japan}

\vskip 0.15in

\begin{abstract}
%We find allowed parameter region in Next-to-Minimal Supersymmetric Standard Model with a long-lived slepton, where dark matter relic density, the Higgs mass, and light elements abundances (including $^7$Li and $^6$Li) are obtained correctly. {\bf ((Shimomura-san))}
We show that the Li problems can be solved in the next-to-minimal supersymmetric
standard model where the slepton as the next-to-lightest SUSY particle is very
long-lived. Such a long-lived slepton induces exotic nuclear reactions in big-bang
nucleosynthesis, and destroys and produces the $^7$Li and $^6$Li nuclei via bound state formation.
We study cases where the lightest SUSY particle is singlino-like neutralino and bino-like neutralino
to present
allowed regions in the parameter space which is consistent with the observations on
the dark matter and the Higgs mass.
\end{abstract}

\maketitle

%%%%%%%%%%%%%%%%%%%%%%%%%%%%%%%%%%%%%%%%%
%%%%%%%%%%%%%%%%%%%%%%%%%%%%%%%%%%%%%%%%%
%%%%%%%%%%%%%%%%%%%%%%%%%%%%%%%%%%%%%%%%%
\section{Introduction}\label{sec:introduction}%%%%%%%%%%%%%%%%%%%
%%%%%%%%%%%%%%%%%%%%%%%%%%%%%%%%%%%%%%%%%
%%%%%%%%%%%%%%%%%%%%%%%%%%%%%%%%%%%%%%%%%
%%%%%%%%%%%%%%%%%%%%%%%%%%%%%%%%%%%%%%%%%

The Standard Model has had enormous successes in describing the
interactions of the elementary particles, predicting almost every
experimental results with accuracy.  
The recent discovery of the Higgs particle finally crowned the
accumulation of successes~\cite{Aad:2012tfa,Chatrchyan:2012ufa}.
On the other hand, it left us a number of questions that suggest the
presence of a more fundamental theory behind.
Among such questions is the nature of dark matter; it became a
compelling question during this decade after the precise observations
of the universe reported their results~\cite{Bennett:2012zja,Ade:2013zuv}.
The theory behind the Standard Model ought to account for this
question.
Supersymmetric models have been attractive candidates for such theories.
The Minimal Supersymmetric Standard Model (MSSM) is the simplest extension
and most analyzed.
The lightest supersymmetric particle should be neutral and stable, and
thus can be the dark matter.
Other extensions are also of interest, one of which is the
Next-to-Minimal Supersymmetric Standard Model (NMSSM).
An extra singlet chiral supermultiplet is introduced in the NMSSM, and
thereby account for the $\mu$-problem~\cite{Kim:1983dt} that complicates the MSSM.
In addition, the NMSSM better reconciles with the observed Higgs mass of
$125~\mathrm{GeV}$.
The MSSM predicts a Higgs mass lighter than that of the $Z$ boson at the
tree level, and employs loop effects to raise it up to the observed
value.
On the other hand, the Higgs mass in the NMSSM has additional terms
contributed from the singlet, and it potentially offers a
straightforward interpretation of the observations.

A series of works by the present authors explored impacts of the
supersymmetry on the nucleosynthesis in the early universe
~\cite{Jittoh:2005pq,Jittoh:2007fr,Jittoh:2008eq,Jittoh:2010wh,Jittoh:2011ni,Kohri:2012gc}.
Focus in these works has been on the case where the next-to-lightest
supersymmetric particle (NLSP) is charged and long-lived so that
it survives until the time of nucleosynthesis after the big-bang.
It takes part in the nuclear reactions and alters the present-day abundance
of the light elements.
%produce the lightest
%supersymmetric particle (LSP), which is stable, neutral, and is
%observed as dark matter today.
%
%Such reactions, if occurred, should be engraved also on the
%present-day abundance of the light elements, which are successfully
%predicted by the standard model of the big-bang nucleosynthesis (BBN)
%without new physics like supersymmetry.
%
Possible disagreement indeed persists on the abundance of lithium
compared with the calculation based on the standard big-bang
nucleosynthesis (BBN) scenario.
The standard calculation predicts the ratio of abundance
${\rm Log}_{10}(\mathrm{^{7}Li/H})$ to be $-9.35\pm 0.06$~\cite{Jittoh:2011ni}, 
while the observation indicates $-9.63 \pm 0.06$~\cite{Melendez:2004ni}.
Lithium 6 provides another possible disagreement; its observed ratio of abundance
$\mathrm{^{6}Li/^{7}Li} = 0.046 \pm 0.022$ is about $10^{2}$--$10^{3}$ larger than the 
theoretical prediction~\cite{Asplund:2005yt}.
These discrepancies can be the trace of the interaction between nuclei
and the NLSP which is absent in the standard BBN scenario.
Our scenario can thereby account for the abundance of the dark matter
and of the lithium in a single framework.
We analyzed if this scenario works within the MSSM with staus as the NLSP
and neutralinos as the lightest supersymmetric particle (LSP), 
and found the parameter region that can
account for these observational handles to the new physics~\cite{Jittoh:2005pq,Jittoh:2007fr,Jittoh:2008eq,Jittoh:2010wh,Jittoh:2011ni}.
Now that the mass of Higgs particle is determined, we are to examine
whether it is compatible with our scenario.
Our previous work analyzed the constrained minimal supersymmetric standard
model~\cite{Konishi:2013gda}.
There we found allowed regions in the parameter space, and presented
phenomenological predictions such as mass spectra and branching
ratios.
In the present paper, we further extend our scenario to the NMSSM with
the flavor violation and search for its further applications.
We demonstrate that the NMSSM under our scenario can simultaneously
account for the three phenomenological clues: the abundance of dark
matter, that of lithium, and Higgs mass.
We explore parameter space and discover parameter points that qualify
the requirements.
Special interest is in the case where the NMSSM singlet is the major
component of the neutralino LSP in expectation of the difference from
the MSSM.
The case of bino-like neutralino is analyzed as well.

This paper is organized as follows.
In Sec.~\ref{sec:NMSSM}, we introduce the NMSSM and define the model of
our interest.
The BBN in the presence of a long-lived slepton is also explained
in this section.
The exotic reactions that are absent from the standard BBN are
introduced.
Section~\ref{sec:strategy} describes our strategy to find out the
parameter point that are consistent with the three phenomenological
clues.
The results are presented in Sec.~\ref{sec:results}.
We show the three benchmark cases characterized by the type of
neutralino LSP.
The first case considers singlino-like neutralinos.
The couplings of the singlet is small and $\tan \beta$ is large in
this case.
The second case also handles singlino-like neutralinos, but the
singlet couplings are large and $\tan \beta$ is small compared with the
first case.
The third case deals with bino-like neutralinos.
Singlet couplings are set large and $\tan \beta$ small as in the second case.
We show that each of these three cases allows the phenomenological
constraints.
We spot benchmark points in the parameter space and exhibit that they
accord with the observations.
The NMSSM parameters are calculated to confirm the adequate Higgs
mass, and the BBN network calculation is carried out to check the
relic abundance of light elements and of the LSP dark matter.
%
%(Section~\ref{sec:DarkMatter} ...)
%
The present work is summarized in Sec.~\ref{sec:summary}.

%LSPの説明が２回されているので注意

%%%%%%%%%%%%%%%%%%%%%%%%%%%%%%%%%%%%%%%%%
%%%%%%%%%%%%%%%%%%%%%%%%%%%%%%%%%%%%%%%%%
%%%%%%%%%%%%%%%%%%%%%%%%%%%%%%%%%%%%%%%%%
\section{The NMSSM and the exotic BBN} \label{sec:NMSSM}%%%%%%%
%%%%%%%%%%%%%%%%%%%%%%%%%%%%%%%%%%%%%%%%%
%%%%%%%%%%%%%%%%%%%%%%%%%%%%%%%%%%%%%%%%%
%%%%%%%%%%%%%%%%%%%%%%%%%%%%%%%%%%%%%%%%%

%%%%%%%%%%%%%%
%%%%%%%%%%%%%%
We recapitulate the Higgs and neutralino sectors, and explain exotic BBN processes with 
a long-lived slepton in the NMSSM.
%%%%%%%%%%%%%%
%%%%%%%%%%%%%%

%%%%%%%%%%%%%%%%%%%%%%%%%%%%%%%%%%%%%%%%%%
%%%%%%%%%%%%%%%%%%%%%%%%%%%%%%%%%%%%%%%%%%
\subsection{Higgs bosons} \label{sec:mhiggs} %%%%%%%%%%%%
%%%%%%%%%%%%%%%%%%%%%%%%%%%%%%%%%%%%%%%%%%
%%%%%%%%%%%%%%%%%%%%%%%%%%%%%%%%%%%%%%%%%%

%%%%%%%%%%%%%%
%%%%%%%%%%%%%%
%We consider the Higgs mass and neutralino sector in the NMSSM with ${\cal Z}_3$
%parity.
%%%%%%%%%%%%%%
%%%%%%%%%%%%%%
%This model has following superpotentials and soft-breaking terms in addition to the MSSM:
%%%%%%%%%%%%%%
%%%%%%%%%%%%%%
The NMSSM specific part of the superpotential and the soft-breaking terms are 
\begin{align}
W^{\rm NMSSM} \supset
&
	 \lambda \hat S \hat H_d \cdot \hat H_u 
	+ \frac{1}{3}\kappa \hat S^3,
\label{eq:superpotential} \\
-{\cal L}_{\rm soft}^{\rm NMSSM} \supset
&
	m_{S}^2 |S|^2 \notag \\
&
	+
	\left(
	 	\lambda A_{\lambda} H_d \cdot H_u S
		+ \frac{1}{3}\kappa A_{\kappa } S^3
		+{\rm h.c.}
	\right),
\label{eq:softterms}
\end{align}
where ${\cal Z}_3$ parity conservation is assumed.
%%%%%%%%%%%%%%
%%%%%%%%%%%%%%
In Eq.~\eqref{eq:superpotential}, $\hat H_d$, $\hat H_u$ and $\hat S$ are
the Higgs superfields and the singlet superfield.
%%%%%%%%%%%%%%
%%%%%%%%%%%%%%
The couplings $\lambda$ and $\kappa$ are dimensionless parameters.
%%%%%%%%%%%%%%
%%%%%%%%%%%%%%
In Eq.~\eqref{eq:softterms},
\begin{align}
H_d = 
	\begin{pmatrix}
		H^0_d \\ H^-_d
	\end{pmatrix},~
H_u = 
	\begin{pmatrix}
		H^+_u \\ H^0_u
	\end{pmatrix},~
S
\end{align}
are two Higgs doublets and a singlet scalar.
%%%%%%%%%%%%%%
%%%%%%%%%%%%%%
We denote vacuum expectation values of their neutral components by 
$v_d, v_u$, and $s$, respectively.
%%%%%%%%%%%%%%
%%%%%%%%%%%%%%
The soft-breaking parameters are $m_{S}, A_{\lambda}$ and $A_{\kappa}$.
%%%%%%%%%%%%%%
%%%%%%%%%%%%%%
The free parameters in the Higgs sector are $\lambda , \kappa , A_{\lambda}, A_{\kappa}, \tan \beta (\equiv v_u/v_d)$, 
and $\mu _{\rm eff} (\equiv \lambda s)$.
%%%%%%%%%%%%%%
%%%%%%%%%%%%%%

%%%%%%%%%%%%%%
%%%%%%%%%%%%%%
The Higgs mass up to dominant 1-loop contribution is
\begin{align}
m_h^2 =
&
	m_Z^2 \cos ^2 2\beta
	+\lambda ^2 v^2 \sin ^2 2\beta
	-\frac{\lambda^2}{\kappa^2}v^2 (\lambda -\kappa \sin 2\beta )^2 \notag \\
&
	+\frac{3m_t^2}{16\pi ^2 v^2}
		\left\{
			\log \left( \frac{m_s^2}{m_t^2} \right)
			+\frac{X_t^2}{m_s^2}
				\left(
				1-\frac{X_t^2}{12m_s^2}
				\right)
		\right\},
\label{eq:mhiggs}
\end{align}
where $m_Z$ is $Z$ boson mass, $m_t$ is the top mass, $m_s\equiv \sqrt{m_{\tilde t_1}m_{\tilde t_2}}$ is 
geometric mean of stop masses, $X_t = A_t -\mu _{\rm eff} \cot \beta$, 
and $v=\sqrt{v_d^2+v_u^2}=174$ GeV.
%%%%%%%%%%%%%%
%%%%%%%%%%%%%%
The second and third tree terms are characteristic in the NMSSM, 
and can lift up the Higgs mass. 
%%%%%%%%%%%%%%
%%%%%%%%%%%%%%
This is one of the attractive features of the model.
%%%%%%%%%%%%%%
%%%%%%%%%%%%%%
In contrast, significant 1-loop contribution (the last term in 
Eq.~\eqref{eq:mhiggs}) is required to obtain the Higgs mass 
around 125 GeV in the MSSM since the tree contribution is at most $m_Z$. 
%%%%%%%%%%%%%%
%%%%%%%%%%%%%%
The 1-loop contribution is maximized by the relation, $X_t=\sqrt 6 m_s$, 
with large stop mass scale.
%%%%%%%%%%%%%%
%%%%%%%%%%%%%%
In the NMSSM we do not need to rely on such conditions to obtain 
observed Higgs mass.
%%%%%%%%%%%%%%
%%%%%%%%%%%%%%

%%%%%%%%%%%%%%%%%%%%%%%%%%%%%%%%%%%%%%%%%%
%%%%%%%%%%%%%%%%%%%%%%%%%%%%%%%%%%%%%%%%%%
\subsection{Neutralinos} \label{sec:neutralinos} %%%%%%%%%%%%
%%%%%%%%%%%%%%%%%%%%%%%%%%%%%%%%%%%%%%%%%%
%%%%%%%%%%%%%%%%%%%%%%%%%%%%%%%%%%%%%%%%%%

%%%%%%%%%%%%%%
%%%%%%%%%%%%%%
%There are five neutralinos in the NMSSM.
%%%%%%%%%%%%%%
%%%%%%%%%%%%%%
In the NMSSM the neutralinos are linear combinations of bino $\tilde B$, 
wino $\tilde W$, neutral higgsinos $\tilde H^0_d$ and $\tilde H^0_u$, and singlino 
$\tilde S$, the supersymmetric partner of the singlet scalar:
%%%%%%%%%%%%%%
%%%%%%%%%%%%%%
\begin{align}
\tilde \chi ^0_i =
	N_{i\tilde B} \tilde B 
	+N_{i\tilde W} \tilde W
	+N_{i\tilde H^0_d} \tilde H^0_d
	+N_{i\tilde H^0_u} \tilde H^0_u
	+N_{i\tilde S} \tilde S,
\label{eq:general-chi}
\end{align}
%%%%%%%%%%%%%%
%%%%%%%%%%%%%%
where $i$ runs from 1 to 5.
%%%%%%%%%%%%%%
%%%%%%%%%%%%%%
The mass matrix of the neutralinos is given by the following symmetric expression 
in the basis $(\tilde B, \tilde W, \tilde H^0_d, \tilde H^0_u, \tilde S)$:
\begin{align}
{\cal M}_{\tilde \chi ^0} =
\begin{pmatrix}
M_1		& 0		& -c_{\beta} s_W m_Z	& s_{\beta} s_W m_Z 	& 0 \\
		& M_2 	& c_{\beta} c_W m_Z	& -s_{\beta} c_W m_Z 	& 0 \\
		&		& 0					& -\mu _{\rm eff} 		& -\mu _{\lambda} s_{\beta} \\
		&		& 					& 0					& -\mu _{\lambda} c_{\beta} \\
		&		&					&					& \mu _{\kappa} 
\end{pmatrix}, \label{eq:massmatrix}
\end{align}
where $M_1$ and $M_2$ are masses of the gauginos, 
$c_{\beta}=\cos \beta, s_{\beta}=\sin \beta, c_W=\cos \theta _W$, 
$s_W=\sin \theta _W$, $\mu _{\lambda}=\lambda v$, 
and $\mu _{\kappa}=2\kappa s$.
%%%%%%%%%%%%%%
%%%%%%%%%%%%%%
Here $\theta _W$ is the Weinberg angle.
%%%%%%%%%%%%%%
%%%%%%%%%%%%%%
We consider the case where the LSP is the lightest neutralino, $\tilde \chi ^0_1$, 
and the NLSP is the lightest slepton, $\tilde l$.
%%%%%%%%%%%%%%
%%%%%%%%%%%%%%

%%%%%%%%%%%%%%
%%%%%%%%%%%%%%
%We require the slepton to be enough long-lived as a result of the requirement 
%that the masses of the slepton and LSP neutarlino are degenerated for obtaining 
%suitable dark matter relic density by the coannihilation mechanism.
%%%%%%%%%%%%%%
%%%%%%%%%%%%%%
%Then we also require that the long-lived slepton affects the BBN through exotic 
%nuclear processes so that suitable light elements abundances to be obtained.
%%%%%%%%%%%%%%
%%%%%%%%%%%%%%
%Not only the mass difference but also the interaction relevant to the decay of 
%the slepton is important in order to realize such long-lived slepton.
%%%%%%%%%%%%%%
%%%%%%%%%%%%%%
As we will see later, the interaction among $\tilde \chi ^0_1, \tilde l$, and a lepton $f$ 
is a key ingredient in this study.
%%%%%%%%%%%%%%
%%%%%%%%%%%%%%
The interaction Lagrangian is written as
\begin{align}
-{\cal L}_{\tilde \chi ^0_1{\mathchar`-}f{\mathchar`-}\tilde l}
	=
		\tilde l^{\ast}\overline{\tilde \chi ^0_1}(G_{{\rm L}f}P_{{\rm L}} +G_{{\rm R}f}P_{{\rm R}}) f + {\rm h.c.},
\label{eq:int-lagrangian}
\end{align}
where
\begin{align}
G_{{\rm L}f} =
& 	c_f \left[  \frac{g_2}{\sqrt 2}\cos \theta _f (\tan \theta _W N_{1\tilde B} +N_{1\tilde W} ) \right. \notag \\
&\quad      \left. -\frac{m_f g_2}{\sqrt 2 m_W \cos\beta} \sin \theta _f N_{1\tilde H_d^0} \right],  
\label{eq:gl} \\
G_{{\rm R}f} = 
&	c_f \left[ \sqrt 2 g_2 \sin\theta _f \tan \theta _W N_{1\tilde B} \right.  \notag \\ 
&\quad      \left. +\frac{m_f g_2}{\sqrt 2 m_W \cos\beta} \cos \theta _f N_{1\tilde H_d^0} \right].
\label{eq:gr}
\end{align}
Here $g_2$ is the gauge coupling constant of SU(2), $m_f$ is the mass of the lepton $f$, and $f=e,~\mu$, and $\tau$.
%%%%%%%%%%%%%%
%%%%%%%%%%%%%%
The slepton is expanded as a linear combination of the flavor eigenstates:
\begin{align}
\tilde l = \sum _{f=e,\mu , \tau} c_f \tilde f, \quad
\tilde f = \cos \theta _f \tilde f_L +\sin \theta _f \tilde f _R,
\label{eq:lfv}
\end{align}
where $\tilde f_L$ and $\tilde f_R$ are the flavor eigenstates, and the coefficients are 
normalized as $c_e^2 +c_{\mu}^2 +c_{\tau}^2 =1$.
%%%%%%%%%%%%%%
%%%%%%%%%%%%%%
Hereafter we assume that 
$\sin \theta _e = \sin \theta _{\mu} = \sin \theta _{\tau}$
and no CP-violating phase exists in the slepton sector.
%%%%%%%%%%%%%%
%%%%%%%%%%%%%%

%%%%%%%%%%%%%%
%%%%%%%%%%%%%%
We study two scenarios where the lightest neutralino is singlino- and bino-like.
%%%%%%%%%%%%%%
%%%%%%%%%%%%%%
The detailed expressions of $\tilde \chi ^0_1$ and the couplings $G_{{\rm L}f}$ and 
$G_{{\rm R}f}$ are different between the two cases.
%%%%%%%%%%%%%%
%%%%%%%%%%%%%%
In the followings, we explicitly write down the mixing of the lightest neutralino 
in each case up to the second order in perturbation theory.
%%%%%%%%%%%%%%
%%%%%%%%%%%%%%

%%%%%%%%%%%%%%%%%%%%%%%%%%%%%%%%%%%%%%%%%%
\subsubsection{Singlino-like neutralino LSP} \label{sec:singlino} %%%%%%%%%%
%%%%%%%%%%%%%%%%%%%%%%%%%%%%%%%%%%%%%%%%%%

%%%%%%%%%%%%%%
%%%%%%%%%%%%%%
%Here we assume that the lightest neutralino is singlino-like, as a specific case in the NMSSM.
%%%%%%%%%%%%%%
%%%%%%%%%%%%%%
%Then the relation $M_1, M_2, \mu _{\rm eff} \gg \mu _{\kappa}$ should be satisfied, 
%and thus the mass of the lightest neutralino is $m_{\tilde \chi ^0_1}\simeq \mu _{\kappa}$.
%%%%%%%%%%%%%%
%%%%%%%%%%%%%%
%Then the lightest neutralino is represented up to the second order in the perturbation theory as 
Up to the second order perturbative expansion, the singlino-like neutralino is
\begin{align}
\tilde \chi^0 =
&	-\frac{m_Z \mu_{\rm eff} \mu _{\lambda} s_W c_{2\beta}}{(\mu _{\kappa} -M_1)(\mu _{\kappa}^2 -\mu _{\rm eff}^2)} \tilde B  \notag  \\
&	+\frac{m_Z \mu_{\rm eff} \mu _{\lambda} c_W c_{2\beta}}{(\mu _{\kappa} -M_2)(\mu _{\kappa}^2 -\mu _{\rm eff}^2)} \tilde W  \notag \\
&	+\frac{\mu _{\lambda} (\mu _{\rm eff} c_{\beta} -\mu _{\kappa} s_{\beta})}{\mu _{\kappa}^2 -\mu _{\rm eff}^2} \tilde H_d^0  \notag \\
&	+\frac{\mu _{\lambda} (\mu _{\rm eff} s_{\beta} -\mu _{\kappa} c_{\beta})}{\mu _{\kappa}^2 -\mu _{\rm eff}^2} \tilde H_u^0  \notag \\
&	+\tilde S,
\label{eq:slike-chi}
\end{align}
where $c_{2\beta} = \cos 2\beta$, and the normalization factor is omitted.
%%%%%%%%%%%%%%
%%%%%%%%%%%%%%
If the neutralino Eq.~\eqref{eq:slike-chi} is the LSP, the mixings 
$N_{1\tilde B}, N_{1\tilde W}$ and $N_{1\tilde H^0_d}$ in Eqs.~\eqref{eq:gl} and \eqref{eq:gr}
can be read off by comparing Eqs.~\eqref{eq:general-chi} and \eqref{eq:slike-chi}.
%%%%%%%%%%%%%%
%%%%%%%%%%%%%%
The free parameters in $G_{{\rm L}f}$ and $G_{{\rm R}f}$ are $M_1, M_2, \lambda, \kappa, \tan\beta, c_f$,
and $\theta _f$.
%%%%%%%%%%%%%%
%%%%%%%%%%%%%%

%%%%%%%%%%%%%%%%%%%%%%%%%%%%%%%%%%%%%%%%%%
\subsubsection{Bino-like neutralino LSP} \label{sec:bino} %%%%%%%%%%%%%%
%%%%%%%%%%%%%%%%%%%%%%%%%%%%%%%%%%%%%%%%%%

%%%%%%%%%%%%%%
%%%%%%%%%%%%%%
%We also consider the case where the lightest neutralino is bino-like in order to compare 
%the result with that in case with pure bino neutralino LSP in the MSSM
%~\cite{Jittoh:2010wh,Jittoh:2011ni,Kohri:2012gc}.
%%%%%%%%%%%%%%
%%%%%%%%%%%%%%
%Then the relation $M_2, \mu _{\rm eff}, \mu _{\kappa} \gg M_1$ should be satisfied, 
%and thus the mass of the lightest neutralino is $m_{\tilde \chi ^0_1}\simeq M_1$.
%%%%%%%%%%%%%%
%%%%%%%%%%%%%%
%In this case, the lightest neutralino is represented up to the second order in perturbation theory as 
Up to the second order perturbative expansion, the bino-like neutralino is
\begin{align}
\tilde \chi^0 =
&	\tilde B  \notag \\
& 	-\frac{m_Z^2 s_W c_W (M_1+ \mu _{\rm eff} s_{2\beta})}{(M_1-M_2)(M_1^2 -\mu _{\rm eff}^2)}\tilde W \notag \\
&	-\frac{m_Z s_W (M_1 c_{\beta} +\mu _{\rm eff} s_{\beta})}{M_1^2 -\mu _{\rm eff}^2}\tilde H_d^0 \notag \\
&	+\frac{m_Z s_W (M_1 s_{\beta} +\mu _{\rm eff} c_{\beta})}{M_1^2 -\mu _{\rm eff}^2}\tilde H_u^0 \notag \\
&	-\frac{m_Z \mu _{\rm eff} \mu _{\lambda} s_W c_{2\beta}}{(M_1 -\mu _{\kappa})(M_1^2 -\mu _{\rm eff}^2 )}\tilde S,
\label{eq:blike-chi}
\end{align}
where $s_{2\beta} =\sin 2\beta$, and the normalization factor is omitted.
%%%%%%%%%%%%%%
%%%%%%%%%%%%%%
If the neutralino Eq.~\eqref{eq:blike-chi} is the LSP, the mixings 
$N_{1\tilde B}, N_{1\tilde W}$ and $N_{1\tilde H^0_d}$ in Eqs.~\eqref{eq:gl} and \eqref{eq:gr}
can be read off by comparing Eqs.~\eqref{eq:general-chi} and \eqref{eq:blike-chi}.
%%%%%%%%%%%%%%
%%%%%%%%%%%%%%
The free parameters in $G_{{\rm L}f}$ and $G_{{\rm R}f}$ are $M_1, M_2, \lambda, \tan\beta, \mu _{\rm eff}, c_f$, 
and $\theta _f$.
%%%%%%%%%%%%%%
%%%%%%%%%%%%%%
The couplings do not depend on $\kappa$.
%%%%%%%%%%%%%%
%%%%%%%%%%%%%%

%%%%%%%%%%%%%%%%%%%%%%%%%%%%%%%%%%%%%%%%%%
%%%%%%%%%%%%%%%%%%%%%%%%%%%%%%%%%%%%%%%%%%
\subsection{BBN with a long-lived slepton} \label{sec:bbn} %%%%%%%%%%%%
%%%%%%%%%%%%%%%%%%%%%%%%%%%%%%%%%%%%%%%%%%
%%%%%%%%%%%%%%%%%%%%%%%%%%%%%%%%%%%%%%%%%%

%%%%%%%%%%%%%%
%%%%%%%%%%%%%%
We explain exotic BBN processes caused by a long-lived slepton.
%%%%%%%%%%%%%%
%%%%%%%%%%%%%%
Then we explain the difference of the couplings $G_{{\rm L}f}$ and $G_{{\rm R}f}$ 
between the NMSSM and the MSSM.
%%%%%%%%%%%%%%
%%%%%%%%%%%%%%

%%%%%%%%%%%%%%%%%%%%%%%%%%%%%%%%%%%%%%%%%%
%%%%%%%%%%%%%%%%%%%%%%%%%%%%%%%%%%%%%%%%%%
\subsubsection{A long-lived slepton} \label{sec:slepton} %%%%%%%%%%%%
%%%%%%%%%%%%%%%%%%%%%%%%%%%%%%%%%%%%%%%%%%
%%%%%%%%%%%%%%%%%%%%%%%%%%%%%%%%%%%%%%%%%%

%%%%%%%%%%%%%%
%%%%%%%%%%%%%%
%We know that observed dark matter relic density and light elements abundances 
%can be obtained in the MSSM~\cite{Jittoh:2010wh,Jittoh:2011ni,Kohri:2012gc}.
%%%%%%%%%%%%%%
%%%%%%%%%%%%%%
%In the scenario the LSP is the lightest neutralino (bino) and the NLSP is the lightest 
%slepton (or lighter stau).
%%%%%%%%%%%%%%
%%%%%%%%%%%%%%
%The tight degeneracy of their masses reproduces observed relic density of dark matter 
%by coannihilation mechanism~\cite{Griest:1990kh,oai:arXiv.org:hep-ph/9704361}.

%%%%%%%%%%%%%%
%%%%%%%%%%%%%%
The relic density of dark mater is well described by the coannihilation mechanism~\cite{Griest:1990kh}.
%%%%%%%%%%%%%%
%%%%%%%%%%%%%%
It requires the small mass difference $\delta m$ between the neutralino LSP and the slepton NLSP.
%%%%%%%%%%%%%%
%%%%%%%%%%%%%%
The two-body decay $\tilde l \to \tilde \chi ^0_1 + \tau$ is kinematically forbidden when $\delta m < m_{\tau}$.
%%%%%%%%%%%%%%
Let us first consider the case where $c_e = c_\mu = 0$ and $c_{\tau} = 1$.
In this case, the flavor is conserved and the slepton is stau.
The only allowed decay channel is three- and four-body decay.
These decay rates are suppressed due to the small phase space and the small couplings, and hence the sleptons become long-lived.
Then it survives until the BBN era to form bound states with nuclei.
Such long-lived slepton can account for the discrepancy of the lithium
by exotic nuclear reactions with the bound state
~\cite{Jittoh:2005pq,Jittoh:2007fr,Jittoh:2008eq,Jittoh:2010wh,Jittoh:2011ni,Kohri:2012gc}
(see also \cite{Kohri:2006cn}).

 %スレプトンの残存量も計算できることを書く。(Jittoh et al 2010

%If the flavor is conserved and the slepton is stau, the only allowed decay channel is three- and four-body decay.
%Such channels are suppressed due to the small phase space and the small couplings, and hence the sleptons become long-lived.
%Then it survives until the BBN era to \blue{form bound states with nuclei}.
%Such long-lived slepton could account for the discrepancy of the lithium by exotic nuclear reactions with the bound state effects.
%%%%%%%%%%%%%%
%\blue{
%The coannihilation mechanism leads the observed relic density of bino-(singlino-)like 
%neutralino dark matter~\cite{Griest:1990kh, }.
%%%%%%%%%%%%%%
%%%%%%%%%%%%%%
%The degeneracy in mass of the neutralino LSP and the slepton NLSP is required 
%for the coannihilation mechanism to successfully work.
%}
%%%%%%%%%%%%%%
%%%%%%%%%%%%%%
%The degeneracy can make the slepton long-lived so that it survives until 
%the BBN era to \blue{form bound states with nuclei}.
%%%%%%%%%%%%%%
%%%%%%%%%%%%%%
%Such long-lived slepton could solve the lithium problems by some exotic nuclear reactions 
%with the bound state effects.
%%%%%%%%%%%%%%
%%%%%%%%%%%%%%
%The longevity of the slepton is realized since the two-body 
%decay $\tilde l \to \tilde \chi ^0_1 + \tau$ is kinematically closed.
%%%%%%%%%%%%%%
%%%%%%%%%%%%%%
%In the situation, tree- and four-body decays are tightly suppressed 
%by small phase space.
%%%%%%%%%%%%%%

This situation radically changes when $c_e$ and/or $c_\mu$ are nonzero as
the flavor violating two-body decay channels
\begin{align}
\tilde l \to \tilde \chi ^0_1 +f, \quad f \ni e, \mu
\label{eq:lfv-decay}
\end{align}
open up.
These channels do not suffer from the phase-space suppressions and
the coupling suppressions, and thus may make the slepton lifetime much shorter.
We note that the lifetime is proportional to $(G_{{\rm L}f} +G_{{\rm R}f})^{-2}$ where $f=\mu$ and $e$.

 %スレプトンの残存量も計算できることを書く。(Kohri et al 2012

%%%%%%%%%%%%%%
%\red{
%The slepton lifetime also 
%When the presence of the flavor mixings affect  in the slepton sector,
%flavor violating two-body decays, 
%\begin{align}
%\tilde l \to \tilde \chi ^0_1 +f, \quad f \ni e, \mu
%\label{eq:lfv-decay}
%\end{align}
%can be opened.
%%%%%%%%%%%%%%%
%%%%%%%%%%%%%%%
%Then, even tiny flavor mixings could make the slepton lifetime much shorter.
%%%%%%%%%%%%%%%
%%%%%%%%%%%%%%%
%This is because the flavor violating two-body decays Eq.~\eqref{eq:lfv-decay} 
%are not so affected by the phase space suppression unless they are kinematically allowed.
%%%%%%%%%%%%%%%
%%%%%%%%%%%%%%%
%The lifetime is proportional to $(G_{{\rm L}f} +G_{{\rm R}f})^{-2}$ where $f=\mu$, and $e$.
%%%%%%%%%%%%%%
%%%%%%%%%%%%%%When the decay modes in Eq.~\eqref{eq:lfv-decay} are dominated
%}

%%%%%%%%%%%%%%%%%%%%%%%%%%%%%%%%%%%%%%%%%%
%%%%%%%%%%%%%%%%%%%%%%%%%%%%%%%%%%%%%%%%%%
\subsubsection{Exotic BBN reactions} \label{sec:bbn} %%%%%%%%%%%%
%%%%%%%%%%%%%%%%%%%%%%%%%%%%%%%%%%%%%%%%%%
%%%%%%%%%%%%%%%%%%%%%%%%%%%%%%%%%%%%%%%%%%

%%%%%%%%%%%%%%
%%%%%%%%%%%%%%
There are three types of exotic BBN reactions with the long-lived slepton.
%%%%%%%%%%%%%%
%%%%%%%%%%%%%%

%%%%%%%%%%%%%%
%%%%%%%%%%%%%%
First one is the internal conversion processes~\cite{Jittoh:2007fr,Bird:2007ge},
\begin{subequations}
\begin{align}
(^7{\rm Be}~\tilde l^-) &\to~^7{\rm Li} + \tilde \chi ^0_1 + \nu _l, \label{eq:icbe} \\
(^7{\rm Li}~\tilde l^-) &\to~^7{\rm He} + \tilde \chi ^0_1 + \nu _l, \label{eq:icli}
\end{align}
\label{eq:ic}
\end{subequations}
where, (X $\tilde l^-$) represents bound state of a nucleus X and the slepton.
Notice that not only $^7$Li but also $^7$Be must be converted since $^7$Be produces $^7$Li by the electron capture in late universe.
%%%%%%%%%%%%%%
%%%%%%%%%%%%%%

%%%%%%%%%%%%%%
%%%%%%%%%%%%%%
The second one is the catalyzed fusion process~\cite{Pospelov:2006sc,Hamaguchi:2007mp,Kawasaki:2007xb}, 
\begin{align}
(^4{\rm He}~\tilde l^-) + {\rm D} \to~^6{\rm Li} + \tilde l^-,
\label{eq:cf}
\end{align}
where D represents deuteron.
%%%%%%%%%%%%%%
%%%%%%%%%%%%%%

%%%%%%%%%%%%%%
%%%%%%%%%%%%%%
The last one is the $^4$He spallation processes~\cite{Jittoh:2011ni}: 
\begin{subequations}
\begin{align}
(^4{\rm He}~\tilde l^-) &\to \tilde \chi ^0_1 + \nu _l + {\rm T} + n,  \label{eq:tn} \\
(^4{\rm He}~\tilde l^-) &\to \tilde \chi ^0_1 + \nu _l + {\rm D} + 2n, \label{eq:dnn} \\
(^4{\rm He}~\tilde l^-) &\to \tilde \chi ^0_1 + \nu _l + p + 3n, \label{eq:pnnn}
\end{align}
\label{eq:spa}
\end{subequations}
where triton (T), deuteron, proton ($p$), and neutron ($n$) are produced from $^4$He.
%%%%%%%%%%%%%%
%%%%%%%%%%%%%%

%%%%%%%%%%%%%%
%%%%%%%%%%%%%%
The timescale to form bound state ($^7$Be~$\tilde l^-$) is ${\cal O}(10^3)$~s~\cite{Jittoh:2008eq}.
This timescale is important because the primordial $^7$Li exists as beryllium-7 in the BBN era.
%%%%%%%%%%%%%%
%%%%%%%%%%%%%%
The internal conversion processes \eqref{eq:ic} proceed much faster and thus 
$^7$Be is very efficiently destroyed.
%%%%%%%%%%%%%%
%%%%%%%%%%%%%%
We note that $^7$Li from \eqref{eq:icbe} is also destroyed by background protons.
%
%Since $^7$Be turns into $^7$Li by the electron capture after the BBN era,
%the abundance of $^7$Li is reduced through these processes \eqref{eq:ic}.
%%%%%%%%%%%%%%(14b) が効く前にリチウムは背景プロトンに壊される？どちらが優勢？（この書き方で問題ないか？ー＞郡さんに確認
%%%%%%%%%%%%%%
Thus if the slepton lifetime is longer than $10^3$~s, we can obtain observed abundance 
of $^7$Li through the internal conversion processes.
%%%%%%%%%%%%%%
%%%%%%%%%%%%%%
We can control the timescales of the internal conversion processes \eqref{eq:ic} 
by changing free parameters in $G_{{\rm L}\tau}$, $G_{{\rm R}\tau}$, and $\delta m$.
%%%%%%%%%%%%%%
%%%%%%%%%%%%%%

%%%%%%%%%%%%%%
%%%%%%%%%%%%%%
The timescale to form bound state $(^4{\rm He}~\tilde l^-)$ is 
${\cal O}(10^4)$~s~\cite{Jittoh:2008eq}.
%%%%%%%%%%%%%%%%%%%%%%%%%
The abundance of $^6$Li has sever upper bound, $\mathrm{^{6}Li/^{7}Li} = 0.046 \pm 0.022$~\cite{Asplund:2005yt}.
%%%%%%%%%%%%%%
%%%%%%%%%%%%%%
The abundance through the catalyzed fusion process \eqref{eq:cf} strongly depends on the slepton lifetime.
%%%%%%%%%%%%%%
%%%%%%%%%%%%%%
Therefore, since the process can overproduce $^6$Li, it gives an upper bound of the slepton lifetime.
%%%%%%%%%%%%%%
%%%%%%%%%%%%%% 
On the other hand, it is argued that
tiny primordial abundance of $^6$Li exists.
This is called the $^6$Li problem~\cite{Asplund:2005yt}.
%Actually, there is the $^6$Li problem~\cite{} where the predicted abundance of 
%$^6$Li by the standard BBN is much smaller than the observed one, though 
%it is controversial whether we regard the discrepancy as the problem in the particle physics.
%%%%%%%%%%%%%%
%%%%%%%%%%%%%% 
We can solve this problem by producing tiny amount of $^6$Li through 
the catalyzed fusion process~\eqref{eq:cf}.
%%%%%%%%%%%%%%
%%%%%%%%%%%%%%
In the situation, the slepton lifetime has to be tuned so that sufficient amount of $^6$Li is produced.
%%%%%%%%%%%%%%
%%%%%%%%%%%%%%
Therefore, the catalyzed fusion process \eqref{eq:cf} can also give 
lower bound of the slepton lifetime.
%though it is longer than the one given by 
%requiring that the $^7$Li problem can be solved by the internal conversion 
%processes \eqref{eq:ic}. 
%%%%%%%%%%%%%%
%%%%%%%%%%%%%%
The amount of $^6$Li can be controlled by changing free parameters 
in $G_{{\rm L}f}$ and $G_{{\rm R}f}$ through the slepton lifetime since 
the timescale of the catalyzed fusion process \eqref{eq:cf} depends on 
neither $G_{{\rm L}f}$ nor $G_{{\rm R}f}$ where $f=\tau, \mu,$ and $e$.
%%%%%%%%%%%%%%
%%%%%%%%%%%%%%

%%%%%%%%%%%%%%
%%%%%%%%%%%%%%
The standard BBN can predict observed abundances of $^3$He and D.
Therefore, the $^4$He spallation processes \eqref{eq:spa} should not be efficient, 
which gives upper bound on their timescales and the slepton lifetime.
%%%%%%%%%%%%%%
We can also control these timescales 
by changing the parameters in $G_{{\rm L}\tau}$, and $G_{{\rm R}\tau}$ so that 
the overproduction of T ($^3$He in later time) and D does not occur.
%%%%%%%%%%%%%%
%%%%%%%%%%%%%%
%It depends on which is more efficient, the catalyzed fusion processes \eqref{eq:cf} 
%or the $^4$He spallation processes \eqref{eq:spa} whether the overproduction 
%of $^3$He and D occur.
%%%%%%%%%%%%%%
%%%%%%%%%%%%%%

%\blue{
%%%%%%%%%%%%%%%
%%%%%%%%%%%%%%%
%Note here that the flavor mixings of the slepton do not affect timescales of all the exotic BBN processes.
%%%%%%%%%%%%%%%
%%%%%%%%%%%%%%%
%}

%%%%%%%%%%%%%%%%%%%%%%%%%%%%%%%%%%%%%%%%%%
%%%%%%%%%%%%%%%%%%%%%%%%%%%%%%%%%%%%%%%%%%
\subsubsection{Difference between the NMSSM and the MSSM} \label{sec:difference} %%%%%%%%%%%%
%%%%%%%%%%%%%%%%%%%%%%%%%%%%%%%%%%%%%%%%%%
%%%%%%%%%%%%%%%%%%%%%%%%%%%%%%%%%%%%%%%%%%

%%%%%%%%%%%%%%%
%%%%%%%%%%%%%%%
In the limit of $\lambda , \kappa \to 0$ fixing $\mu _{\rm eff}$, the NMSSM 
with bino-like LSP is reduced to the MSSM as long as we consider the exotic 
BBN processes~\cite{Jittoh:2010wh,Jittoh:2011ni,Kohri:2012gc}. 
%%%%%%%%%%%%%%%
In the singlino-like LSP scenario, the couplings $G_{{\rm L}f}$ and $G_{{\rm R}f}$ 
are in general smaller than those in the MSSM, and hence the slepton lifetime tends to be much longer than that in the MSSM.
%%%%%%%%%%%%%%%
The couplings $G_{{\rm L}\tau}$ and $G_{{\rm R}\tau}$ need to be large 
so that the timescales of the internal conversion processes \eqref{eq:ic} 
are sufficiently short to solve the $^7$Li problem.
%%%%%%%%%%%%%%%
%%%%%%%%%%%%%%%
In such situation, however, the timescales of the $^4$He spallation 
processes \eqref{eq:spa} are also short, and thus $^3$He and D can 
be overproduced.
%%%%%%%%%%%%%%%
%%%%%%%%%%%%%%%
We have to adjust the slepton lifetime to avoid the overproduction
taking the flavor mixings of the slepton into account.
Taking these facts into account, we will search for parameter sets which can solve 
the lithium problems along the strategy that we show in the next section.
%%%%%%%%%%%%%%
%%%%%%%%%%%%%%

%%%%%%%%%%%%%%%%%%%%%%%%%%%%%%%%%%%%%%%%%%
%%%%%%%%%%%%%%%%%%%%%%%%%%%%%%%%%%%%%%%%%%
%%%%%%%%%%%%%%%%%%%%%%%%%%%%%%%%%%%%%%%%%%
\section{Strategy} \label{sec:strategy} %%%%%%%%%%%%%%%%%%%%
%%%%%%%%%%%%%%%%%%%%%%%%%%%%%%%%%%%%%%%%%%
%%%%%%%%%%%%%%%%%%%%%%%%%%%%%%%%%%%%%%%%%%
%%%%%%%%%%%%%%%%%%%%%%%%%%%%%%%%%%%%%%%%%%

%%%%%%%%%%%%%%
%%%%%%%%%%%%%%
We acquire parameter sets giving observed light element abundances, 
the Higgs mass, and dark matter relic density according to the following strategy.
%%%%%%%%%%%%%%
%%%%%%%%%%%%%%

%%%%%%%%%%%%%%%%%%%%%%%%%%%%%%%%%%%%%%%%%%
\subsection{Search for candidate region on $\lambda$-$\kappa$ plane} \label{sec:lk-const} %%%
%%%%%%%%%%%%%%%%%%%%%%%%%%%%%%%%%%%%%%%%%%

%%%%%%%%%%%%%%
%%%%%%%%%%%%%%
%\blue{
%First, we constrain $\lambda$-$\kappa$ plane 
%}
%%%%%%%%%%%%%%
%%%%%%%%%%%%%%
We narrow the parameter space on $\lambda$-$\kappa$ plane requiring 
that the slepton lifetime, $\tau _{\tilde l}$, and the timescale of 
the internal conversion processes, $\tau _{\rm IC}$, are in range where the lithium 
problems can be solved.
%%%%%%%%%%%%%%
%%%%%%%%%%%%%%
Then the parameter region on $\lambda$-$\kappa$ plane can be constrained since 
the lifetime and the timescale depends on the couplings $G_{{\rm L}f}$ and $G_{{\rm R}f}$.
%%%%%%%%%%%%%%
%%%%%%%%%%%%%%
We require the following conditions to be satisfied when $\delta m$ is
around $0.1$~GeV which is favored in our scenario
~\cite{Jittoh:2007fr,Jittoh:2008eq,Jittoh:2010wh,Jittoh:2011ni,Kohri:2012gc}.
%%%%%%%%%%%%%%
%%%%%%%%%%%%%%

%%%%%%%%%%%%%%
%%%%%%%%%%%%%%
The requirement for the slepton lifetime is
\begin{align}
10^3~{\rm s} <& \tau _{\tilde l} < 10^5~{\rm s}. \label{eq:cond-sleptonlife}
\end{align}
%%%%%%%%%%%%%%
%%%%%%%%%%%%%%
As we mentioned in Sec.~\ref{sec:bbn}, the slepton lifetime must be at least 
longer than the timescale of bound-state formation for the internal conversion processes 
\eqref{eq:ic}, ${\cal O}(10^3)$~s, in order to obtain observed $^7$Li abundance.
%%%%%%%%%%%%%%
%%%%%%%%%%%%%%
The upper bound $10^5$~s comes from the requirement that the slepton has sufficient 
longevity to produce sufficient amount of $^6$Li avoiding the overproduction
through the catalyzed fusion process \eqref{eq:cf}.
%%%%%%%%%%%%%%
%%%%%%%%%%%%%%

%%%%%%%%%%%%%%
%%%%%%%%%%%%%%
We require the timescale of the internal conversion processes to be much shorter than
the slepton lifetime so that the internal conversion \eqref{eq:ic} works sufficiently and then solve 
the $^7$Li problem,
\begin{align}
\tau _{\rm IC} < 0.1 \tau _{\tilde l}. \label{eq:cond-ictime}
\end{align}
%%%%%%%%%%%%%%
%%%%%%%%%%%%%%
%The parameter dependence of the timescale is different from that of the slepton lifetime, 
%so we should require this condition separately from Eq.~\eqref{eq:cond-sleptonlife}.
Since the parameter dependence on the timescale is different from that of 
the slepton lifetime, we introduce the requirement Eq.~\eqref{eq:cond-ictime} 
independently from the requirement of the lifetime Eq.~\eqref{eq:cond-sleptonlife}.
%%%%%%%%%%%%%%
%%%%%%%%%%%%%%

%%%%%%%%%%%%%%
%%%%%%%%%%%%%%
In addition, we require that the fraction of singlino in the neutralino LSP is larger than 90$\%$,
\begin{align}
N_{1\tilde S}^2 > 0.9, \label{eq:cond-s-rate}
\end{align}
for the singlino-like LSP scenario and the fraction of bino in the neutralino LSP is larger than 90$\%$,
\begin{align}
N_{1\tilde B}^2 > 0.9, \label{eq:cond-b-rate}
\end{align}
for the bino-like LSP scenario.
%%%%%%%%%%%%%%
%%%%%%%%%%%%%%

%%%%%%%%%%%%%%
%%%%%%%%%%%%%%
%We find the region on $\lambda$-$\kappa$ plane where the above three conditions, 
%Eqs.~\eqref{eq:cond-sleptonlife}, \eqref{eq:cond-ictime}, and \eqref{eq:cond-s-rate} 
%(for singlino-like LSP) or \eqref{eq:cond-b-rate} (for bino-like LSP) are satisfied.
%%%%%%%%%%%%%%
%%%%%%%%%%%%%%

%%%%%%%%%%%%%%%%%%%%%%%%%%%%%%%%%%%%%%%%%%
\subsection{Selection of parameter sets} \label{sec:tools} %%%
%%%%%%%%%%%%%%%%%%%%%%%%%%%%%%%%%%%%%%%%%%

%%%%%%%%%%%%%%
%%%%%%%%%%%%%%
We impose constraints from the recent results of observational relic density 
of dark matter and experimental value of the Higgs mass in addition to 
those in Sec.~\ref{sec:lk-const}.
%%%%%%%%%%%%%%
%%%%%%%%%%%%%%
We use NMSSMTools 4.1.1
~\cite{Ellwanger:2004xm, Ellwanger:2005dv, Belanger:2005kh, Ellwanger:2006rn, Muhlleitner:2003vg}.
to calculate dark matter relic density and the Higgs mass.
%%%%%%%%%%%%%%
%%%%%%%%%%%%%%
\footnote{
In the neutralino sector, 
NMSSMTools includes 1-loop radiative corrections for $M_1, M_2$, 
and $\mu_{\rm eff}$ when it numerically diagonalize the mass matrix 
of the neutralino Eq.~\eqref{eq:massmatrix}.
%%%%%%%
%%%%%%%
On the other hand, we do not include the loop contributions in calculations 
of the neutralino masses and mixings for simplicity since the loop effects 
are negligible in our discussion.
}
%%%%%%%%%%%%%%
%%%%%%%%%%%%%%
Then we select several parameter sets from the region we obtained in 
the previous step.
%%%%%%%%%%%%%%
%%%%%%%%%%%%%%

%%%%%%%%%%%%%%
%%%%%%%%%%%%%%
The latest result for observed abundance of dark matter is reported by 
the Planck Collaboration
\begin{align}
0.1118 \leq \Omega _{\rm DM}h^2 \leq 0.1280
\label{eq:obs-dm}
\end{align}
at the $3\sigma$ level~\cite{Ade:2013zuv}, and we apply this result as a 
constraint.
%%%%%%%%%%%%%%
%%%%%%%%%%%%%%
%MicrOMEGAs~\cite{Belanger:2013oya} is implemented in NMSSMTools 
%for calculating the relic density.
In the calculation, we use MicrOMEGAs~\cite{Belanger:2013oya} included in NMSSMTools.
%%%%%%%%%%%%%%
%%%%%%%%%%%%%%

%%%%%%%%%%%%%%
%%%%%%%%%%%%%%
The latest experimental value of the Higgs mass is 
\begin{align}
m_h = 125.7\pm 0.3(\text{stat.}) \pm 0.3(\text{syst.})~{\rm GeV}
\end{align}
by the CMS Collaboration~\cite{CMS:yva}, and
\begin{align}
m_h = 125.5\pm 0.2(\text{stat.}) ^{+0.5}_{-0.6} (\text{syst.})~{\rm GeV}
\end{align}
by the ATLAS Collaboration~\cite{ATLAS:2013mma}, respectively.
%%%%%%%%%%%%%%
%%%%%%%%%%%%%%
There are several public codes for calculation of the Higgs mass, 
and it is known that uncertainty about $\pm 3$ GeV exists among their 
calculations
~\cite{Allanach:2001hm,Djouadi:2002nh,Allanach:2003jw,Allanach:2004rh,Cao:2012fz}.
%%%%%%%%%%%%%%
%%%%%%%%%%%%%%
Then we take the uncertainty into account in this study and require that
\begin{align}
m_h = 125.6 \pm 3.0~{\rm GeV}.
\label{eq:const-mhiggs}
\end{align}
%%%%%%%%%%%%%%
%%%%%%%%%%%%%%

%%%%%%%%%%%%%%%%%%%%%%%%%%%%%%%%%%%%%%%%%%
\subsection{Constraints from BBN} \label{sec:bbncalc} %%%
%%%%%%%%%%%%%%%%%%%%%%%%%%%%%%%%%%%%%%%%%%

We perform reaction network calculations for light elements including
the exotic nuclear-reactions with the bound-state effects.  The baryon
to photon ratio is taken to be $\eta =(6.04 \pm 0.08) \times
10^{-10}$~\cite{Ade:2013zuv}.  Then we obtain allowed regions in the
parameter space as a function of $\delta m$ to fit observational light
element abundances.

We adopt following observational bounds on light element abundances:
the lithium 7 to hydrogen ratio ${\rm Log}_{10}(^7\text{Li}/\text{H})
= -9.63 \pm 0.06$~\cite{Melendez:2004ni}, the lithium 6 to lithium 7
ratio $^6$Li/$^7$Li $= 0.046 \pm 0.022$~\cite{Asplund:2005yt}, the
deuterium to hydrogen ratio D/H = $(2.80 \pm 0.20) \times
10^{-5}$~\cite{Pettini:2008mq}, and the helium 3 to deuterium ratio
$^{3}$He/D $<$ 0.87 + 0.27~\cite{GG03}. It is notable that our
scenario does not change the abundance of $^4$He significantly.

For the moment, in order to obtain milder bounds as conservatively as
possible we do not adopt newer observational bounds on the abundance
of deuterium such as D/H = $(2.54 \pm 0.05) \times 10^{-5}$ and D/H =
$(2.53 \pm 0.04) \times 10^{-5}$. They were quite-recently reported by
Pettini and Cooke 2012~\cite{Pettini:2012ph} and Cooke et al
2013~\cite{Cooke:2013cba}, respectively. About such small errors, we
need third-party verifications as soon as possible.

%%% from Kohri et al. 2012 %%%%%%%%%%%%%%%%%%%%%%%%
%
%The $^7$Li  problem is
%a discrepancy between the observed abundance  of $^7$Li, e.g., ${\rm
%Log}_{10}(^7\text{Li}/\text{H}) = -9.63 \pm
%0.06$~\cite{Melendez:2004ni},  and a theoretical one predicted in the
%standard BBN,  ${\rm Log}_{10}(^7\text{Li}/\text{H}) = -9.35 \pm
%0.06$~\cite{Jittoh:2011ni}. 
%
% A fit by the observational abundance
%$^6$Li/$^7$Li $= 0.046 \pm 0.022$~\cite{Asplund:2005yt}   gives an
%allowed region in a parameter space of the mixing parameters. In
%addition,  both deuterium (D) and tritium (T) (or $^{3}$He after its
%decay) are produced nonthermally  by the latter processes.  By
%adopting recent observational  constraints   D/H = $(2.80 \pm 0.20)
%\times 10^{-5}$~\cite{Pettini:2008mq} and $^{3}$He/D $<$ 0.87 +
%0.27~\cite{GG03},  an upper bound on the number density of the slepton
%is obtained as a function of its lifetime.
%
%%%%%%%%%%%%%%%%%%%%%%%%%%%%%%%%%%%%%%

%%%%%%%%%%%%%%%%%%%%%%%%%%%%%%%%%%%%%%%%%%%%%%%%%%%%
%%%%%%%%%%%%%%%%%%%%%%%%%%%%%%%%%%%%%%%%%%%%%%%%%%%%
%%%%%%%%%%%%%%%%%%%%%%%%%%%%%%%%%%%%%%%%%%%%%%%%%%%%
\section{Results} \label{sec:results} %%%%%%%%%%%%%%%%%%%%%%%%%%%%%%%%%%
%%%%%%%%%%%%%%%%%%%%%%%%%%%%%%%%%%%%%%%%%%%%%%%%%%%%
%%%%%%%%%%%%%%%%%%%%%%%%%%%%%%%%%%%%%%%%%%%%%%%%%%%%
%%%%%%%%%%%%%%%%%%%%%%%%%%%%%%%%%%%%%%%%%%%%%%%%%%%%

%%%%%%%%%%%%%%
%%%%%%%%%%%%%%
Now we show the results following the strategy shown in the previous section.
%%%%%%%%%%%%%%
%%%%%%%%%%%%%%
The neutralino LSP is either singlino- or bino-like.
%We consider two scenarios where the neutralino LSP is singlino- and bino-like.
%%%%%%%%%%%%%%
%%%%%%%%%%%%%%
We have two types of phenomenologically favored parameter spaces in each case: 
one is the region where $\lambda$ and $\kappa$ are relatively small, and $\tan\beta$ 
is large; the other is the region where $\lambda$ and $\kappa$ are relatively large, 
and $\tan\beta$ is small.
%%%%%%%%%%%%%%
%%%%%%%%%%%%%%
Therefore, we have four cases.
%%%%%%%%%%%%%%
%%%%%%%%%%%%%%
However, we do not consider the case where the neutralino LSP is bino-like, 
$\lambda$ and $\kappa$ are relatively small, and $\tan\beta$ is large.
%%%%%%%%%%%%%%
%%%%%%%%%%%%%%
In the limit of $\lambda , \kappa \to 0$ with fixed $\mu _{\rm eff}$, the NMSSM is 
reduced to the MSSM, so the results are same as that 
in the MSSM~\cite{Jittoh:2010wh,Jittoh:2011ni,Kohri:2012gc}. 
%%%%%%%%%%%%%%
%%%%%%%%%%%%%%
We consider the other three cases in the followings.
%%%%%%%%%%%%%%
%%%%%%%%%%%%%%
%We consider three cases including the two different regions.
%%%%%%%%%%%%%%
%%%%%%%%%%%%%%

%%%%%%%%%%%%%%
%%%%%%%%%%%%%%
First we consider the case where the neutralino LSP is singlino-like, $\lambda$ and $\kappa$ are relatively small 
($\kappa \lesssim \lambda \ll 1$), and $\tan\beta$ is large~$(\gtrsim 30)$~\cite{Belanger:2005kh}.
%%%%%%%%%%%%%%
%%%%%%%%%%%%%%
%\red{
%In this case observed dark matter relic density can be obtained by 
%the coannihilation mechanism~\cite{Belanger:2005kh}.
%}
%%%%%%%%%%%%%%
%%%%%%%%%%%%%%
We take four points of parameter sets and denote them by SS-1, SS-2, SS-3, and SS-4, respectively.
%%%%%%%%%%%%%%
%%%%%%%%%%%%%%
Here ``SS" stands for the ``S"inglino-like neutralino LSP and ``S"mall couplings region.
%%%%%%%%%%%%%%
%%%%%%%%%%%%%%

%%%%%%%%%%%%%%
%%%%%%%%%%%%%%
Next we consider the case where the neutralino LSP is singlino-like,  $\lambda$ and $\kappa$ are relatively large
($0.5 \lesssim \lambda \lesssim 0.7$ and at most same order value of $\kappa \lesssim \lambda$) 
and $\tan\beta$ is small~$(\sim 2)$.
%%%%%%%%%%%%%%
%%%%%%%%%%%%%%
Large $\lambda$ with small $\tan\beta$ leads to large tree contributions to the Higgs mass 
(the second and the third terms in Eq.~\eqref{eq:mhiggs}) while it is small in the region in 
the first case.
%%%%%%%%%%%%%%
%%%%%%%%%%%%%%
Actually, $\lambda$ has its maximal value $\sim 0.7$ though it depends on 
$\kappa$ (see for example Table 1 in~\cite{King:2012is}).
%%%%%%%%%%%%%%
%%%%%%%%%%%%%%
This comes from the requirement to avoid the Landau pole up to the GUT scale.
%%%%%%%%%%%%%%
%%%%%%%%%%%%%%
%We should note that $\lambda$ has its maximal value when we require to avoid Landau 
%pole up to the GUT scale, and the value depends on $\kappa$ and $\tan \beta$.
%%%%%%%%%%%%%%
%%%%%%%%%%%%%%
%The maximal value tends to be large with decreasing $\kappa$ 
%(see for example Table 1 in~\cite{King:2012is}) and is at most $\sim 0.7$.
%%%%%%%%%%%%%%
%%%%%%%%%%%%%%
The value of $\kappa$ should not be much smaller than $\lambda$ in order to obtain large tree contributions to the Higgs mass;
in this case the negative contribution (the third term in Eq.~\eqref{eq:mhiggs}) do not become significant.
%%%%%%%%%%%%%%
%%%%%%%%%%%%%%
Therefore, we investigate the range from at least $0.5$ to $0.7$ for $\lambda$, 
and the same order of magnitude but relatively small range of $\kappa$. 
%%%%%%%%%%%%%%
%%%%%%%%%%%%%%
We take $\tan\beta \sim 2$ since the tree contributions in Eq.~\eqref{eq:mhiggs} are 
maximized around the value with large $\lambda$.
%%%%%%%%%%%%%%
%%%%%%%%%%%%%%
We take four points of parameter sets and denote them by SL-1, SL-2, SL-3, and SL-4, respectively. 
%%%%%%%%%%%%%%
%%%%%%%%%%%%%%
Here ``SL" stands for the ``S"inglino-like neutralino LSP and ``L"arge couplings region.
%%%%%%%%%%%%%%
%%%%%%%%%%%%%%

%%%%%%%%%%%%%%
%%%%%%%%%%%%%%
In the third case we consider the bino-like neutralino LSP and search the same 
region on $\lambda$-$\kappa$ plane as in the second case~\cite{Belanger:2005kh, King:2012is}.
%%%%%%%%%%%%%%
%%%%%%%%%%%%%%
In the NMSSM the tree contributions to the Higgs mass are large, which
is different from those in the MSSM where well tuned 1-loop contribution is required.
%%%%%%%%%%%%%%
%%%%%%%%%%%%%%
We take four points of parameter sets and denote them by BL-1, BL-2, BL-3, and BL-4, respectively.
%%%%%%%%%%%%%%
%%%%%%%%%%%%%%
Here ``BL" stands for the ``B"ino-like neutralino LSP and ``L"arge couplings region.
%%%%%%%%%%%%%%
%%%%%%%%%%%%%%

%%%%%%%%%%%%%%
%%%%%%%%%%%%%%
In the following results, we fix the parameters as $m_{\tilde \chi ^0_1}=350$~GeV, 
and $\sin \theta _f=0.8$ for $f=e,\mu$, and $\tau$.
%%%%%%%%%%%%%%
%%%%%%%%%%%%%%

%%%%%%%%%%%%%%%%%%%%%%%%%%%%%%%%%%%%%%%%%%%%%%%%
%%%%%%%%%%%%%%%%%%%%%%%%%%%%%%%%%%%%%%%%%%%%%%%%
\subsection{Singlino-like neutralino LSP; small $\lambda$-$\kappa$ region with large $\tan\beta$} \label{sec:lk-s-small}
 %%%%%%%%%%%%%%%%%%%%%%%%%%%%%%%%%%%%%%%%%%%%%%%%
%%%%%%%%%%%%%%%%%%%%%%%%%%%%%%%%%%%%%%%%%%%%%%%%

%%%%%%%%%%%%%%%
%%%%%%%%%%%%%%%
%\red{We show the results of the first case shown in the previous section.}
%%%%%%%%%%%%%%%
%%%%%%%%%%%%%%%
%We see that observed Higgs mass is reproduced by the
%significant contribution of the 1-loop term as in the MSSM. 
%%%%%%%%%%%%%%%
%%%%%%%%%%%%%%%

%%%%%%%%%%%%%%%%%%%%%%%%%%%%%%%%%%%%%%%%%%%%%%%%
\subsubsection{Benchmark points} \label{sec:bmp-s-small}%%%%%%%%%%%%%%%
%%%%%%%%%%%%%%%%%%%%%%%%%%%%%%%%%%%%%%%%%%%%%%%%

%%%%%%%%%%%%%%
%%%%%%%%%%%%%%
Figure~\ref{fig:allowed-lk-s-small} shows the region on $\lambda$-$\kappa$ plane 
with large $\tan \beta$ where the requirements, Eqs.~\eqref{eq:cond-sleptonlife}-\eqref{eq:cond-s-rate}, 
are satisfied.
%%%%%%%%%%%%%%
%%%%%%%%%%%%%%
The left and right panels show the results for $c_e=2\times 10^{-9}$ and  $10^{-9}$, 
respectively.
%%%%%%%%%%%%%%
%%%%%%%%%%%%%%
We discuss only the case for $\tan\beta=30$ in each panel 
since we obtain almost same results for $\tan\beta>30$.
%%%%%%%%%%%%%%
%%%%%%%%%%%%%%
The mass of the lightest neutralino, $m_{\tilde \chi ^0_1}\simeq \mu_{\kappa}$ ,
is almost equal to that of the next-to-lightest one, $m_{\tilde \chi^0_2} \simeq \mu_{\rm eff}$, 
on the dotted line.
%%%%%%%%%%%%%%
%%%%%%%%%%%%%%
Above the dotted line, the singlino-like neutralino is no longer the lightest one.
%%%%%%%%%%%%%%
%%%%%%%%%%%%%%

%%%%%%%%%%%%%%
%%%%%%%%%%%%%%
We can see that larger $\lambda$ and smaller $\kappa$ are allowed by 
larger $c_e$ with fixed $\tan \beta$.
%%%%%%%%%%%%%%
%%%%%%%%%%%%%%
This is explained by the requirement Eq.~\eqref{eq:cond-sleptonlife} and the 
dependence of the couplings, Eqs.~\eqref{eq:gl} and \eqref{eq:gr}, on 
$\lambda$ and $\kappa$.
%%%%%%%%%%%%%%
%%%%%%%%%%%%%%
These couplings become large as $\kappa$ increase, and/or $\lambda$ decreases.
%%%%%%%%%%%%%%
%%%%%%%%%%%%%%
The slepton lifetime becomes short as the couplings and/or $c_e$ increase.
%%%%%%%%%%%%%%
%%%%%%%%%%%%%%
Therefore, larger $c_e$ gives shorter slepton lifetime and 
allows the region with larger $\lambda$ and smaller $\kappa$.
%%%%%%%%%%%%%%
%%%%%%%%%%%%%%
%Similarly, larger $\lambda$ and smaller $\kappa$ are allowed by larger 
%$c_e$ with fixed $\tan\beta$ by the same reason.
%%%%%%%%%%%%%%
%%%%%%%%%%%%%%

To check which requirements determine the favored region in Fig.~\ref{fig:allowed-lk-s-small},
we draw Fig.~\ref{fig:allowed-lk-s-small-grad} to show the distribution of the quantities which are 
relevant to the requirements, Eqs.~\eqref{eq:cond-sleptonlife}-\eqref{eq:cond-s-rate}.
The parameters used are  $c_e=2\times 10^{-9}$, $m_{\tilde \chi^0_1}=350$~GeV, 
$\delta m = 0.1$~GeV, $\sin \theta _f=0.8, \tan\beta =30, M_1=500$~GeV, and $M_2=1000$~GeV.
% (the red region in the left panel of Fig.~\ref{fig:allowed-lk-s-small}).
%%%%%%%%%%%%%%
%%%%%%%%%%%%%%
From the result we can see the favored region is determined by the requirement for the slepton lifetime of Eq.~\eqref{eq:cond-sleptonlife}.
%%%%%%%%%%%%%%
%%%%%%%%%%%%%%

%
%The top-left panel shows the distribution of the slepton lifetime, $\tau _{\tilde l}$, 
%in the region.
%%%%%%%%%%%%%%%
%%%%%%%%%%%%%%%
%We can see that both top and bottom side edges of the region are determined by 
%the requirement for the slepton lifetime of Eq.~\eqref{eq:cond-sleptonlife}.
%%%%%%%%%%%%%%% 
%%%%%%%%%%%%%%%
%The top-right and bottom-left panels show that the requirement Eq.~\eqref{eq:cond-ictime} is satisfied.
%%%%%%%%%%%%%%%条件を満たすー＞internal conversion works well\UTF{00A0}%
%%%%%%%%%%%%%%%
%%If the both values are larger than unity, Eq.~\eqref{eq:cond-ictime} is satisfied,
%%and actually it is realized at all points in the region.
%%%%%%%%%%%%%%%
%%%%%%%%%%%%%%%
%The two panels are almost same, since timescales of the internal conversion 
%processes are almost same.
%%%%%%%%%%%%%%%
%%%%%%%%%%%%%%%
%The bottom-right panel shows the \blue{fraction} of singlino component in the lightest 
%neutralino, $N_{1\tilde S}^2$.
%%%%%%%%%%%%%%%
%%%%%%%%%%%%%%%
%The requirement Eq.~\eqref{eq:cond-s-rate} is also satisfied at all points in the region.
%%%%%%%%%%%%%%%
%%%%%%%%%%%%%%

%%%%%%%%%%%%%%
%%%%%%%%%%%%%%
We take four reference points in the favored region for $\tan\beta=30$
(red region in Fig.~\ref{fig:allowed-lk-s-small}) as shown in Table~\ref{tab:points1}.
%%%%%%%%%%%%%%
%%%%%%%%%%%%%%
Table~\ref{tab:points-s-small} shows the spectra and observables at 
these points.
%%%%%%%%%%%%%%
%%%%%%%%%%%%%%
We omit small flavor mixing of the slepton in this calculation, and thus 
SS-1 and SS-2 give same results.
%%%%%%%%%%%%%%
%%%%%%%%%%%%%%
All the dimensionful values are shown in GeV.
%%%%%%%%%%%%%%
%%%%%%%%%%%%%%
The top rows show input parameters.
%%%%%%%%%%%%%%
%%%%%%%%%%%%%%
We assume the relations for gaugino masses
\begin{align}
M_1=M_2/2, M_3=3M_2,
\end{align}
similar to the GUT relation, and, for simplicity, universal input soft-masses for each squark and slepton, 
\begin{align}
&m_{\tilde Q_{1,2}} =m_{\tilde Q_{3}}=m_{\tilde U_{1,2}}=m_{\tilde U_{3}}=m_{\tilde D_{1,2}}=m_{\tilde D_{3}}, \notag \\
&m_{\tilde L_{1,2}}=m_{\tilde L_{3}},
~m_{\tilde E_{1,2}}=m_{\tilde E_{3}}.
\end{align}
%%%%%%%%%%%%%%
%%%%%%%%%%%%%%
The middle rows show output spectra.
%%%%%%%%%%%%%%
%%%%%%%%%%%%%%
Every points give the observed Higgs mass~Eq.\eqref{eq:const-mhiggs}.
%%%%%%%%%%%%%%
%%%%%%%%%%%%%%
At these points, $m_h$ receives significant contribution from the 1-loop 
correction (the fourth term in Eq.~\eqref{eq:mhiggs}) by the maximal mixing 
and large stop masses because the tree contributions (the second and third terms in 
Eq.~\eqref{eq:mhiggs}) are small.
%%%%%%%%%%%%%%
%%%%%%%%%%%%%%
%In this sense this case is the MSSM-like though the lightest neutralino is singlino-like.
%%%%%%%%%%%%%%
%%%%%%%%%%%%%%

%%%%%%%%%%%%%%
%%%%%%%%%%%%%%
In the bottom rows, we show relic density of the lightest neutralino, spin-independent 
cross section between the lightest neutralino and nucleon, the SUSY contribution 
to the muon anomalous magnetic moment, and the branching ratios of rare decays 
$B_s \to \mu ^+ \mu ^-$ and $b\to s\gamma$, from top to bottom.
%%%%%%%%%%%%%%
%%%%%%%%%%%%%%
At each point, the dark matter relic density is in range of the measured value~\cite{Ade:2013zuv}.
%%%%%%%%%%%%%%
%%%%%%%%%%%%%%
The spin-independent cross section at each point is about four orders of magnitude 
smaller than the present experimental bound~\cite{Akerib:2013tjd}. 
%%%%%%%%%%%%%%
%%%%%%%%%%%%%%
The latest experimental result for the SUSY contribution to the anomalous magnetic 
moment is, 
\begin{align}
\delta a_{\mu}
= a_{\mu}^{\rm exp} -a_{\mu}^{\rm SM}
= (26.1\pm 8.0)\times 10^{-10},
\label{eq:b-2}
\end{align}
where $a_{\mu} \equiv (g-2)_{\mu}/2$~\cite{Bennett:2006fi, Davier:2010nc, Hagiwara:2011af}.
%%%%%%%%%%%%%%
%%%%%%%%%%%%%%
For the branching ratio of $B_s \to \mu ^+ \mu ^-$, the CMS Collaboration recently reported in
Ref.~\cite{Chatrchyan:2013bka} as
\begin{align}
{\rm BR}(B_s \to \mu ^+ \mu ^-)
 = 3.0^{+1.0}_{-0.9} \times 10 ^{-9}
\label{eq:bsmumu-cms}
\end{align}
and also the LHCb Collaboration reported in Ref.~\cite{Aaij:2013aka} as
\begin{align}
{\rm BR}(B_s \to \mu ^+ \mu ^-)
 = 2.9^{+1.1}_{-1.0}({\rm stat.})^{+0.3}_{-0.1}({\rm syst.})\times 10^{-9}.
\label{eq:bsmumu-lhcb}
\end{align}
The branching ratio of $B\to X_s \gamma$~\cite{Amhis:2012bh} is given by
\begin{align}
{\rm BR}(B \to X_s\gamma)
 = (3.43 \pm 0.21 \pm 0.07) \times 10^{-4}.
\label{eq:bsgamma}
\end{align}
%%%%%%%%%%%%%%
%%%%%%%%%%%%%%
At each point, $\delta a_{\mu}$ and ${\rm BR}(B_s \to \mu ^+ \mu ^-)$ are 
in the ranges of 2$\sigma$ and 1$\sigma$.
%%%%%%%%%%%%%%
%%%%%%%%%%%%%%
The branching ratio of $b \to s\gamma$ is in the 2$\sigma$ range at SS-3 and SS-4.
%%%%%%%%%%%%%%
%%%%%%%%%%%%%%

%%%%%%%%%%%%%%%%%%%%%%%%%%%%%%%%%%%%%%%%%%%%%%%%
\subsubsection{BBN results at the benchmark points} \label{sec:bbn-s-small}%%%%%%
%%%%%%%%%%%%%%%%%%%%%%%%%%%%%%%%%%%%%%%%%%%%%%%%

%%%%%%%%%%%%%%
%%%%%%%%%%%%%%
The left panels in Fig.~\ref{fig:small-lk-s_bbn1} show 
the slepton lifetime $\tau _{\tilde l}$~(red-solid line; ``slepton lifetime"), 
the timescales of the internal conversion processes
\eqref{eq:icbe}~(blue-solid line; ``$^7$Be$\to$$^7$Li"), 
\eqref{eq:icli}~(blue-dash-dotted line; ``$^7$Li$\to$$^7$He"), 
the $^4$He spallation processes 
\eqref{eq:tn}~(brown-solid line; ``tn"), 
\eqref{eq:dnn}~(brown-dashed line; ``dnn"), and 
\eqref{eq:pnnn}~(brown-dash-dotted line; ``pnnn"), 
as a function of the mass difference between the slepton and the neutralino 
at SS-1, SS-2, SS-3, and SS-4 from top to bottom, respectively.
%%%%%%%%%%%%%%
%%%%%%%%%%%%%%
The horizontal black-dashed line represents the timescale of the catalyzed 
fusion process \eqref{eq:cf}~\cite{Hamaguchi:2007mp} at the temperature $T=5$~keV 
($5\times10^4$~s) when ($^4$He~$\tilde l^-$) is formed.
%%%%%%%%%%%%%%
%%%%%%%%%%%%%%

%%%%%%%%%%%%%%
%%%%%%%%%%%%%%
We show the allowed regions in the right panels of Fig.~\ref{fig:small-lk-s_bbn1} 
which we obtain by comparing theoretical values to observational ones for light element abundances
at SS-1, SS-2, SS-3, and SS-4 from top to bottom, respectively.
%%%%%%%%%%%%%%
%%%%%%%%%%%%%%
Horizontal axis is the mass difference between the slepton NLSP and 
the neutralino LSP, and vertical axis is the yield value of the slepton at the beginning of the BBN, 
$Y_{\tilde l^-}=n_{\tilde l^-}/s$, where $n_{\tilde l^-}$ is the number density
of the slepton and $s$ is the entropy density.
%%%%%%%%%%%%%%
%%%%%%%%%%%%%%
The regions surrounded by magenta-dotted(-solid) lines are allowed by observed 
$^7$Li/H abundance at 2$\sigma$(3$\sigma$).
%%%%%%%%%%%%%%
%%%%%%%%%%%%%%
The regions between the blue-solid line and the blue region are allowed by 
observed $^6$Li/$^7$Li abundance at 2$\sigma$.
%%%%%%%%%%%%%%
%%%%%%%%%%%%%%
The orange-solid lines (``Theoretical") represent the yield value of the slepton calculated with the parameters in Table.~\ref{tab:points-s-small}.
%%%%%%%%%%%%%%
%%%%%%%%%%%%%%
The colored regions are excluded for 
$^6$Li/$^7$Li (blue region; ``$^6$Li/$^7$Li excluded"), 
$^3$He/D (red region; ``$^3$He/D excluded"), and 
D/H (cyan region; ``D/H excluded"), respectively. 
%%%%%%%%%%%%%%
%%%%%%%%%%%%%%
The shaded and dotted regions are allowed by only $^7$Li/H (3$\sigma$) 
and both $^7$Li/H (3$\sigma$) and $^6$Li/$^7$Li (2$\sigma$), respectively,
since $Y_{\tilde l^-}$ can be reduced by changing $c_{\mu}$ and $\theta _f$. 
%%%%%%%%%%%%%%
%%%%%%%%%%%%%%
At each point, we obtain allowed region from  
$^7$Li/H (3$\sigma$) and $^6$Li/$^7$Li (2$\sigma$) simultaneously.
%%%%%%%%%%%%%%
%%%%%%%%%%%%%%

%%%%%%%%%%%%%%%%%%%%%%%%%%%%%%%%%%%%%%%%%%%%%%%%%%%
%%% Sugai modified below sentence %%%%%%%%%%%%%%%%%
%%%%%%%%%%%%%%%%%%%%%%%%%%%%%%%%%%%%%%%%%%%%%%%%%%%

We see in the right panels of Fig.~\ref{fig:small-lk-s_bbn1} that the allowed region 
from $^7$Li/H on the curve of the slepton
yield value is in the range of $\delta m \sim (0.04, 0.07)$, $(0.07, 0.1)$, $(0.07, 0.1)$, and $0.1$~GeV at the points SS-1, 
SS-2, SS-3, and SS-4 respectively.
%We also see that the bounds from $^3$He and D as well as
%$^6$Li exclude larger parameter space at SS-2 than at SS-1.
%%%%%%%%%%%%%%
%%%%%%%%%%%%%%
%
We note that the catalyzed fusion process affects the abundance of the $^6$Li in these allowed regions 
even if the timescale of the reaction is much longer than 
the slepton lifetime and the timescales of the $^4$He spallation processes.  
This is because the yield value of the slepton is much larger than observed abundance of the $^6$Li.  
We roughly calculate a yield value of the $^6$Li through the catalyzed fusion as follows;
\begin{equation}
  \Delta Y _{^{6}\text{Li}} = 
  Y _{\tilde l^-} e ^{- \tau _{\text{B.F.}}/ \tau _{\tilde l} } 
  \frac{\Gamma _{\text{C.F.}}}{\Gamma _{\text{Sp.}} + \Gamma _{\tilde l} + \Gamma _{\text{C.F.}}}, \label{eq:produced}
\end{equation}
where 
$\Delta Y _{^{6} \text{Li}}$ is the yield value of the $^{6}$Li, 
$\tau _{\text{B.F.}}$ is a timescale of ($^{4}\text{He}~\tilde l^-$) formation,
$\Gamma _{\tilde l}$ is a decay rate of the slepton, 
$\Gamma _{\text{Sp.}}$ and $\Gamma _{\text{C.F.}}$ are a reaction rate of the $^{4}$He spallation processes and the catalyzed fusion process, respectively.  
In Eq~\eqref{eq:produced}, $Y _{\tilde l^-} e ^{- \tau _{\text{B.F.}} / \tau _{\tilde l}}$ represents a yield value of the bound state, 
and the factor at the last of the right hand side is the branching ratio of the bound state to the catalyzed fusion reaction.  
Here we consider the predicted value of the $^{6}\text{Li}$ along with the slepton yield value which we calculate with the parameters shown in Table~\ref{tab:points-s-small}.  
We put the constraint from the observed value on the calculated value as follows;
\begin{equation}
  \frac{Y _{^{6}\text{Li,Ob.}}}{Y _{\tilde l^-}} 
  \simeq 10 ^{-8} 
  {\gsim} 
  e ^{- \tau _{\text{B.F.}}/ \tau _{\tilde l} } \frac{\Gamma _{\text{C.F.}}}{\Gamma _{\text{Sp.}} + \Gamma _{\tilde l} + \Gamma _{\text{C.F.}}}, \label{eq:deltaLi6}
\end{equation}
where we choose the yield value as $Y _{\tilde l} \simeq 10 ^{-13}$ and $Y _{^{6}\text{Li,Ob.}} \simeq \mathcal{O}(10^{-21})$ which is the observed yield value of $^6$Li~\cite{Asplund:2005yt}.  
We take the timescale of the bound state formation $\tau_{\text{B.F.}}\simeq 5 \times 10 ^{4}$ s (see the Fig.~1 of Ref.~\cite{Jittoh:2008eq}), 
and $\Gamma _{\text{C.F.}} \simeq \mathcal{O}(10 ^{-10})$~s$^{-1}$.  
We see that $\tau _{\tilde l} \simeq 10 ^{4}$~s and $ \Gamma _{\text{Sp.}} \simeq 10^{-5}$~s$^{-1}$ in the allowed region 
at the left panels of Fig.~\ref{fig:small-lk-s_bbn1}.  
We confirm that the calculated yield value of $^6$Li is consistent with that of the observational value at the allowed region.

%%%%%%%%%%%%%%
%%%%%%%%%%%%%%
We also see that the allowed range of the mass difference is different in each point.
This difference can be explained as follows.
The selectron mixing is smaller at SS-2 than at SS-1.
The smaller mixing results in smaller coupling $G_{L,Re}$ and hence longer slepton lifetimes than
that at SS-1
shown in the left panels in Fig.~\ref{fig:small-lk-s_bbn1}.
The longer the lifetime is, the more the exotic BBN reactions occur because a larger number of the
sleptons
remains until they form bound state with $^7$Be.
More $^7$Be are destroyed by the internal conversion processes at SS-2 than at SS-1.
In such a situation, the yield value of the slepton required by observed $^7$Li abundance can be
smaller for
the same $\delta m$. In fact, comparing the right panels of Fig.~\ref{fig:small-lk-s_bbn1} of SS-1
and SS-2,
we see that the $2$ and $3~\sigma$ lines of $^7$Li/H allows lower yield values and larger
$\delta m$ at SS-2.
%In general, the longer the lifetime is, the more the slepton survives until the BBN era.
%Then, the exotic BBN reactions occur more because a larger number of nuclei is
%captured by the sleptons.
This result also can be understood in terms of the slepton lifetime.
The lifetime must be adjusted to a certain range to solve the Li problems.
As mentioned in Sec.~\ref{sec:slepton}, the lifetime is determined by both $G_{L,Re}$ and $\delta m$.
 For a fixed lifetime that can explain the observed $^7$Li abundance,  $\delta m$ is larger when
$G_{L,Re}$ is smaller.
As a result, the allowed region extends to larger $\delta m$ at SS-2 than at SS-1.
For a longer slepton lifetime, on the other hand, more $^3$He and D are produced by the $^4$He spallation processes, and more $^6$Li is produced by the catalyzed fusion process.
%it means that the internal conversion processes
%difference
%is required to obtain observed abundance of $^7$Li/H
%through the internal conversion processes.
%%%%%%%%%%%%%%
%%%%%%%%%%%%%%
%Thus the slepton lifetime at SS-2 is longer than the one at SS-1, and
%the allowed region from $^7$Li/H is obtained for larger $\delta m$.
%%%%%%%%%%%%%%
%%%%%%%%%%%%%%
Thus, the excluded regions due to the $^4$He spallation processes
(red and cyan regions) and the catalyzed fusion process (blue region) are large compared to those at SS-1.
%
%
%It is explained by the difference of the slepton lifetime for SS-1 and SS-2
%because the timescales of the processes are almost the same at both points
%shown in the left panels of Fig.~\ref{fig:small-lk-s_bbn1}.
%%%%%%%%%%%%%%%
%%%%%%%%%%%%%%%
%%The longer the slepton lifetime, the larger number of the sleptons can form bound
%%state with nuclei.
%%%%%%%%%%%%%%%
%%%%%%%%%%%%%%%
%%Then the number of spallation products tends to be large.
%%%%%%%%%%%%%%%
%%%%%%%%%%%%%%%
%Therefore, D, $^3$He and $^6$Li are overproduced for same slepton
%yield values but  longer lifetimes. Then, the constraints from these nuclei abundances
%exclude a larger region at SS-2.
%%%%%%%%%%%%%%
%%%%%%%%%%%%%%

%%%%%%%%%%%%%%
%%%%%%%%%%%%%%
At the point SS-3, $\lambda$ is taken slightly larger than the one at SS-1 while $c_e$ and $\kappa$ are
taken the same.
We see that the allowed region is in the range of $\delta m \sim (0.07, 0.1)$ GeV and is larger than that at SS-1.
It is because the lifetime is longer than that at SS-1.
However, the reason to make the lifetime longer is different from the case of SS-2.
%%%%%%%%%%%%%%
%%%%%%%%%%%%%%
At SS-3, larger $\lambda$ gives longer slepton lifetime (see top-left panel of
Fig.~\ref{fig:allowed-lk-s-small}),
because $G_{L,Re}$ become smaller than those at SS-1 as shown in Table~\ref{tab:points-s-small}, due to the small mixing weights $N_{1}$'s in the lightest neutralino.
In Eq.~\eqref{eq:slike-chi}, the mixing weights are inversely proportional to $\mu_{\kappa}^2 - \mu_{\rm eff}^2$.
In the singlino-like LSP case, $\mu_{\kappa}$ and $\mu_{\rm eff}$, hence $\kappa$ and $\lambda$,
must be tuned so that $G_{L,R}$ are enough large.
 % to obtain short timescales of the internal conversion processes. 
 In fact, $\kappa$ and $\lambda$ are tuned well at SS-1. 
 Taking $\lambda$ larger than that at SS-1 even by a few \%, 
$G_{L,Re}$ become smaller and hence the lifetime becomes longer.
 At the same time,  $G_{L,R\tau}$ become smaller. 
It is still enough large to reduce $^7$Be by the internal conversion processes 
but is small for the $^4$He spallation processes. Thus, the excluded regions by 
$^3$He and D are narrower than those at SS-2. 
 
%  and $1$ \% of  changes
% in these parameters results in the increase of the lifetime by a factor.
% The excluded regions by D and $^3$He are not so large

%%%%%%%%%%%%%%
%%%%%%%%%%%%%%
%On the other hand, the slepton lifetime at SS-3 is shorter than the one at SS-2.
%%%%%%%%%%%%%%
%%%%%%%%%%%%%%
%Therefore, the allowed region from $^7$Li/H appears in the range of
%smaller $\delta m$, and the excluded regions for $^3$He/D and D/H are
%not extended downward compared to those at SS-2.
%%%%%%%%%%%%%%
%%%%%%%%%%%%%%

%%%%%%%%%%%%%%
%%%%%%%%%%%%%%
At SS-4, $\kappa$ is taken smaller by $1$ \% than the one at SS-1 while the other parameters
are taken the same. 
Again, due to the smaller $\kappa$,  the lifetime of the slepton is longer than at SS-1, and indeed
is the longest among
the four parameters (see the left panels in Fig.~\ref{fig:small-lk-s_bbn1}).
%%%%%%%%%%%%%%
%%%%%%%%%%%%%%
Then $^3$He and D are too much produced even for smaller slepton yield values
so that the excluded region due to D/H narrows the allowed region.
%%%%%%%%%%%%%%
%%%%%%%%%%%%%%
%As a result, the excluded region for D/H overlaps all the allowed region
%for $^7$Li/H on the theoretical curve for the slepton yield value.
%%%%%%%%%%%%%%
%%%%%%%%%%%%%%
Furthermore, the timescales of the internal conversion processes
\eqref{eq:ic} are longer than those at the other points.
%%%%%%%%%%%%%%
%%%%%%%%%%%%%%
Thus more slepton yield value is necessary to obtain the observed abundance of $^7$Li, 
and hence the allowed region at SS-4 is narrower.
%, even if the timescales of
%the $^4$He spallation processes \eqref{eq:cf} are long enough to
%lift up the excluded regions for $^3$He/D and D/H.
%%%%%%%%%%%%%%
%%%%%%%%%%%%%%

In the end, it is important to notice that in the singlino-like LSP case, $\kappa$ and $\lambda$
must be well tuned
to obtain fast internal conversion processes. As we have shown, only a few \% of difference in the
parameters
drastically change the results. Our scenario is thus very predictable this tuning.

\clearpage
%%%%%%%%%%%%%%
%%%%%%%%%%%%%%
\begin{widetext}
\begin{center}
\begin{figure}[h]
%%%%%%
%%%%%%
\begin{tabular}{l}
\hspace{-10mm}
\vspace{0mm}
\begin{minipage}{83mm}\vspace{-28mm}
\begin{center}
 \includegraphics[width=9.2cm,clip]{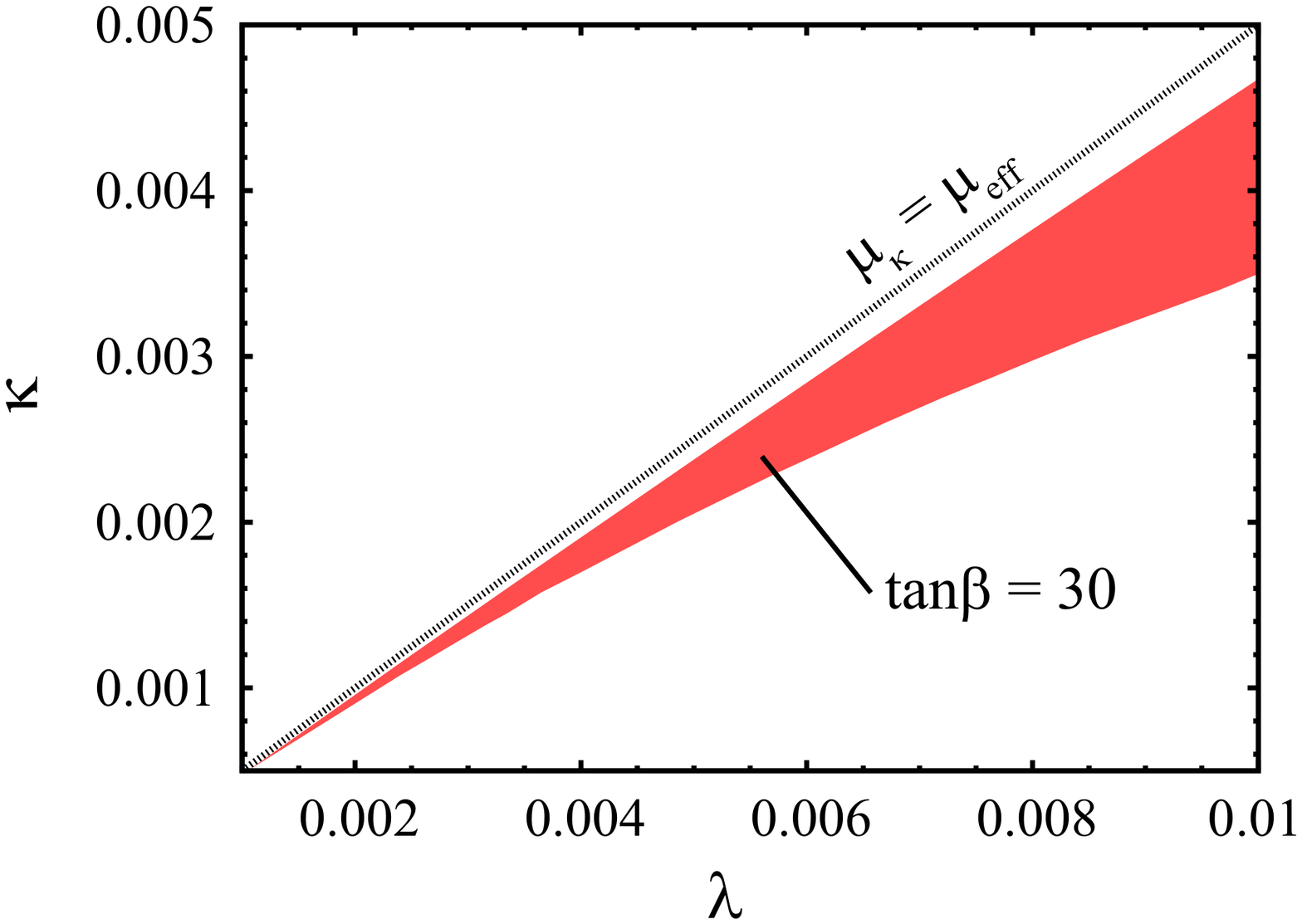} 
\end{center}
\end{minipage}
\hspace{2mm}
\begin{minipage}{83mm}\vspace{-28mm}
\begin{center}
 \includegraphics[width=9cm,clip]{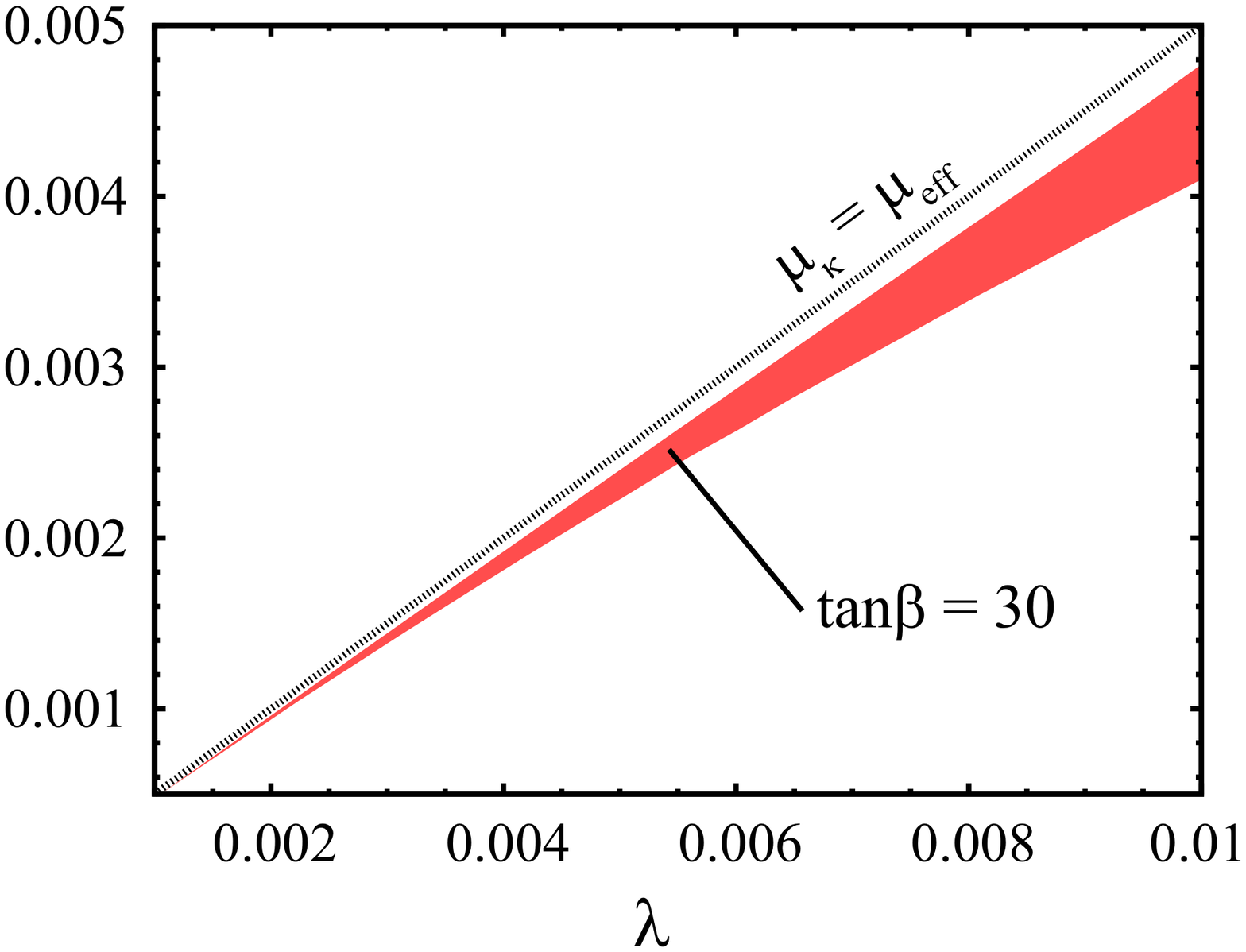}
\end{center}
\end{minipage}
\end{tabular}
\vspace{-33mm}
\caption{
Favored region (red region; ``$\tan \beta =30$") in terms of requirements Eqs.~\eqref{eq:cond-sleptonlife}-\eqref{eq:cond-s-rate} 
in $\lambda$-$\kappa$ plane.
%%%%%%%%%%%%%%
%%%%%%%%%%%%%%
Parameters are taken as $\tan \beta=30$ and $c_e=2\times10^{-9}$(left panel) and $c_e=10^{-9}$(right panel).
%%%%%%%%%%%%%%
%%%%%%%%%%%%%%
Other parameters are fixed as $m_{\tilde \chi ^0_1}=350$~GeV, $\delta m=0.1$~GeV, 
$\sin \theta _f=0.8$, $M_1=500$~GeV, and $M_2=1000$~GeV.
%%%%%%%%%%%%%%
%%%%%%%%%%%%%%
%Red region is the allowed region.
%%%%%%%%%%%%%%
%%%%%%%%%%%%%%
The singlino-like neutralino is no longer the lightest one above the dotted line.
}
\label{fig:allowed-lk-s-small}
%%%%%%
%%%%%%
\begin{tabular}{l}
\hspace{-9mm}
\begin{minipage}{80mm}\vspace{-38mm}
\begin{center}
 \includegraphics[width=9.4cm,clip]{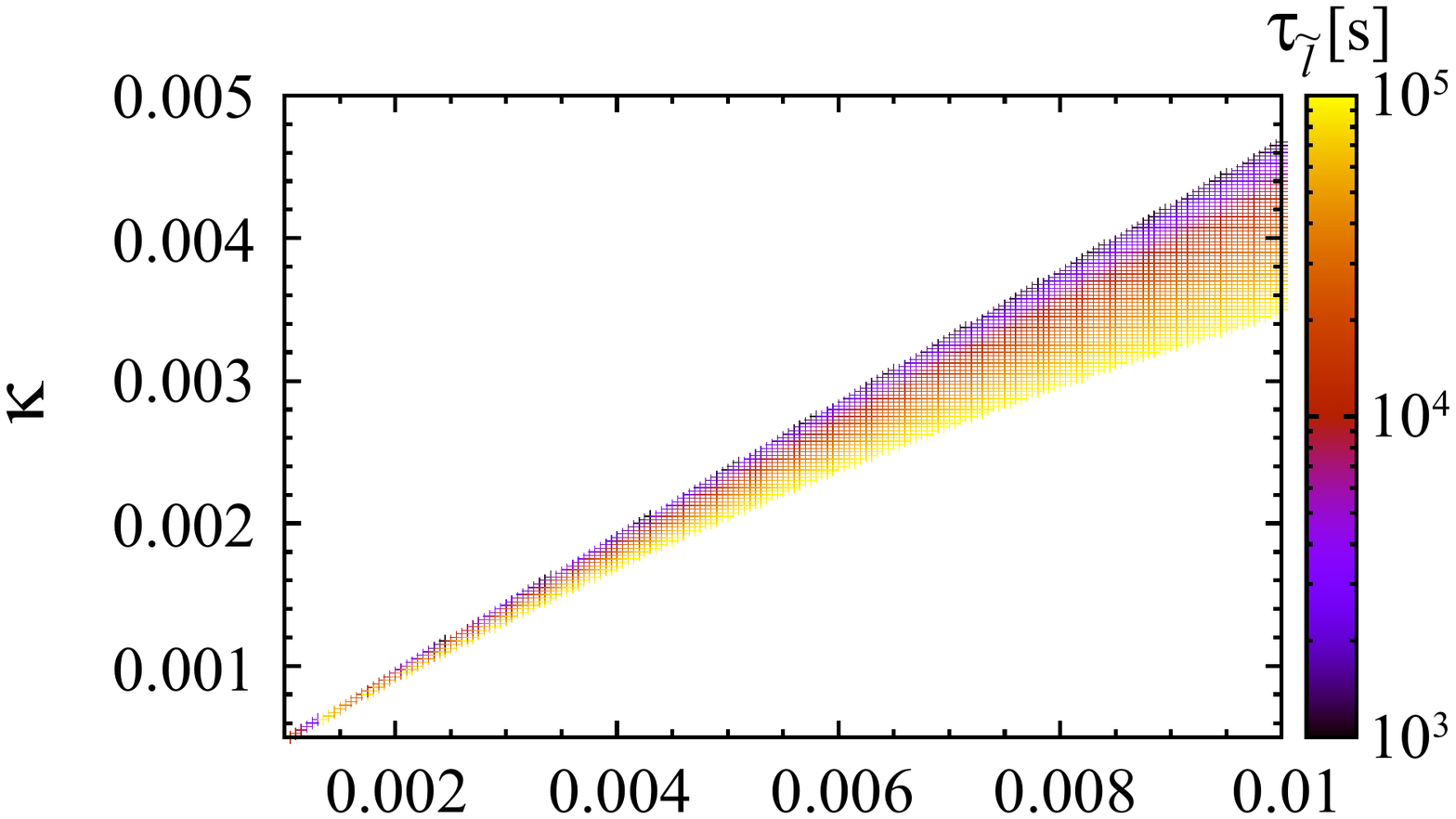}
 \label{fig:}
\end{center}
\end{minipage}
\hspace{8mm}
\begin{minipage}{80mm}\vspace{-38mm}
\begin{center}
 \includegraphics[width=8.9cm,clip]{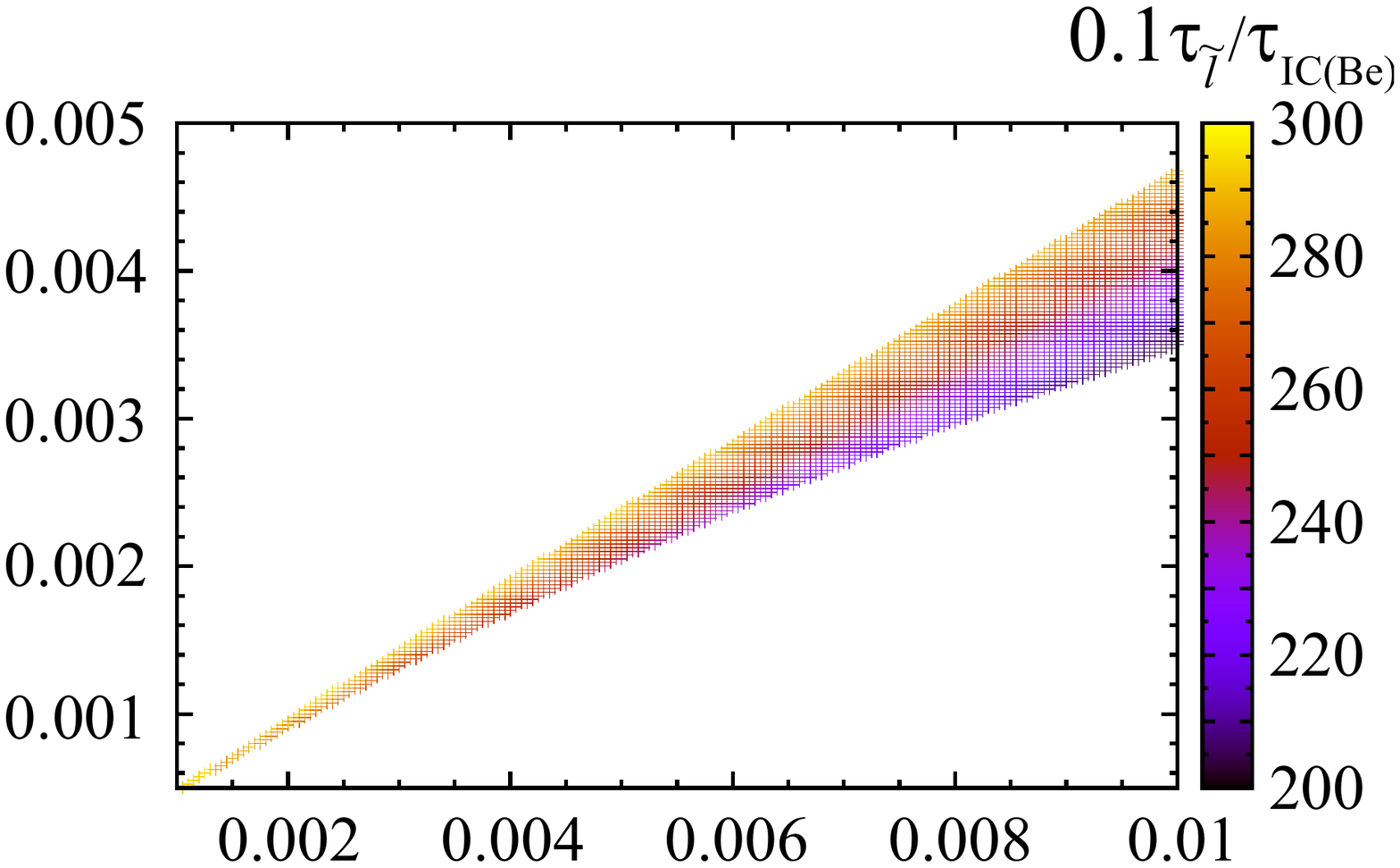}
 \label{fig:}
\end{center}
\end{minipage}
\\[-70mm]
\hspace{-9mm}
\begin{minipage}{79mm}\vspace{-12mm}
\begin{center}
  \includegraphics[width=9.4cm,clip]{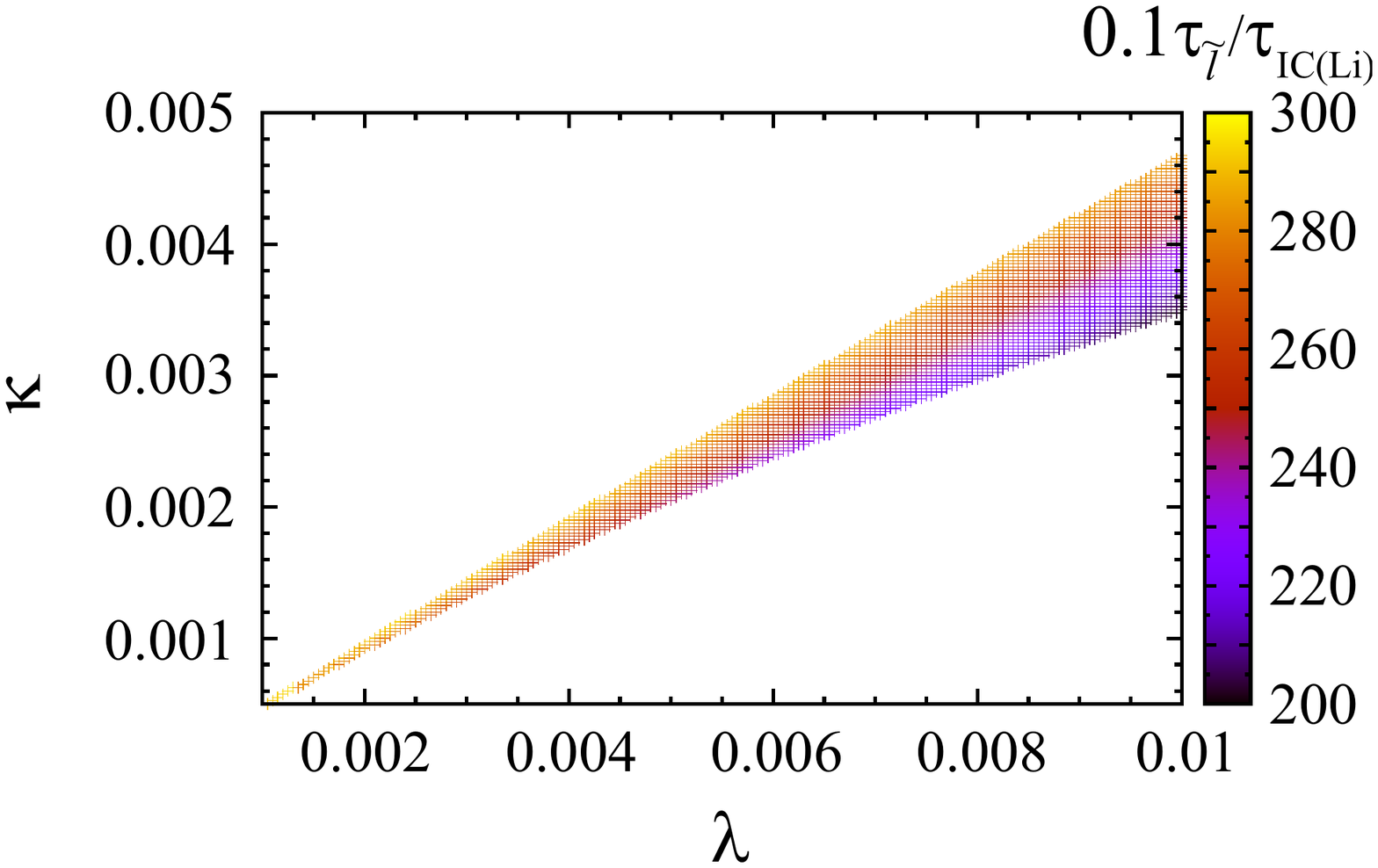}
  \label{fig:}
\end{center}
\end{minipage}
\hspace{13mm}
\begin{minipage}{79.8mm}\vspace{-12mm}
\begin{center}
 \includegraphics[width=8.7cm,clip]{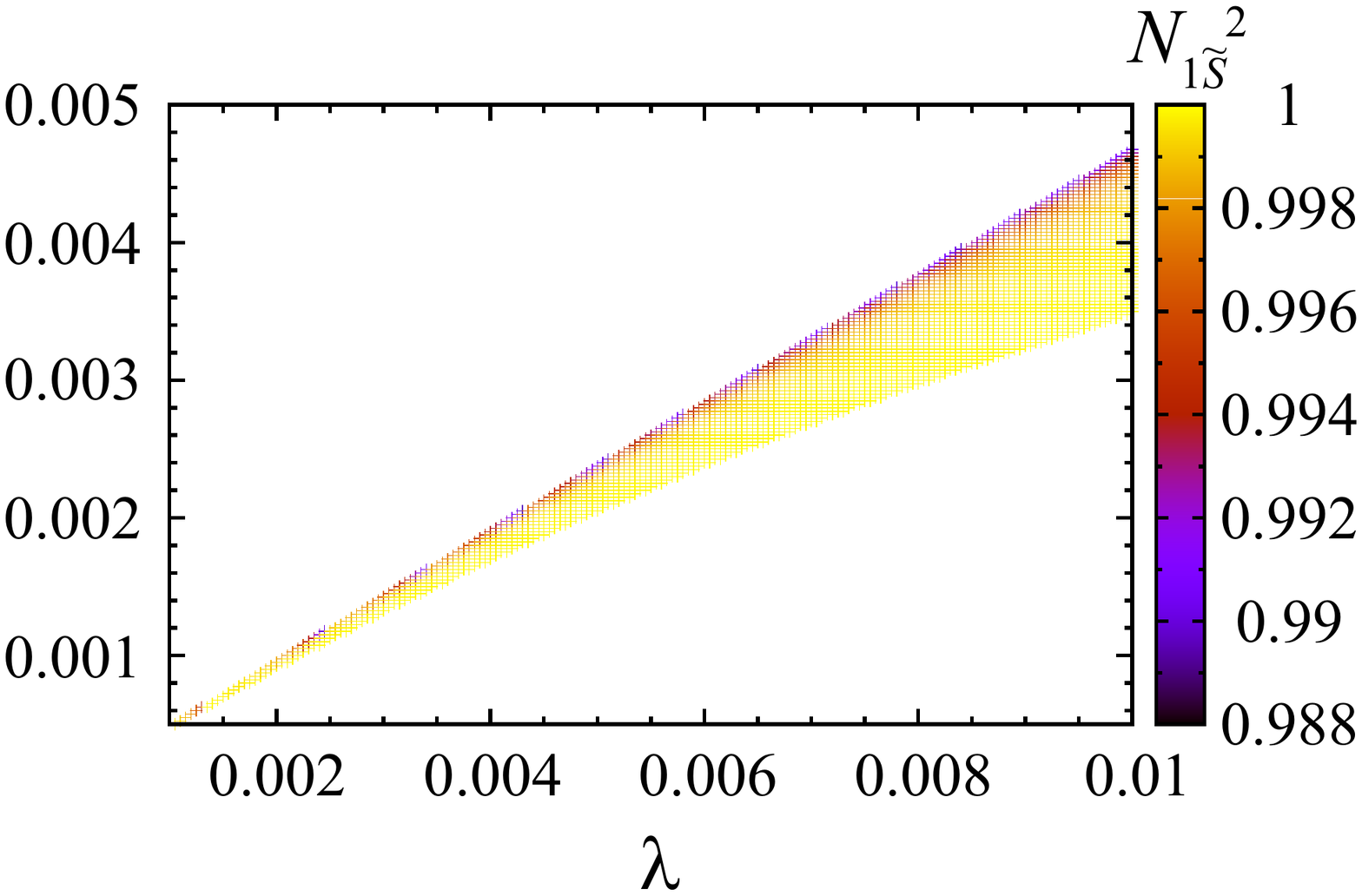}
 \label{fig:}
\end{center}
\end{minipage}
\end{tabular}
\vspace{-38mm}
\caption{
The distribution of the quantities relevant to the requirements, Eqs.~\eqref{eq:cond-sleptonlife}-\eqref{eq:cond-s-rate}, 
in the favored region from $c_e=2\times 10^{-9}$ and $\tan\beta=30$. 
%%%%%%%%%%%%%%
%%%%%%%%%%%%%%
The color bar in each panel shows 
$\tau _{\tilde l}$ (top-left), 
$0.1\tau _{\tilde l}/\tau _{\rm IC}$ for $(^7{\rm Be}~\tilde l^-) \to~^7{\rm Li} + \tilde \chi ^0_1 + \nu _l$ (top-right), 
$0.1\tau _{\tilde l}/\tau _{\rm IC}$ for $(^7{\rm Li}~\tilde l^-) \to~^7{\rm He} + \tilde \chi ^0_1 + \nu _l$ (bottom-left), 
and $N_{1\tilde S}^2$ (bottom-right).
}
\label{fig:allowed-lk-s-small-grad}
%%%%%%
%%%%%%
\end{figure}
\end{center}
\clearpage
\end{widetext}
%%%%%%%%%%%%%%
%%%%%%%%%%%%%%

\begin{widetext}
\clearpage

%%%%%%%%%%%%%%
%%%%%%%%%%%%%%
\begin{table}[t]
\begin{center}
\caption{
Benchmark points in the favored region of Fig.~\ref{fig:allowed-lk-s-small}.
}
\begin{tabular}{ccccc} \hline\hline
~~Parameters~~	& ~~~SS-1~~		& ~~SS-2~~	& ~~SS-3~~		& ~~SS-4~~		\\ \hline %\hline
$c_e$			& $2\times 10^{-9}$	& $10^{-9}$	& $2\times 10^{-9}$	& $2\times 10^{-9}$	\\ %\hline
$\lambda$			& 0.0042			& 0.0042		& 0.00425			& 0.0042	 		\\ %\hline
$\kappa$			& 0.002 			& 0.002		& 0.002			& 0.00195	 		\\ \hline\hline
\end{tabular}
\label{tab:points1}
\end{center}
\end{table}
%%%%%%%%%%%%%%
%%%%%%%%%%%%%%

\begin{table}[tbl]
\begin{center}
\caption{
Spectra and observables at each point (see Tab.~\ref{tab:points1}).
All the dimensionful values are shown in GeV.
The top rows show input parameters.
%We assume the relations for gauging masses, $M_1=M_2/2$ and $M_3=3M_2$, 
%and for squarks and sleptons, 
%$m_{\tilde Q_{1,2}} =m_{\tilde Q_{3}}$,
%$m_{\tilde U_{1,2}}=m_{\tilde U_{3}}$,
%$m_{\tilde D_{1,2}}=m_{\tilde D_{3}}$, 
%$m_{\tilde L_{1,2}}=m_{\tilde L_{3}}$,
%and 
%$m_{\tilde E_{1,2}}=m_{\tilde E_{3}}$.
SS-1 and SS-2 give common results since we omit small flavor mixing of the slepton.
The middle rows show output spectra.
The bottom rows show relic density of the lightest neutralino, spin-independent cross 
section between the lightest neutralino and nucleon, the SUSY contribution to the muon 
anomalous magnetic moment, and the branching ratios of rare decays 
$B_s \to \mu ^+ \mu ^-$ and $b\to s\gamma$, and couplings Eqs.~\eqref{eq:gl} and \eqref{eq:gr} from top to bottom.
}
\begin{tabular}{cccc} \hline\hline
~~{\bf Input}~~	&~~{\bf SS-1, SS-2}~~	&~~{\bf SS-3}~~	&~~{\bf SS-4}~~	\\ \hline %\hline
%$M_1$				& 500.00		& 500.00		& 500.00		\\ %\hline
$M_2$				& 1000.0		& 1000.0		& 1000.0		\\ %\hline
%$M_3$				& 3000.0		& 3000.0		& 3000.0		\\ \hline
$A_t$				& $-3000.0$	& $-4800.0$	& $-7500.0$	\\ %\hline
%$A_b$				& $-3000.0$	& $-4800.0$	& $-7500.0$	\\ \hline
%$A_{\tau}$				& $-3000.0$	& $-4800.0$	& $-7500.0$	\\ %\hline
%$A_{\mu}$			& $-3000.0$	& $-4800.0$	& $-7500.0$	\\ \hline
%$m_{\tilde L_1}$		& 391.30		& 397.06		& 405.43		\\ \hline
%$m_{\tilde L_2}$		& 391.30		& 397.06		& 405.43		\\ \hline
$m_{\tilde L_3}$		& 391.30		& 397.06		& 405.43		\\ %\hline
%$m_{\tilde E_1}$		& 372.99		& 376.34		& 381.29		\\ \hline
%$m_{\tilde E_2}$		& 372.99		& 376.34		& 381.29		\\ \hline	
$m_{\tilde E_3}$		& 372.99		& 376.34		& 381.29		\\ %\hline
%$m_{\tilde Q_1}$		& 1500.0		& 2000.0 		& 3000.0		\\ \hline	
%$m_{\tilde Q_2}$		& 1500.0		& 2000.0		& 3000.0		\\ \hline
$m_{\tilde Q_3}$		& 1500.0		& 2000.0		& 3000.0		\\ %\hline
%$m_{\tilde U_1}$		& 1500.0		& 2000.0		& 3000.0		\\ \hline
%$m_{\tilde U_2}$		& 1500.0		& 2000.0		& 3000.0		\\ \hline
%$m_{\tilde U_3}$		& 1500.0		& 2000.0		& 3000.0		\\ %\hline
%$m_{\tilde D_1}$		& 1500.0		& 2000.0		& 3000.0		\\ \hline
%$m_{\tilde D_2}$		& 1500.0		& 2000.0		& 3000.0		\\ \hline
%$m_{\tilde D_3}$		& 1500.0		& 2000.0		& 3000.0		\\ \hline
$\lambda$				& 0.0042		& 0.00425		& 0.0042		\\ %\hline
$\kappa$				& 0.0020		& 0.0020		& 0.00195		\\ %\hline
$A_{\lambda}$			& 1000.0		& 1300.0		& 1000.00		\\ %\hline
$A_{\kappa}$			& $-100.00$	& $-100.00$	& $-100.00$	\\ %\hline
$\mu _{\rm eff}$		& 359.82		& 363.07		& 366.76		\\ %\hline 
$\tan \beta $			& 30.000		& 30.000		& 30.000		\\ \hline %\hline \hline
~{\bf Output}~			&			&			&			\\ \hline
$h^0_1$				& 123.89		& 126.16		& 127.30 		\\ %\hline
$h^0_2$				& 316.70		& 317.05		& 314.58		\\ %\hline
$h^0_3$				& 3513.4		& 3939.4		& 3499.3		\\ %\hline
$a^0_1$				& 226.72		& 226.40		& 226.02		\\ %\hline
$a^0_2$				& 3513.4		& 3939.4		& 3499.3		\\ %\hline
$H^{\pm}$				& 3514.2		& 3940.1		& 3500.0		\\ %\hline
%$\tilde d_L$			& 1510.7 		& 2070.0		& 3100.8		\\ \hline
%$\tilde d_R$			& 1509.8 		& 2069.3		& 3100.4		\\ \hline
%$\tilde u_L$			& 1508.6		& 2068.5		& 3099.8		\\ \hline
%$\tilde u_R$			& 1509.1		& 2068.9		& 3100.1		\\ \hline
%$\tilde s_L$			& 1510.7		& 2070.0		& 3100.8		\\ \hline
%$\tilde s_R$			& 1509.8		& 2069.3		& 3100.4		\\ \hline
%$\tilde c_L$			& 1508.6		& 2068.5		& 3099.8		\\ \hline
%$\tilde c_R$			& 1509.1		& 2068.9		& 3100.1		\\ \hline
%$\tilde b_1$			& 1499.0		& 2060.5		& 3093.7		\\ %\hline
%$\tilde b_2$			& 1521.4		& 2078.8		& 3107.5		\\ %\hline
$\tilde t_1$			& 1367.9		& 1897.7		& 2925.2		\\ %\hline
$\tilde t_2$			& 1650.6		& 2235.9		& 3271.5		\\ %\hline
%$\tilde e_L$			& 393.96		& 399.66		& 407.94		\\ \hline
%$\tilde e_R$			& 375.46		& 378.78		& 383.69		\\ \hline
%$\tilde \nu _{eL}$		& 386.24		& 392.10		& 400.62		\\ \hline
%$\tilde \mu_L$		& 393.96		& 399.66		& 407.94		\\ \hline
%$\tilde \mu_R$		& 375.46		& 378.78		& 383.69		\\ \hline
%$\tilde \nu _{\mu L}$	& 386.24		& 392.10		& 400.62		\\ \hline
$\tilde \tau _1$			& 350.10		& 350.10 		& 350.10  		\\ %\hline
$\tilde \tau _2$			& 416.67		& 425.01		& 437.11		\\ %\hline
%$\tilde \nu _{\tau L}$	& 386.24		& 392.10		& 400.62		\\ \hline
%$\tilde g$			& 2913.0		& 2978.6		& 3193.6		\\ %\hline
$\tilde \chi ^0_1$		& 350.00		& 350.00 		& 350.00 		\\ %\hline
$\tilde \chi ^0_2$		& 355.27		& 359.35		& 364.03		\\ %\hline
$\tilde \chi ^0_3$		& 370.74		& 375.08		& 380.10		\\ %\hline
$\tilde \chi ^0_4$		& 498.84		& 498.04		& 496.76		\\ %\hline
$\tilde \chi ^0_5$		& 1021.1		& 1027.8		& 1035.7		\\ \hline
%$\tilde \chi ^{\pm}_1$	& 364.43		& 368.80		& 373.83		\\ %\hline
%$\tilde \chi ^{\pm}_2$	& 1021.1		& 1027.8		& 1035.7		\\ \hline
$\Omega _{\tilde \chi ^0_1}h^2$	
					& 0.11236		& 0.12656		& 0.12490  	\\ %\hline
%$N_{5\tilde S}$		& 0.99557		& 0.99857		& 0.99938  	\\ %\hline
%$\sin \theta _{\tau}$	& 0.79968		& 0.79998		&  0.80009  	\\ \hline %\hline
$\sigma _{\rm SI}$[cm$^2$]
					&  $9.4085\times 10^{-49}$
					&  $1.4012\times 10^{-49}$
					&  $3.8721\times 10^{-50}$  \\
$\delta a_{\mu}$
					&  $1.1967\times 10^{-9} (2\sigma )$
					&  $1.2656\times 10^{-9} (2\sigma )$
					&  $1.2347\times 10^{-9} (2\sigma )$ \\ %\hline
Br$(B_s^0\to \mu^+\mu^-)$
					&  $3.3924\times 10^{-9} (1\sigma )$
					&  $3.4349\times 10^{-9} (1\sigma )$
					&  $3.4442\times 10^{-9} (1\sigma )$ \\ %\hline
Br$(b \to s \gamma)$
					&  $2.5984\times 10^{-4} (>3\sigma )$
					&  $2.8109\times 10^{-4} (2\sigma )$
					&  $3.0311\times 10^{-4} (2\sigma )$ \\  
$G_{{\rm L}\tau}/c_{\tau}$	& $-0.014882$	& $-0.0084607$	& $-0.0055185$	\\
$G_{{\rm R}\tau}/c_{\tau}$	& 0.019332	& 0.011050	& 0.0072524	\\
$G_{{\rm L}e}/c_{e}$		& 0.00066840	& 0.00038490	&  0.00025476	\\
$G_{{\rm R}e}/c_{e}$		& 0.0076689	& 0.0044158	& 0.0029225	\\
\hline\hline			
\end{tabular}
\label{tab:points-s-small}
\end{center}
\end{table}
\clearpage
\end{widetext}

%%%%%%%%%%%%%%
%%%%%%%%%%%%%%
%%%%%%%%%%%%%%
%%%%%%%%%%%%%%

\begin{widetext}
\begin{center}
\begin{figure}[h]
%%%%%%
%%%%%%
\begin{tabular}{l}
\hspace{-24mm}
\begin{minipage}{80mm}\vspace{-28mm}
\begin{center}
 \includegraphics[width=8.75cm,clip]{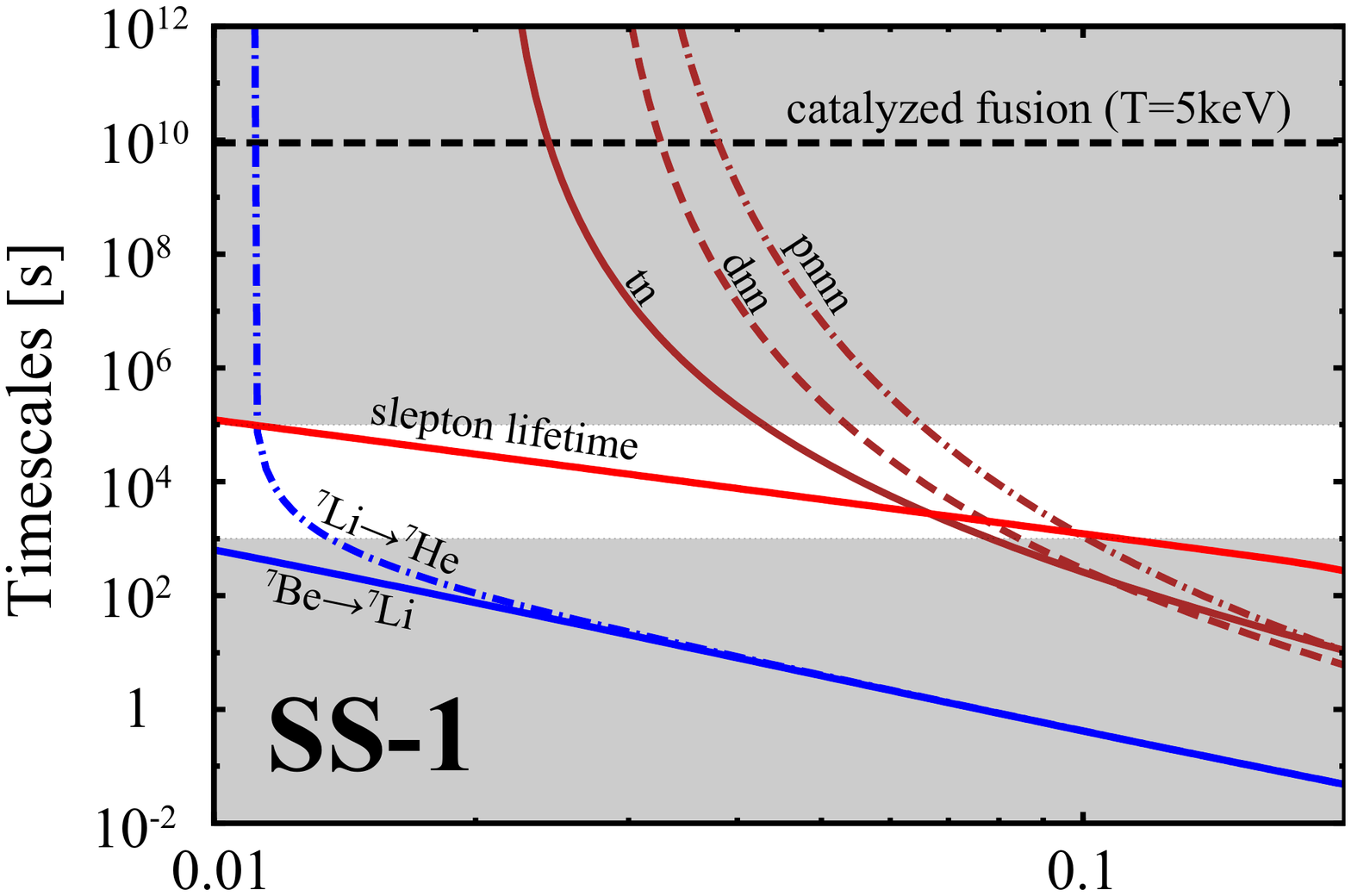}
 \label{fig:}
\end{center}
\end{minipage}
\hspace{-5.5mm}\vspace{-1.5cm}
\begin{minipage}{80mm}\vspace{-28mm}
\begin{center}
 \includegraphics[width=10.3cm,clip]{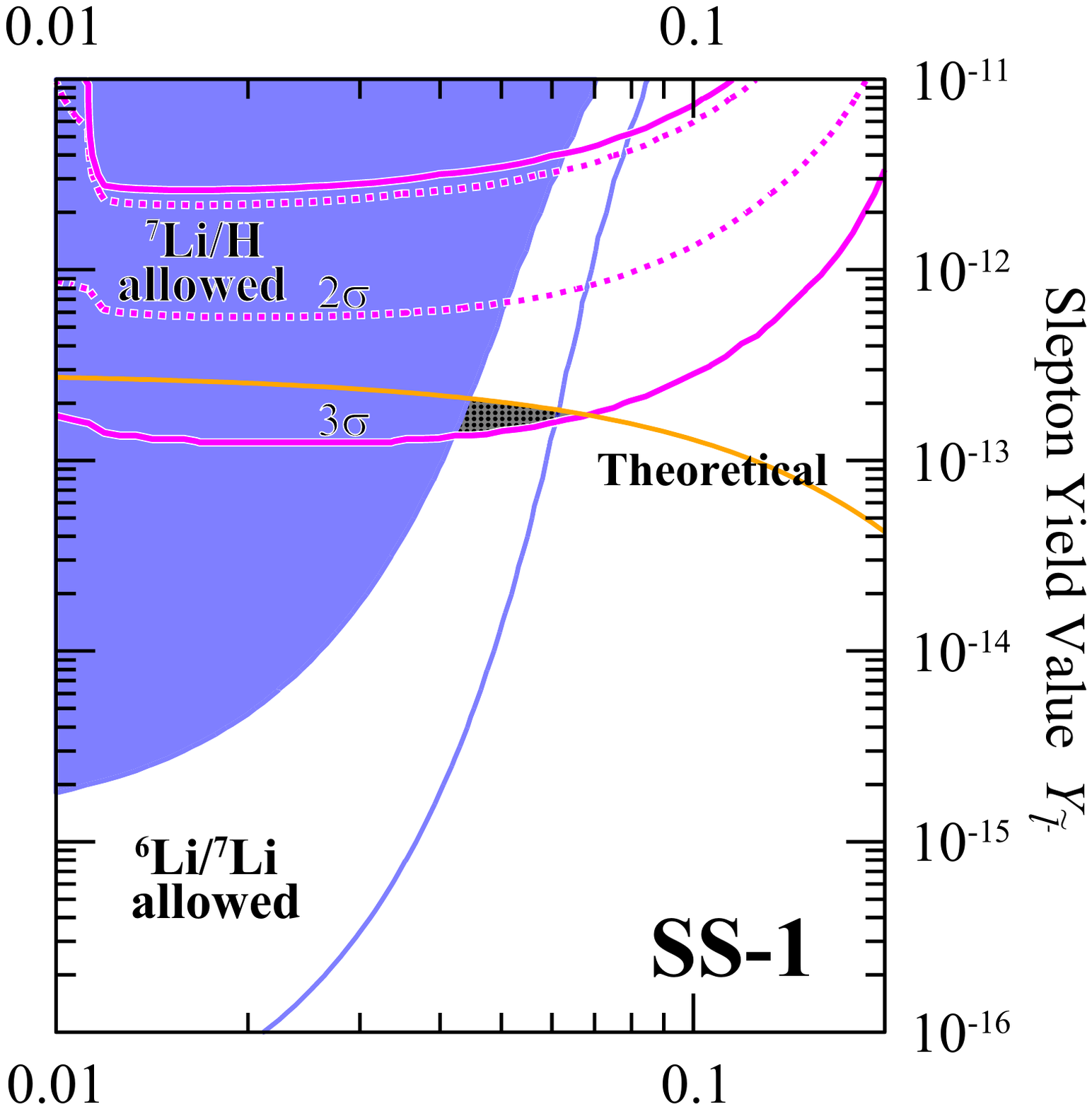}
 \label{fig:}
\end{center}
\end{minipage}
\\[-3mm]
\hspace{-24mm}
\begin{minipage}{79mm}\vspace{-0.8mm}\vspace{-48mm}
\begin{center}
  \includegraphics[width=9.1cm,clip]{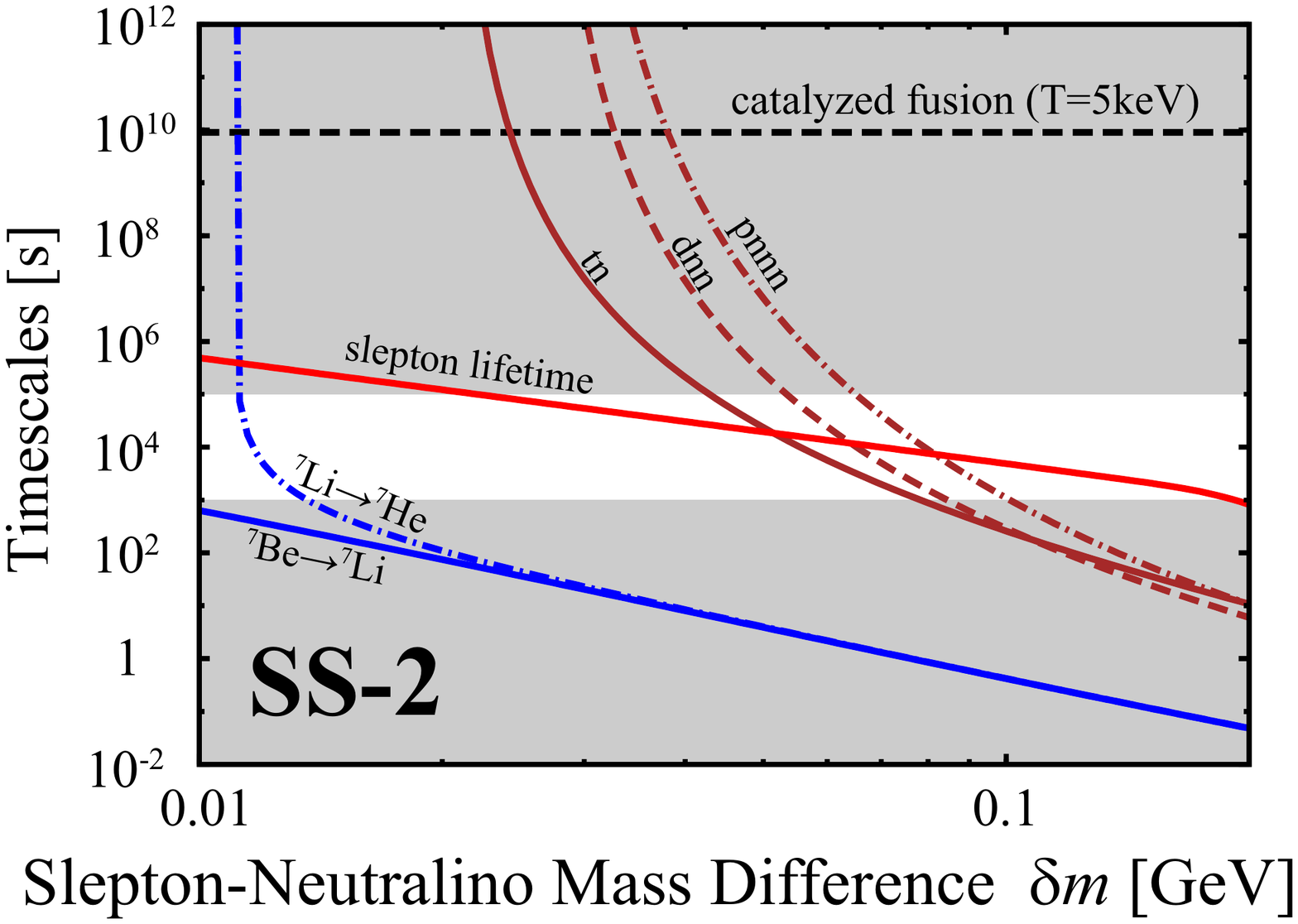}
  \label{fig:}
\end{center}
\end{minipage}
\hspace{-4mm}
\begin{minipage}{79.8mm}\vspace{0.5mm}\vspace{-48mm}
\begin{center}
 \includegraphics[width=10.3cm,clip]{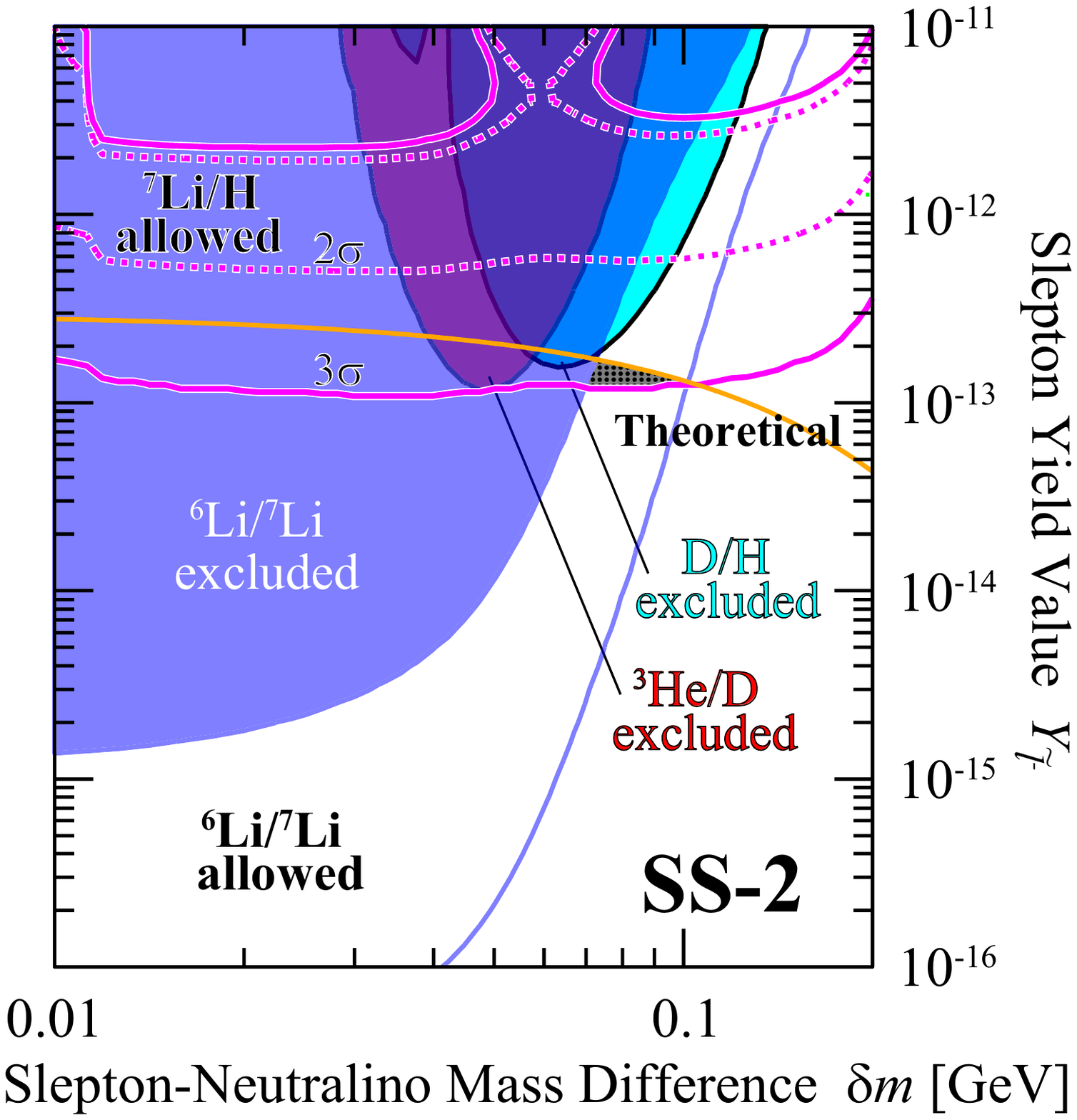}
 \label{fig:}
\end{center}
\end{minipage}
\end{tabular}
\vspace{-28mm}
\caption{
The left panels show 
the slepton lifetime $\tau _{\tilde l}$~(red-solid line; ``slepton lifetime"), 
the timescales of the internal conversion processes \eqref{eq:icbe}~(blue-solid line; ``$^7$Be$\to$$^7$Li"), \eqref{eq:icli}~(blue-dash-dotted line; ``$^7$Li$\to$$^7$He"), 
the $^4$He spallation processes \eqref{eq:tn}~(brown-solid line; ``tn"), \eqref{eq:dnn}~(brown-dashed line; ``dnn"), and \eqref{eq:pnnn}~(brown-dash-dotted line; ``pnnn"), 
as a function of the mass difference between the slepton and the neutralino 
at SS-1 (top panel) and SS-2 (bottom panel).
We also show the timescale of the catalyzed fusion \eqref{eq:cf} at the temperature $T=5$~keV ($5\times10^4$~s) when ($^4$He~$\tilde l^-$) is formed as horizontal black-dashed line.
In shaded regions Eq.~\eqref{eq:cond-sleptonlife} is not satisfied.
The right panels show the allowed regions from observational light element abundances on $\delta m$-$Y_{\tilde l^-}$ plane at SS-1 (top panel) and SS-2 (bottom panel).
%%%%%%%%%%%%%%
%%%%%%%%%%%%%%
The regions surrounded by magenta-dotted(-solid) lines are allowed by observed 
$^7$Li/H abundance at 2$\sigma$(3$\sigma$).
%%%%%%%%%%%%%%
%%%%%%%%%%%%%%
The regions between the blue-solid line and the blue region are allowed by 
observed $^6$Li/$^7$Li abundance at 2$\sigma$.
%%%%%%%%%%%%%%
%%%%%%%%%%%%%%
The orange-solid lines (``Theoretical") represent the yield value of the slepton 
at the begging of the BBN as a function of the mass difference.
%%%%%%%%%%%%%%
%%%%%%%%%%%%%%
The colored regions are excluded for 
$^6$Li/$^7$Li (blue region; ``$^6$Li/$^7$Li excluded"), 
$^3$He/D (red region; ``$^3$He/D excluded"), and 
D/H (cyan region; ``D/H excluded"), respectively. 
%%%%%%%%%%%%%%
%%%%%%%%%%%%%%
The shaded and dotted regions are allowed by only $^7$Li/H (3$\sigma$) 
and both $^7$Li/H (3$\sigma$) and $^6$Li/$^7$Li (2$\sigma$), respectively.
}
\label{fig:small-lk-s_bbn1}
%%%%%%
%%%%%%
\end{figure}
\end{center}
\clearpage
\end{widetext}
%%%%%%%%%%%%%%
%%%%%%%%%%%%%%

\setcounter{figure}{2}

\begin{widetext}
\begin{center}
\begin{figure}[h]
%%%%%%
%%%%%%
\begin{tabular}{l}
\hspace{-18mm}
\begin{minipage}{80mm}\vspace{-28mm}
\begin{center}
 \includegraphics[width=8.75cm,clip]{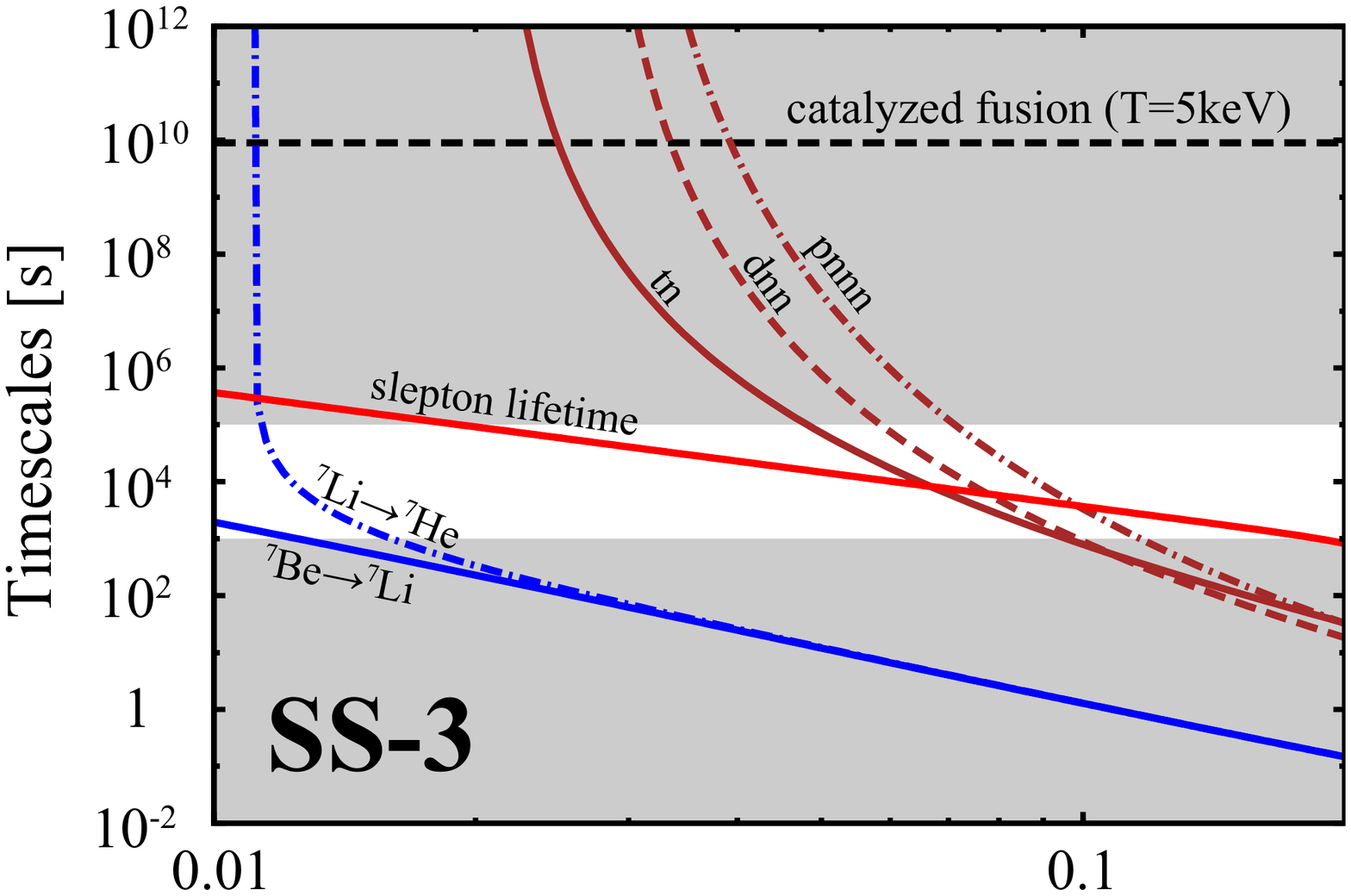}
 \label{fig:}
\end{center}
\end{minipage}
\hspace{-3.5mm}\vspace{-1.5cm}
\begin{minipage}{80mm}\vspace{-28mm}
\begin{center}
 \includegraphics[width=10.1cm,clip]{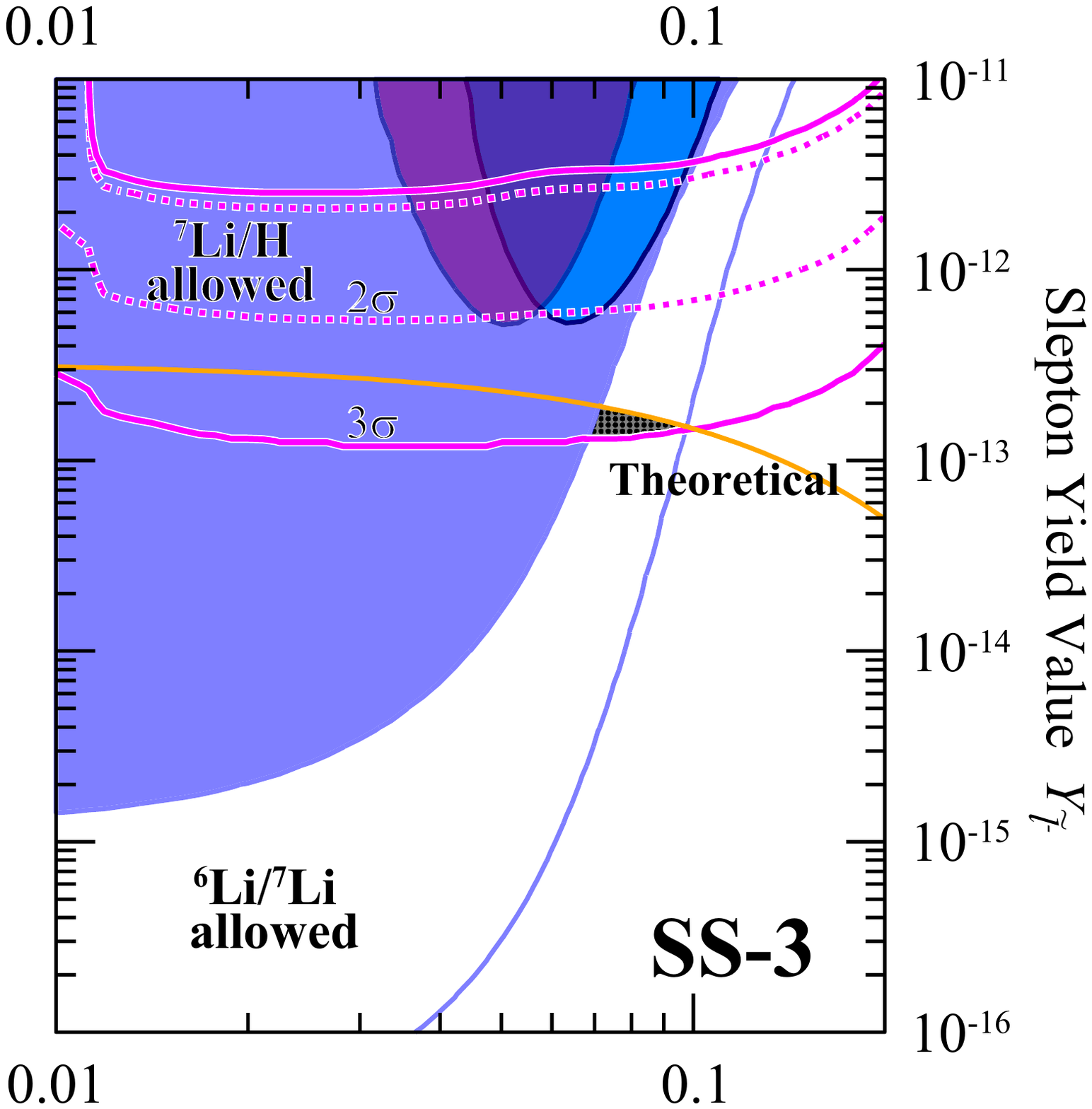}
 \label{fig:}
\end{center}
\end{minipage}
\\[-3mm]
\hspace{-18.5mm}
\begin{minipage}{79mm}\vspace{-48mm}
\begin{center}
  \includegraphics[width=9.1cm,clip]{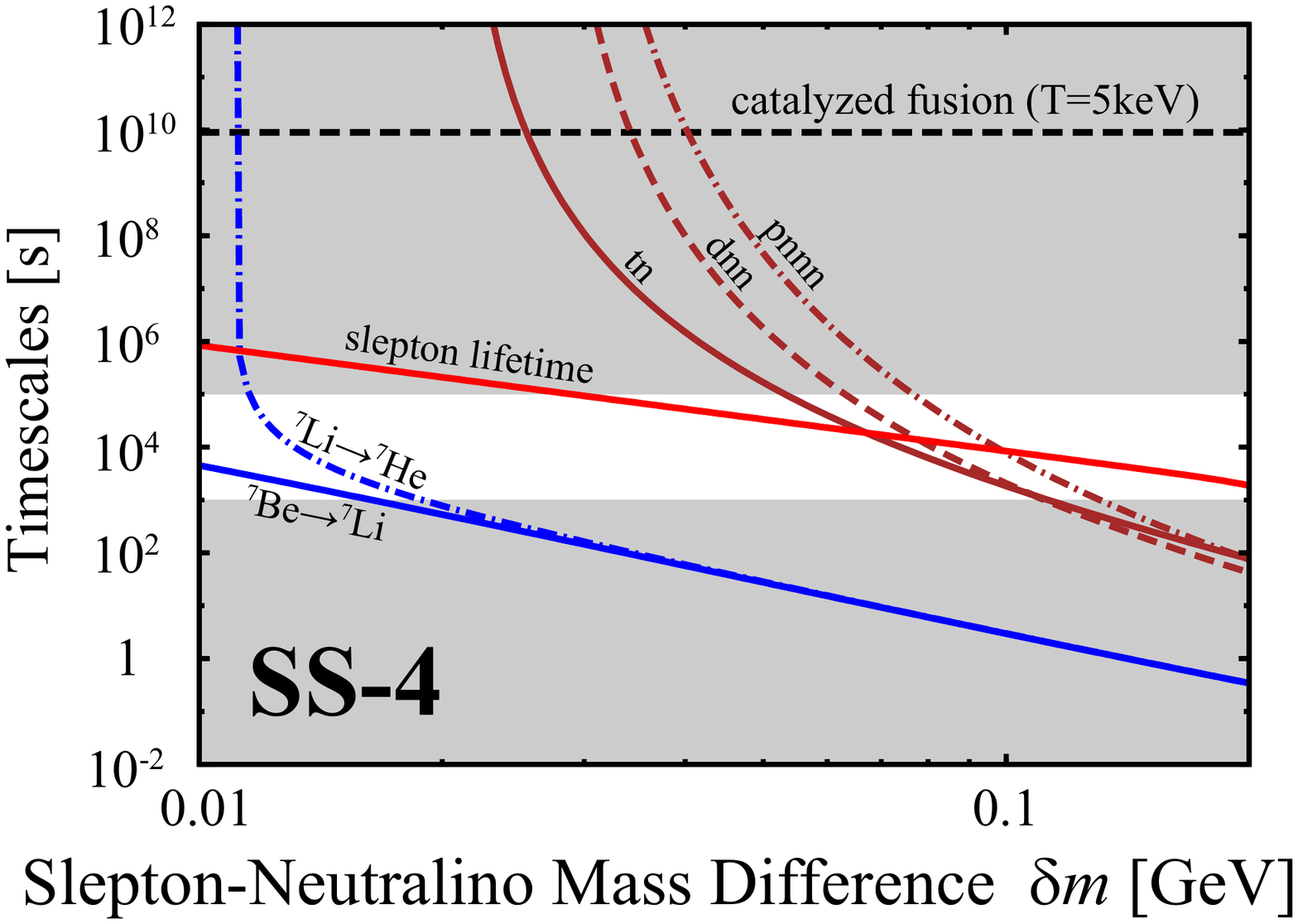}
  \label{fig:}
\end{center}
\end{minipage}
\hspace{-2mm}
\begin{minipage}{79.8mm}\vspace{-48mm}
\begin{center}
 \includegraphics[width=10cm,clip]{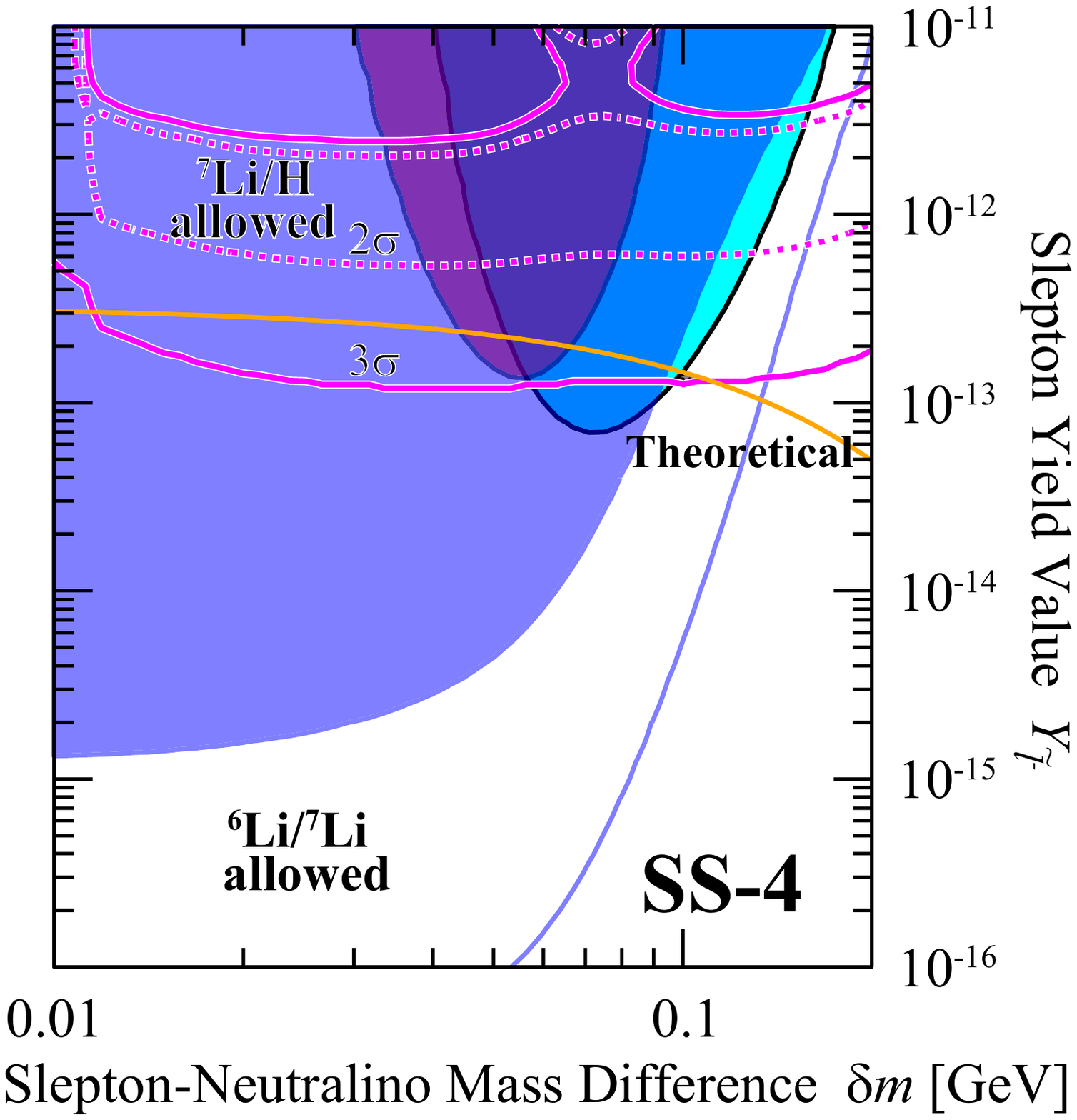}
 \label{fig:}
\end{center}
\end{minipage}
\end{tabular}
\vspace{-28mm}
\caption{
The results at SS-3 (top panels) and SS-4 (bottom panels) are shown.
The lines and regions stand for the same meanings as those in the previous page.
}
\label{fig:small-lk-s_bbn2}
%%%%%%
%%%%%%
\end{figure}
\end{center}
\clearpage
\end{widetext}
%%%%%%%%%%%%%%
%%%%%%%%%%%%%%

%%%%%%%%%%%%%%%%%%%%%%%%%%%%%%%%%%%%%%%%%%%%%%%%
%%%%%%%%%%%%%%%%%%%%%%%%%%%%%%%%%%%%%%%%%%%%%%%%
\subsection{Singlino-like neutralino LSP; large $\lambda$-$\kappa$ region with small $\tan\beta$} \label{sec:lk-s-large}
%%%%%%%%%%%%%%%%%%%%%%%%%%%%%%%%%%%%%%%%%%%%%%%%
%%%%%%%%%%%%%%%%%%%%%%%%%%%%%%%%%%%%%%%%%%%%%%%%

%%%%%%%%%%%%%%
%%%%%%%%%%%%%%
%In this large $\lambda$-$\kappa$ case, the condition Eq.~\eqref{eq:cond-s-rate} 
%and tree contributions to the Higgs mass (the second and the third terms in 
%Eq.~\eqref{eq:mhiggs}) play important roles to obtain desirable parameter sets.
%%%%%%%%%%%%%%
%%%%%%%%%%%%%%

%%%%%%%%%%%%%%
%%%%%%%%%%%%%%
%Different from the previous case, not only Eq.~\eqref{eq:cond-sleptonlife} 
%but also Eq.~\eqref{eq:cond-s-rate} constrain $\lambda$-$\kappa$ plane.
%We consider the second case where $\lambda$ and $\kappa$ are relatively large 
%and $\tan\beta$ is small, and the neutralino LSP is singlino-like.
%%%%%%%%%%%%%%%
%%%%%%%%%%%%%%%
%In this case the NMSSM-specific tree terms have significant contribution in 
%the Higgs mass Eq.~\eqref{eq:mhiggs}.
%%%%%%%%%%%%%%%
%%%%%%%%%%%%%%%
%This is completely different from the case of the MSSM where the 1-loop 
%term must have significant contribution.
%%%%%%%%%%%%%%%
%%%%%%%%%%%%%%%

%%%%%%%%%%%%%%%%%%%%%%%%%%%%%%%%%%%%%%%%%%%%%%%%
\subsubsection{Benchmark points} \label{sec:bmp-s-large}%%%%%%%%%%%%%%%
%%%%%%%%%%%%%%%%%%%%%%%%%%%%%%%%%%%%%%%%%%%%%%%%

%%%%%%%%%%%%%%
%%%%%%%%%%%%%%
Figure~\ref{fig:allowed-lk-s-large} shows the favored region on $\lambda$-$\kappa$ 
plane with small $\tan \beta$ where the requirements, Eqs.~\eqref{eq:cond-sleptonlife}-\eqref{eq:cond-s-rate}, 
are satisfied.
%%%%%%%%%%%%%%
%%%%%%%%%%%%%%
The left and right panels show the results for $c_e=5\times 10^{-10}$ and  
$2\times 10^{-10}$, respectively.
%%%%%%%%%%%%%%
%%%%%%%%%%%%%%
The red and hatched regions are allowed by $\tan\beta =2$ and $2.5$, 
respectively.
%%%%%%%%%%%%%%
%%%%%%%%%%%%%%

%%%%%%%%%%%%%%
%%%%%%%%%%%%%%
To check which requirements determine the favored region in Fig.~\ref{fig:allowed-lk-s-large}, 
we draw Fig.~\ref{fig:allowed-lk-s-large-grad} to show the distribution of the quantities 
which are relevant to Eqs.~\eqref{eq:cond-sleptonlife}-\eqref{eq:cond-s-rate}.
The parameters are $c_e=5\times 10^{-10}, m_{\tilde \chi ^0_1}=350$~GeV, 
$\delta m=0.1$~GeV, $\sin \theta _f=0.8, \tan\beta =2, M_1=500$~GeV, and $M_2=1000$~GeV.
From the result we see the favored region is determined by the requirement for 
the slepton lifetime Eq.~\eqref{eq:cond-sleptonlife}.
%(the red region in the right panel of Fig.~\ref{fig:allowed-lk-s-large}).
%%%%%%%%%%%%%%
%%%%%%%%%%%%%%
%The top-left panel shows the distribution of the slepton lifetime. 
%%%%%%%%%%%%%%%
%%%%%%%%%%%%%%%
%\red{
%The bottom and top edges of the region is determined by Eq.~\eqref{eq:cond-sleptonlife}.
%}
%%%%%%%%%%%%%%%
%%%%%%%%%%%%%%%
%The top-right and bottom-left panels show the ratio of ($0.1\times$ slepton lifetime) 
%and the internal conversion timescales.
%%%%%%%%%%%%%%%
%%%%%%%%%%%%%%%
%The bottom-right panel shows the fraction of singlino component in the lightest 
%neutralino, $N_{1\tilde S}^2$.
%%%%%%%%%%%%%%%
%%%%%%%%%%%%%%%
%The conditions Eq.~\eqref{eq:cond-ictime} and \eqref{eq:cond-s-rate} are satisfied 
%everywhere in the favored region.
%%%%%%%%%%%%%%%
%%%%%%%%%%%%%%

%%%%%%%%%%%%%%
%%%%%%%%%%%%%%
We take four reference points in the favored region for $\tan\beta=2$ 
(red regions in Fig.~\ref{fig:allowed-lk-s-large}) as shown in Table~\ref{tab:points2}. 
%%%%%%%%%%%%%%
%%%%%%%%%%%%%%
Table~\ref{tab:points-s-large} shows the spectra and observables at these points.
%%%%%%%%%%%%%%
%%%%%%%%%%%%%%
We omit small flavor mixing of the slepton in the calculation, and show 
the dimensionful values in GeV.
%%%%%%%%%%%%%%
%%%%%%%%%%%%%%
Every points give the observed Higgs mass.
%%%%%%%%%%%%%%
%%%%%%%%%%%%%%
In this case where $\lambda$ and $\kappa$ are large and $\tan\beta$ is small, 
the tree contributions in Eq.~\eqref{eq:mhiggs} (the second and the third terms) 
are significant, and 1-loop contribution (the forth term) is not so large. 
%%%%%%%%%%%%%%
%%%%%%%%%%%%%%

%%%%%%%%%%%%%%
%%%%%%%%%%%%%%
In the bottom rows, we show relic density of the lightest neutralino, spin-independent cross 
section between the lightest neutralino and nucleon, the SUSY contribution to the muon 
anomalous magnetic moment, and the branching ratios of rare decays 
$B_s \to \mu ^+ \mu ^-$ and $b\to s\gamma$, from top to bottom.
%%%%%%%%%%%%%%
%%%%%%%%%%%%%%
At each point, the dark matter relic density is in range of the measured value~\cite{Ade:2013zuv}.
%%%%%%%%%%%%%%
%%%%%%%%%%%%%%
The spin-independent cross section at each point is much larger than those at the points 
in the previous section and just below the present experimental bound.
%%%%%%%%%%%%%%
%%%%%%%%%%%%%%
The calculated values of $\delta a_{\mu}$ at the points are out of 3$\sigma$ range, 
since $\tan\beta$ is relatively small in this case.
%%%%%%%%%%%%%%
%%%%%%%%%%%%%%
For the branching ratio of $B_s\to \mu^+\mu^-$ and $b\to s\gamma$, we obtained the 
values within 1$\sigma$ and 2$\sigma$, respectively.
%%%%%%%%%%%%%%
%%%%%%%%%%%%%%

%%%%%%%%%%%%%%%%%%%%%%%%%%%%%%%%%%%%%%%%%%%%%%%%
\subsubsection{BBN results at the benchmark points} \label{sec:bbn-s-large} %%%%%%
%%%%%%%%%%%%%%%%%%%%%%%%%%%%%%%%%%%%%%%%%%%%%%%%

%%%%%%%%%%%%%%
%%%%%%%%%%%%%%
The left panels in Fig.~\ref{fig:large-lk-s_bbn1} show the slepton lifetime 
$\tau _{\tilde l}$~(red-solid line; ``slepton lifetime"), 
the timescales of the internal conversion processes
\eqref{eq:icbe}~(blue-solid line; ``$^7$Be$\to$$^7$Li"), 
\eqref{eq:icli}~(blue-dash-dotted line; ``$^7$Li$\to$$^7$He"), 
the $^4$He spallation processes 
\eqref{eq:tn}~(brown-solid line; ``tn"), 
\eqref{eq:dnn}~(brown-dashed line; ``dnn"), and 
\eqref{eq:pnnn}~(brown-dash-dotted line; ``pnnn"), 
as a function of the mass difference between the slepton and the neutralino 
at SL-1, SL-2, SL-3, and SL-4 from top to bottom, respectively.
%%%%%%%%%%%%%%
%%%%%%%%%%%%%%
The horizontal black-dashed line represents the timescale of the catalyzed 
fusion process \eqref{eq:cf}~\cite{Hamaguchi:2007mp} at the temperature $T=5$~keV 
($5\times10^4$~s) when ($^4$He~$\tilde l^-$) is formed.
%%%%%%%%%%%%%%
%%%%%%%%%%%%%%
In the right panels
horizontal axis is the mass difference between the slepton NLSP and 
the neutralino LSP, and vertical axis is the yield value of the slepton at the beginning of the BBN. 
%$Y_{\tilde l^-}=n_{\tilde l^-}/s$, where $n_{\tilde l^-}$ is the number density
%of the slepton and $s$ is the entropy density.
%%%%%%%%%%%%%%%
%%%%%%%%%%%%%%%

%%%%%%%%%%%%%%
%%%%%%%%%%%%%%
We show the allowed regions in the right panels of Fig.~\ref{fig:large-lk-s_bbn1} 
which we obtain by comparing theoretical values to observational ones for light 
element abundances at SL-1, SL-2, SL-3, and SL-4 from top to bottom, 
respectively.
%%%%%%%%%%%%%%
%%%%%%%%%%%%%%
The lines and regions are the same as in Fig.~\ref{fig:small-lk-s_bbn1}.
%%%%%%%%%%%%%%
%%%%%%%%%%%%%%
Accordingly we obtain allowed region from $^7$Li/H (3$\sigma$) and
$^6$Li/$^7$Li (2$\sigma$) simultaneously at each point.
%%%%%%%%%%%%%%
%%%%%%%%%%%%%%
%At SL-2, we obtain allowed region only for $^7$Li/H (3$\sigma$).
%%%%%%%%%%%%%%
%%%%%%%%%%%%%%

We can get large $G_{L,R}$ without tuning at the SL points in contrast to SS case, 
since $\lambda$ and $\kappa$ are large enough.
%%%%%%%%%%%%%%
%%%%%%%%%%%%%%
Therefore, the couplings $G_{L,R}$ are almost same as each other
at the points SL-3 and SL-4 though $\lambda$ is larger at SL-3 and $\kappa$ is smaller at SL-4 than those at SL-1, respectively.
Furthermore $c_e$ is same at each point and the slepton lifetime is almost same.
%%%%%%%%%%%%%%
%%%%%%%%%%%%%%
Thus we get similar result at these points.
%%%%%%%%%%%%%%
%%%%%%%%%%%%%%
%These results are similar to the one at SS-1 since the slepton lifetime 
%and the timescales of the internal conversion and the $^4$He spallation
%processes at the three points are almost same as those at SS-1.
%%%%%%%%%%%%%%
%%%%%%%%%%%%%%

%%%%%%%%%%%%%%
%%%%%%%%%%%%%%
On the contrary, at SL-2 the selectron mixing is small compared with
that of SL-1, and hence the slepton lifetime is longer than the other
points.
%between the results at the two points is same as that between the 
%results at SS-1 and SS-2.
%%%%%%%%%%%%%%
%%%%%%%%%%%%%%
This result in more spallation processes and make the allowed region
narrower.

%The result at SL-2 is similar to the one at SS-4 where allowed region 
%neither for $^7$Li/H nor for $^6$Li/$^7$Li exists.
%%%%%%%%%%%%%%%
%%%%%%%%%%%%%%%
%However, the timescales of the internal conversion processes at SL-2 
%are about one order of magnitude shorter than those at SS-4.
%%%%%%%%%%%%%%
%%%%%%%%%%%%%%
%\red{
%Thus at SL-2, observed abundance of $^7$Li/H can be reproduced 
%for smaller slepton yield value compared with the result at SS-4, and 
%then wider allowed region is obtained.
%}
%%%%%%%%%%%%%%%
%%%%%%%%%%%%%%%

\clearpage
%%%%%%%%%%%%%%
%%%%%%%%%%%%%%
\begin{widetext}
\begin{center}
\begin{figure}[h]
%%%%%%
%%%%%%
\begin{tabular}{l}
\hspace{-10mm}
\begin{minipage}{83mm}\vspace{-28mm}
\begin{center}
 \includegraphics[width=9.3cm,clip]{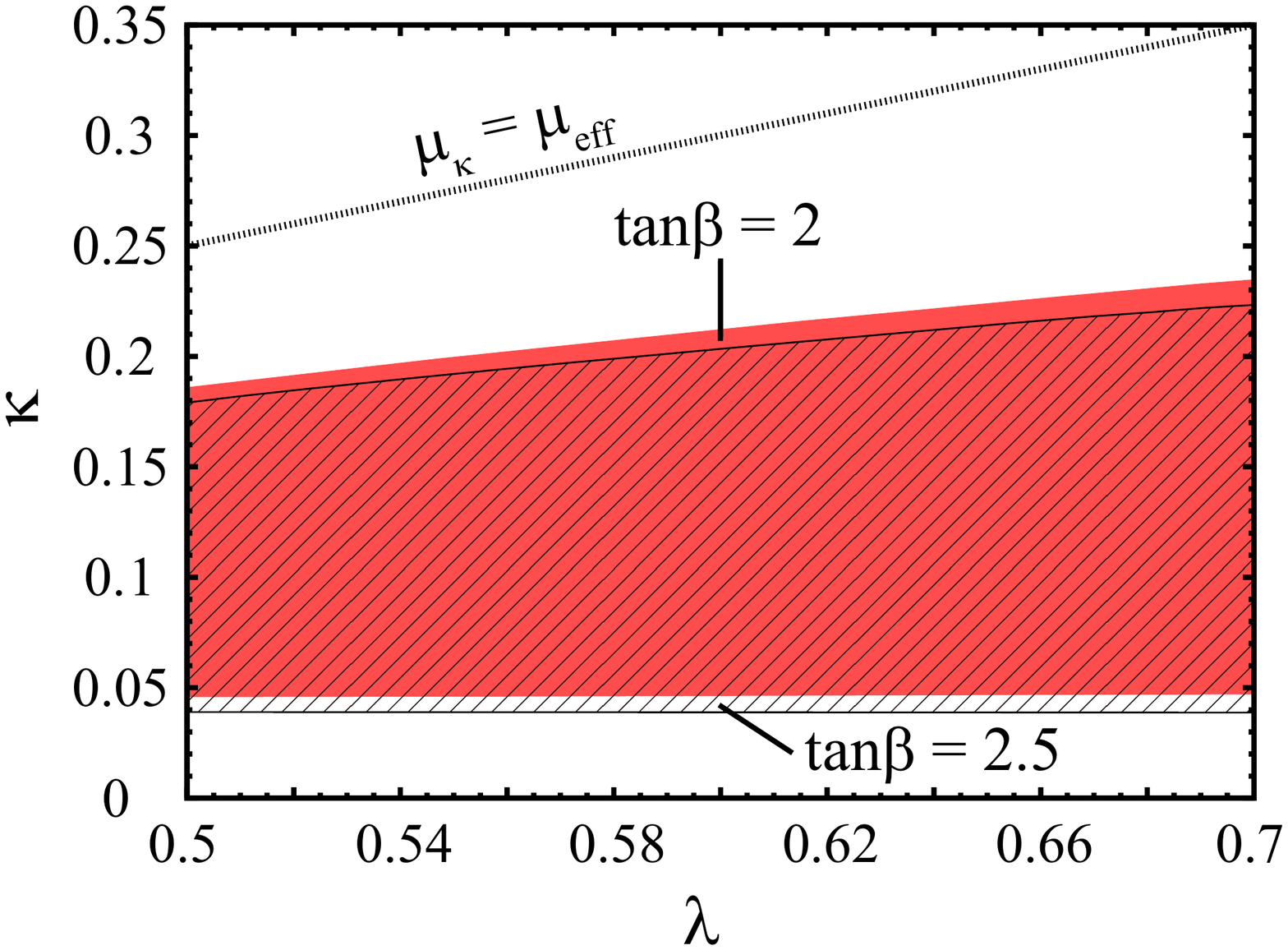} 
\end{center}
\end{minipage}
\hspace{6mm}
\begin{minipage}{83mm}\vspace{-28mm}
\begin{center}
 \includegraphics[width=8.75cm,clip]{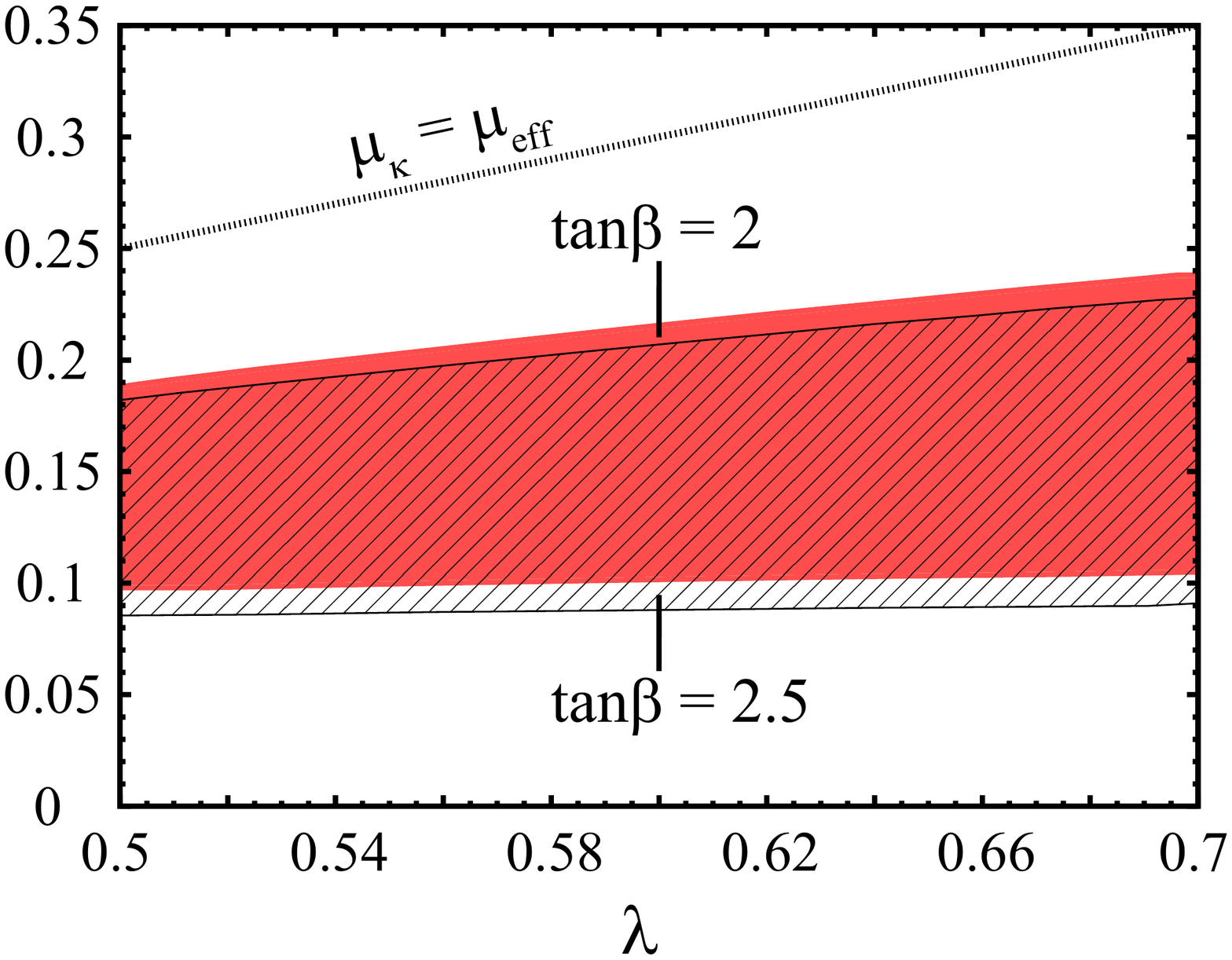}
\end{center}
\end{minipage}
\end{tabular}
\vspace{-33mm}
\caption{%
Favored region in terms of requirements Eqs.~\eqref{eq:cond-sleptonlife}-\eqref{eq:cond-s-rate}
in $\lambda$-$\kappa$ plane.
%%%%%%%%%%%%%%
%%%%%%%%%%%%%%
We took $\tan\beta = 2$ and 2.5 and $c_e=5\times 10^{-10}$ in left panel 
($c_e=2\times 10^{-10}$ in right panel).
%%%%%%%%%%%%%%
%%%%%%%%%%%%%%
Fixed parameters are %
$m_{\tilde \chi ^0_1}=350 \, \mathrm{GeV}$, %
$\delta m = 0.1 \, \mathrm{GeV}$, %
$\sin \theta_f = 0.8$, %
$M_1 = 500 \, \mathrm{GeV}$, and %
$M_2 = 1000 \, \mathrm{GeV}$.
%%%%%%%%%%%%%%
%%%%%%%%%%%%%%
Red and shaded regions are the favored region from $\tan\beta =2$ and $2.5$, 
respectively.
%%%%%%%%%%%%%%
%%%%%%%%%%%%%%
The singlino-like neutralino is no longer the lightest one above the
dotted line.  }
\label{fig:allowed-lk-s-large}
%%%%%%
%%%%%%
\begin{tabular}{l}
\hspace{-10mm}
\begin{minipage}{80mm}\vspace{-38mm}
\begin{center}
 \includegraphics[width=9.3cm,clip]{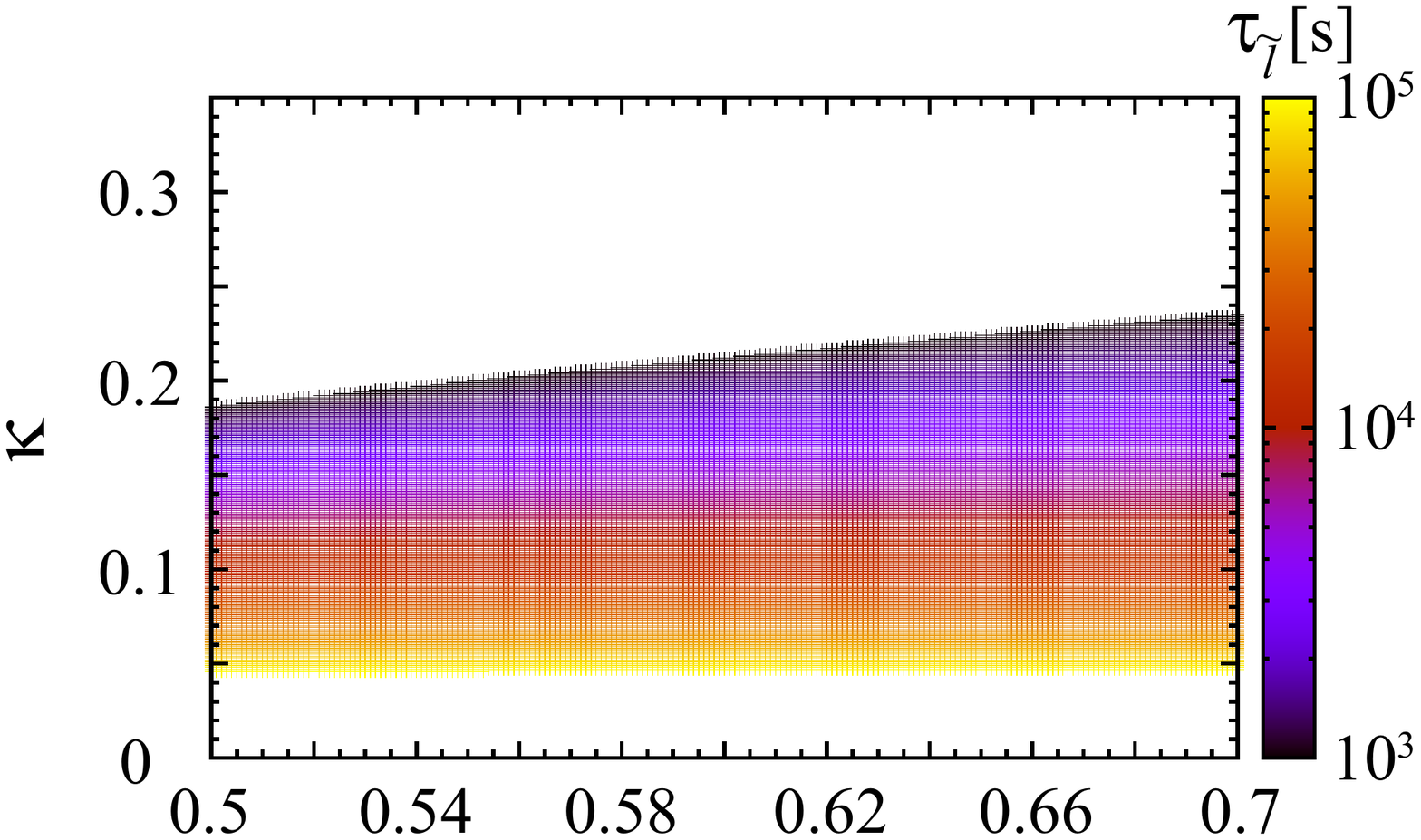}
 \label{fig:}
\end{center}
\end{minipage}
\hspace{12mm}
\begin{minipage}{80mm}\vspace{-38mm}
\begin{center}
 \includegraphics[width=8.8cm,clip]{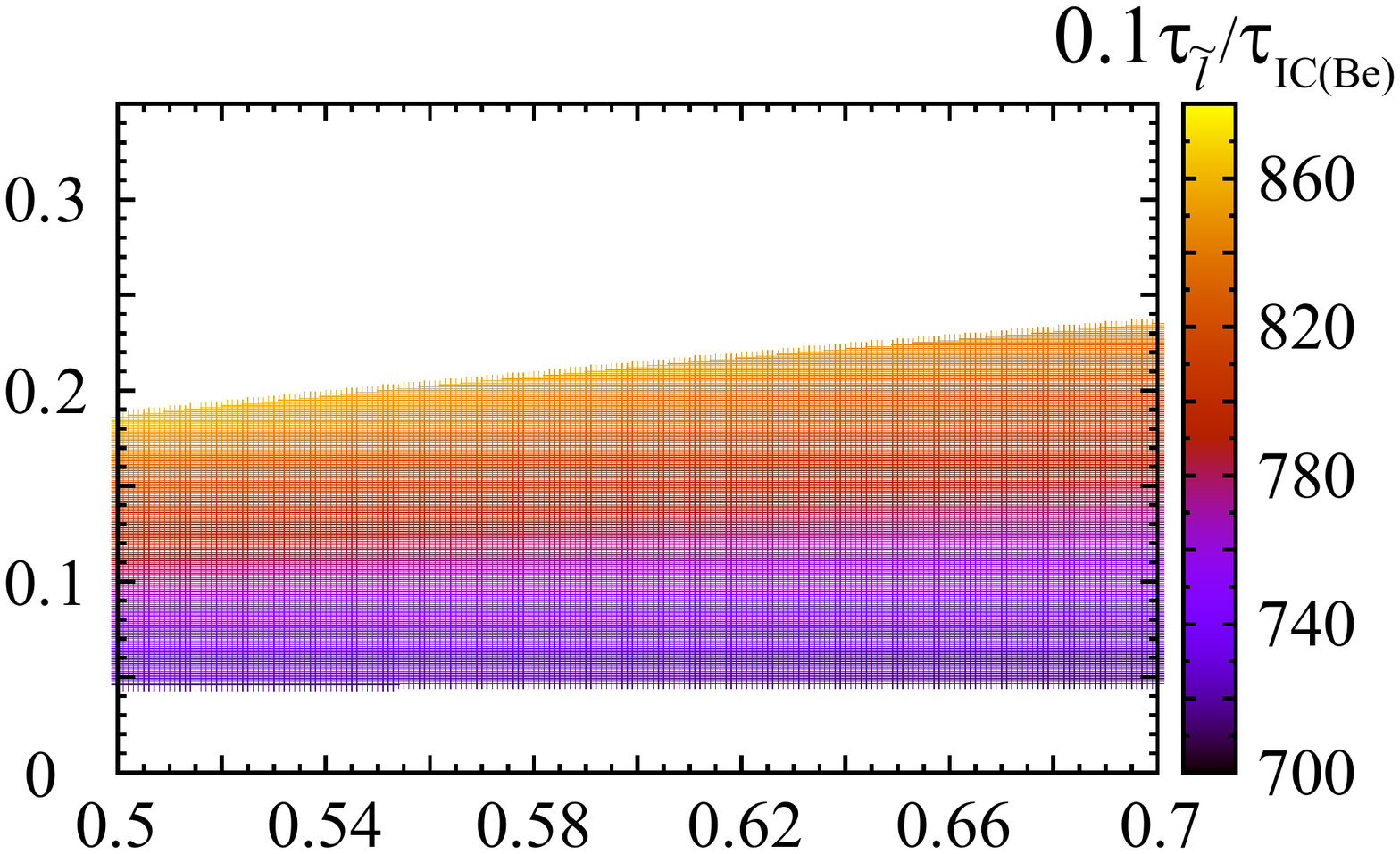}
 \label{fig:}
\end{center}
\end{minipage}
\\[-66mm]
\hspace{-10mm}
\begin{minipage}{79mm}\vspace{-12mm}
\begin{center}
  \includegraphics[width=9.6cm,clip]{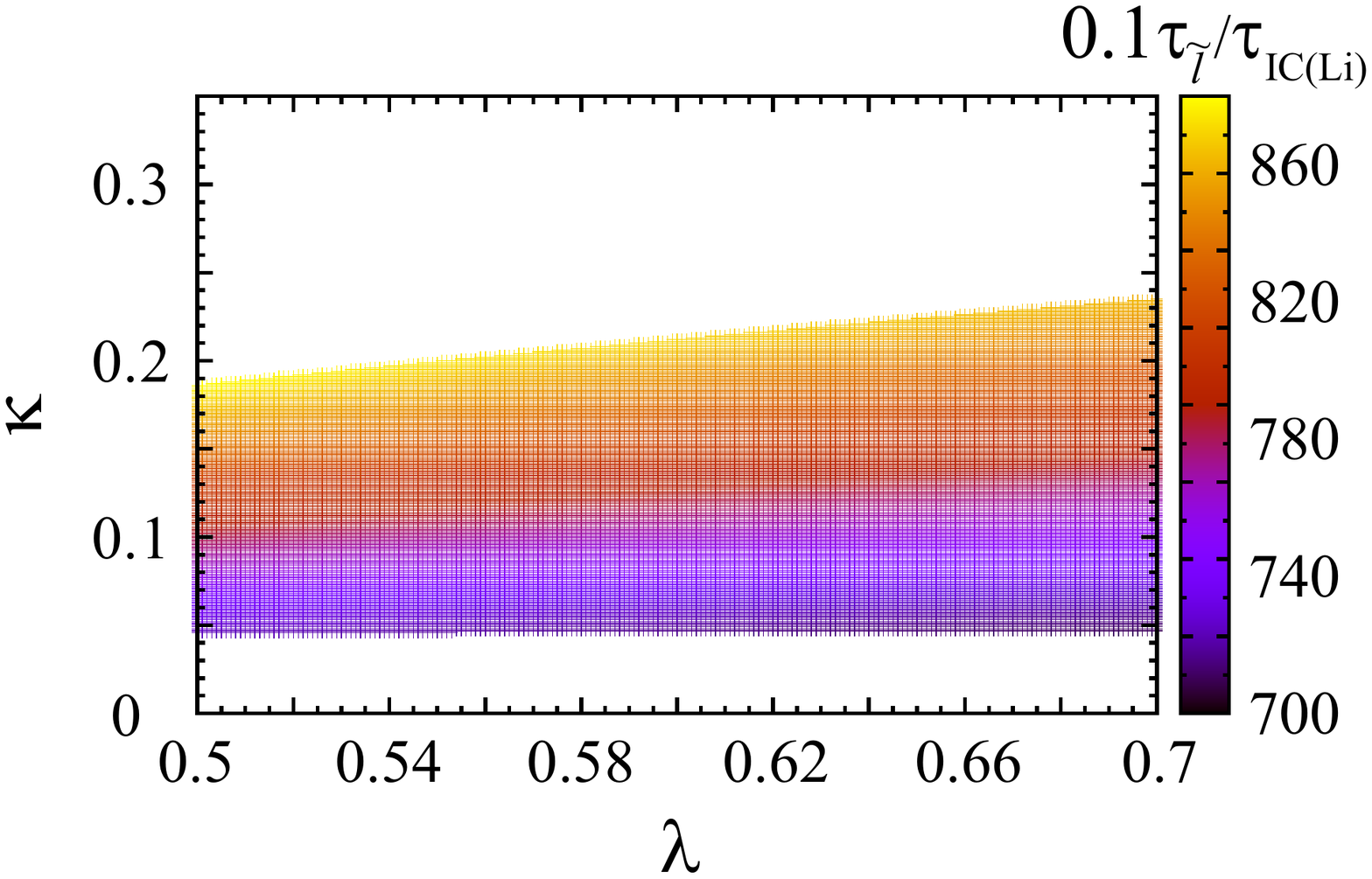}
  \label{fig:}
\end{center}
\end{minipage}
\hspace{14mm}\vspace{-2mm}
\begin{minipage}{79.8mm}\vspace{-12mm}
\begin{center}
 \includegraphics[width=8.7cm,clip]{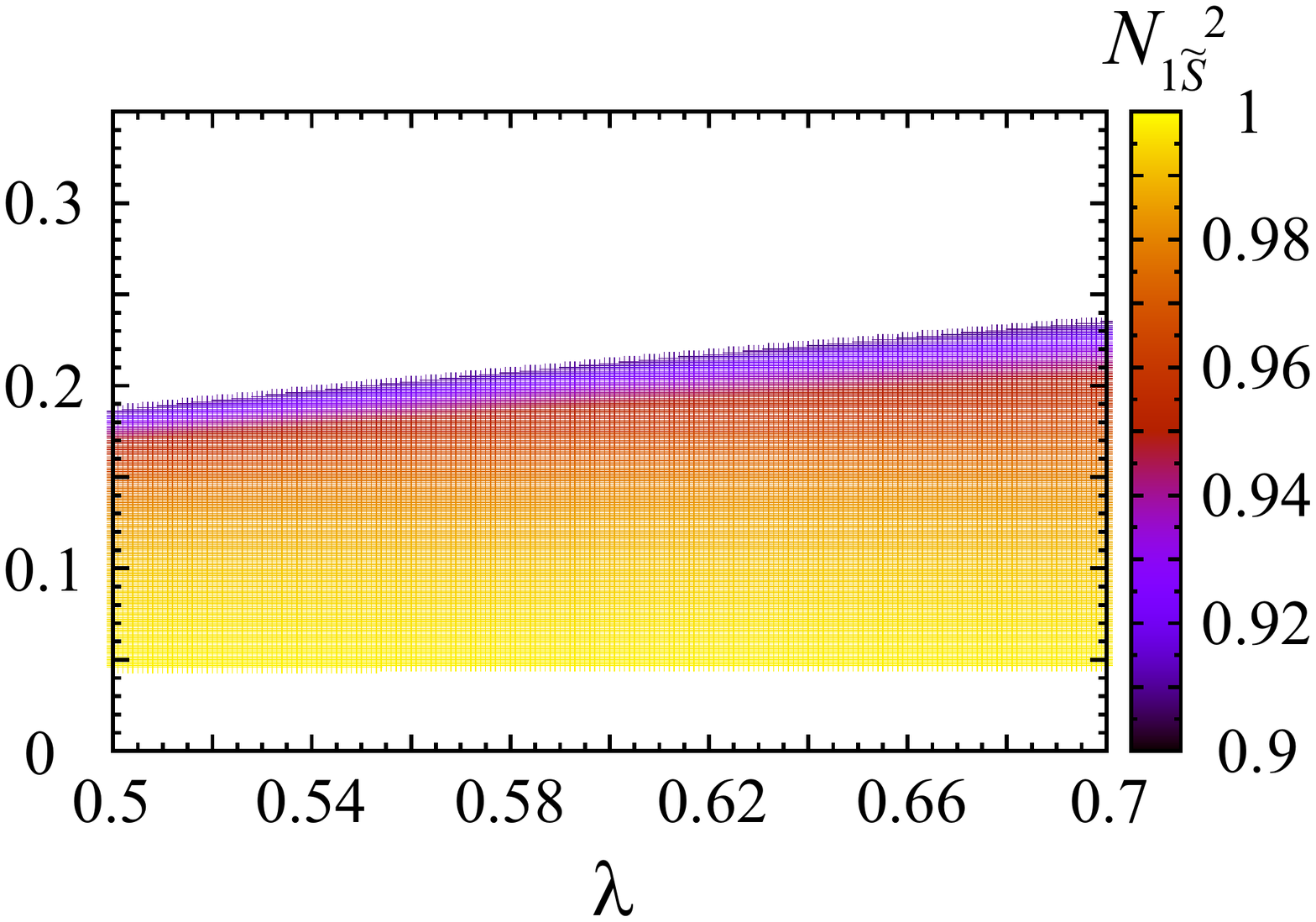}
 \label{fig:}
\end{center}
\end{minipage}
\end{tabular}
\vspace{-38mm}
\caption{
The distribution of the quantities relevant to the requirements, Eqs.~\eqref{eq:cond-sleptonlife}-\eqref{eq:cond-s-rate}, 
in the favored region from $c_e=5\times 10^{-10}$ and $\tan\beta=2$. 
%%%%%%%%%%%%%%
%%%%%%%%%%%%%%
The color bar in each panel shows 
$\tau _{\tilde l}$ (top-left), 
$0.1\tau _{\tilde l}/\tau _{\rm IC}$ for $(^7{\rm Be}~\tilde l^-) \to~^7{\rm Li} + \tilde \chi ^0_1 + \nu _l$ (top-right), 
$0.1\tau _{\tilde l}/\tau _{\rm IC}$ for $(^7{\rm Li}~\tilde l^-) \to~^7{\rm He} + \tilde \chi ^0_1 + \nu _l$ (bottom-left), 
and $N_{1\tilde S}^2$ (bottom-right).
}
\label{fig:allowed-lk-s-large-grad}
%%%%%%
%%%%%%
\end{figure}
\end{center}
\clearpage
\end{widetext}
%%%%%%%%%%%%%%
%%%%%%%%%%%%%%

\begin{widetext}
\clearpage

%%%%%%%%%%%%%%
%%%%%%%%%%%%%%
\begin{table}[tp]
\begin{center}
\caption{
Benchmark points in the favored region of Fig.~\ref{fig:allowed-lk-s-large} for $\tan\beta =2$.
}
\begin{tabular}{ccccc} \hline\hline
~~Parameters~~	& ~~~SL-1~~		& ~~SL-2~~		& ~~SL-3~~		& ~~SL-4~~		\\ \hline %\hline
$c_e$			& $5 \times 10^{-10}$	& $2\times10^{-10}$	& $5\times10^{-10}$	& $5\times10^{-10}$	\\ %\hline
$\lambda$			& 0.68			& 0.68			& 0.695			& 0.68			\\ %\hline
$\kappa$			& 0.22 			& 0.22			& 0.22			& 0.215	 	\\ \hline\hline
\end{tabular}
\label{tab:points2}
\end{center}
\end{table}
%%%%%%%%%%%%%%
%%%%%%%%%%%%%%

\begin{table}[htb]
\begin{center}
\caption{
Spectra and observables at each point (see Tab.~\ref{tab:points2}).
All the dimensionful values are shown in GeV.
The top rows show input parameters.
%We assume the relations for gauging masses, $M_1=M_2/2$ and $M_3=3M_2$, 
%and for squarks and sleptons, 
%$m_{\tilde Q_{1,2}} =m_{\tilde Q_{3}}$,
%$m_{\tilde U_{1,2}}=m_{\tilde U_{3}}$,
%$m_{\tilde D_{1,2}}=m_{\tilde D_{3}}$, 
%$m_{\tilde L_{1,2}}=m_{\tilde L_{3}}$,
%and 
%$m_{\tilde E_{1,2}}=m_{\tilde E_{3}}$.
SL-1 and SL-2 give common results since we omit small flavor mixing of the slepton.
The middle rows show output spectra.
The bottom rows show relic density of the lightest neutralino, spin-independent cross 
section between the lightest neutralino and nucleon, the SUSY contribution to the muon 
anomalous magnetic moment, and the branching ratios of rare decays 
$B_s \to \mu ^+ \mu ^-$ and $b\to s\gamma$, and couplings Eqs.~\eqref{eq:gl} and \eqref{eq:gr} from top to bottom.
}
\begin{tabular}{cccc} \hline\hline
~~{\bf Input}~~	&~~{\bf SL-1, SL-2}~~	&~~{\bf SL-3}~~	&~~{\bf SL-4}~~	\\ \hline 
%$M_1$				& 500.00			& 500.00		& 500.00		\\ \hline
$M_2$				& 1000.0			& 1000.0		& 1000.0		\\ %\hline
%$M_3$				& 3000.0			& 3000.0		& 3000.0		\\ \hline
$A_t$				& $-1500.0$		& $-1800.0$	& $-2800.0$	\\ %\hline
%$A_b$				& $-1500.0$		& $-1800.0$	& $-2800.0$	\\ \hline
%$A_{\tau}$				& $-1500.0$		& $-1800.0$	& $-2800.0$	\\ %\hline
%$A_{\mu}$			& $-1500.0$		& $-1800.0$	& $-2800.0$	\\ \hline
%$m_{\tilde L_1}$		& 356.86			& 357.93		& 361.23		\\ \hline
%$m_{\tilde L_2}$		& 356.86			& 357.93		& 361.23		\\ \hline
$m_{\tilde L_3}$		& 356.86			& 357.93		& 361.23		\\ %\hline
%$m_{\tilde E_1}$		& 353.36			& 353.96		& 355.84		\\ \hline
%$m_{\tilde E_2}$		& 353.36			& 353.96		& 355.84		\\ \hline	
$m_{\tilde E_3}$		& 353.36			& 353.96		& 355.84		\\ %\hline
%$m_{\tilde Q_1}$		& 1000.0			& 1000.0 		& 1000.0		\\ \hline	
%$m_{\tilde Q_2}$		& 1000.0			& 1000.0		& 1000.0		\\ \hline
$m_{\tilde Q_3}$		& 1000.0			& 1000.0		& 1000.0		\\ %\hline
%$m_{\tilde U_1}$		& 1000.0			& 1000.0		& 1000.0		\\ \hline
%$m_{\tilde U_2}$		& 1000.0			& 1000.0		& 1000.0		\\ \hline
%$m_{\tilde U_3}$		& 1000.0			& 1000.0		& 1000.0		\\ %\hline
%$m_{\tilde D_1}$		& 1000.0			& 1000.0		& 1000.0		\\ \hline
%$m_{\tilde D_2}$		& 1000.0			& 1000.0		& 1000.0		\\ \hline
%$m_{\tilde D_3}$		& 1000.0			& 1000.0		& 1000.0		\\ %\hline
$\lambda$				& 0.6800			& 0.6950		& 0.6800		\\ %\hline
$\kappa$				& 0.2200			& 0.2200		& 0.2150		\\ %\hline
$A_{\lambda}$			& 1120.0			& 1180.0		& 1100.0		\\ %\hline
$A_{\kappa}$			& $-10.000$		& $-10.000$	& $-10.000$	\\ %\hline
$\mu _{\rm eff}$		& 527.17			& 537.90		& 538.91		\\ %\hline 
$\tan \beta $			& 2.0000			& 2.0000		& 2.0000		\\ \hline 
~{\bf Output}~	&				&			&			\\ \hline
$h^0_1$				& 127.13			& 126.66		& 124.36 		\\ %\hline
$h^0_2$				& 370.96			& 354.77		& 369.96		\\ %\hline
$h^0_3$				& 1303.6			& 1348.4		& 1313.9		\\ %\hline
$a^0_1$				& 152.72			& 262.92		& 150.73		\\ %\hline
$a^0_2$				& 1303.8			& 1348.8		& 1313.8		\\ %\hline
$H^{\pm}$				& 1295.0			& 1339.0		& 1304.3		\\ %\hline
%$\tilde d_L$			& 827.06			& 827.06		& 827.06		\\ \hline
%$\tilde d_R$			& 826.00 			& 826.00		& 826.00		\\ \hline
%$\tilde u_L$			& 824.70			& 824.70		& 824.70		\\ \hline
%$\tilde u_R$			& 825.28			& 825.28		& 825.28		\\ \hline
%$\tilde s_L$			& 827.06			& 827.06		& 827.06		\\ \hline
%$\tilde s_R$			& 826.00			& 826.00		& 826.00		\\ \hline
%$\tilde c_L$			& 824.70			& 824.70		& 824.70		\\ \hline
%$\tilde c_R$			& 825.28			& 825.28		& 825.28		\\ \hline
%$\tilde b_1$			& 823.06			& 822.55		& 820.98		\\ %\hline
%$\tilde b_2$			& 830.00			& 830.49		& 832.05		\\ %\hline
$\tilde t_1$			& 696.15			& 660.83		& 528.37		\\ %\hline
$\tilde t_2$			& 959.18			& 983.66		& 1059.9		\\ %\hline
%$\tilde e_L$			& 358.63			& 359.70		& 362.98		\\ \hline
%$\tilde e_R$			& 354.93			& 355.52		& 357.40		\\ \hline
%$\tilde \nu _{eL}$		& 353.51			& 354.59		& 357.92		\\ \hline
%$\tilde \mu_L$		& 358.63			& 359.70		& 362.98		\\ \hline
%$\tilde \mu_R$		& 354.93			& 355.52		& 357.40		\\ \hline
%$\tilde \nu _{\mu L}$	& 353.51			& 354.59		& 357.92		\\ \hline
$\tilde \tau _1$			& 350.10			& 350.10 		& 350.10  		\\ %\hline
$\tilde \tau _2$			& 363.35			& 364.99		& 370.03		\\ %\hline
%$\tilde \nu _{\tau L}$	& 353.51			& 354.59		& 357.92		\\ \hline
%$\tilde g$			& 2881.7			& 2881.8		& 2882.2		\\ %\hline
$\tilde \chi ^0_1$		& 350.00			& 350.00 		& 350.00 		\\ %\hline
$\tilde \chi ^0_2$		& 468.58			& 471.30		& 471.37		\\ %\hline
$\tilde \chi ^0_3$		& 547.60			& 558.78		& 559.13		\\ %\hline
$\tilde \chi ^0_4$		& 553.96			& 561.55		& 562.02		\\ %\hline
$\tilde \chi ^0_5$		& 1013.7			& 1014.1		& 1014.1		\\ \hline
%$\tilde \chi ^{\pm}_1$	& 521.17			& 531.63		& 532.58		\\ %\hline
%$\tilde \chi ^{\pm}_2$	& 1013.6			& 1014.1		& 1014.0		\\ \hline%\hline
$\Omega  _{\tilde \chi ^0_1}h^2$	
					& 0.12684			& 0.12672		& 0.12489  	\\ %\hline
%$N_{5\singlino}$		& 0.96284			& 0.96494		& 0.96665  	\\ %\hline
%$\sin \theta _{\tau}$	& 0.79990			& 0.80011		& 0.80008  	\\ \hline  %\hline
$\sigma _{\rm SI}$[cm$^2$]
					&  $1.8983\times 10^{-45}$
					&  $3.1914\times 10^{-45}$
					&  $1.7546\times 10^{-45}$  \\
$\delta a_{\mu}$
					&  $1.0707\times 10^{-10} (>3\sigma )$
					&  $1.1346\times 10^{-10} (>3\sigma )$
					&  $1.3675\times 10^{-10} (>3\sigma )$ \\ %\hline
Br$(B_s^0\to \mu^+\mu^-)$
					&  $3.5373\times 10^{-9} (1\sigma )$
					&  $3.5374\times 10^{-9} (1\sigma )$
					&  $3.5365\times 10^{-9} (1\sigma )$ \\ %\hline
Br$(b \to s \gamma)$
					&  $3.2030\times 10^{-4} (2\sigma )$
					&  $3.1658\times 10^{-4} (2\sigma )$
					&  $3.0507\times 10^{-4} (2\sigma )$  \\ 
$G_{{\rm L}\tau}/c_{\tau}$	& 0.0012549	& 0.0013068	& 0.0012826	\\
$G_{{\rm R}\tau}/c_{\tau}$	& 0.030877	& 0.029833	& 0.029090	\\
$G_{{\rm L}e}/c_{e}$		& 0.0026200	& 0.0025373	& 0.0024746	\\
$G_{{\rm R}e}/c_{e}$		& 0.029853	& 0.028910	& 0.028196	\\
\hline \hline
\end{tabular}
\label{tab:points-s-large}
\end{center}
\end{table}
\clearpage
\end{widetext}

\begin{widetext}
\begin{center}
\begin{figure}[h]
%%%%%%
%%%%%%
\begin{tabular}{l}
\hspace{-19mm}
\begin{minipage}{80mm}\vspace{-16mm}
\begin{center}
 \includegraphics[width=8.7cm,clip]{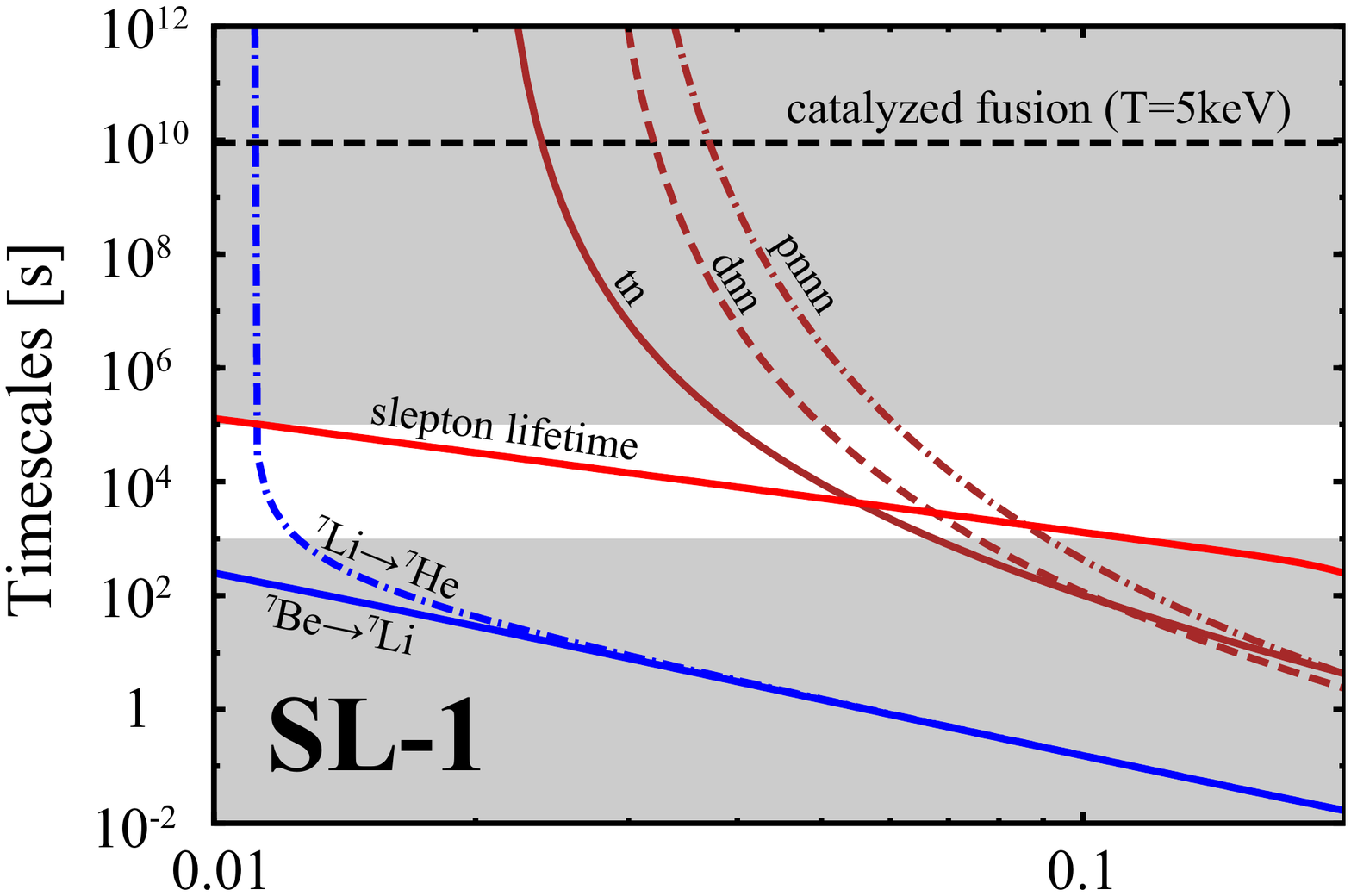}
 \label{fig:}
\end{center}
\end{minipage}
\hspace{1.5mm}\vspace{-1.5cm}
\begin{minipage}{80mm}\vspace{-16mm}
\begin{center}
 \includegraphics[width=9.7cm,clip]{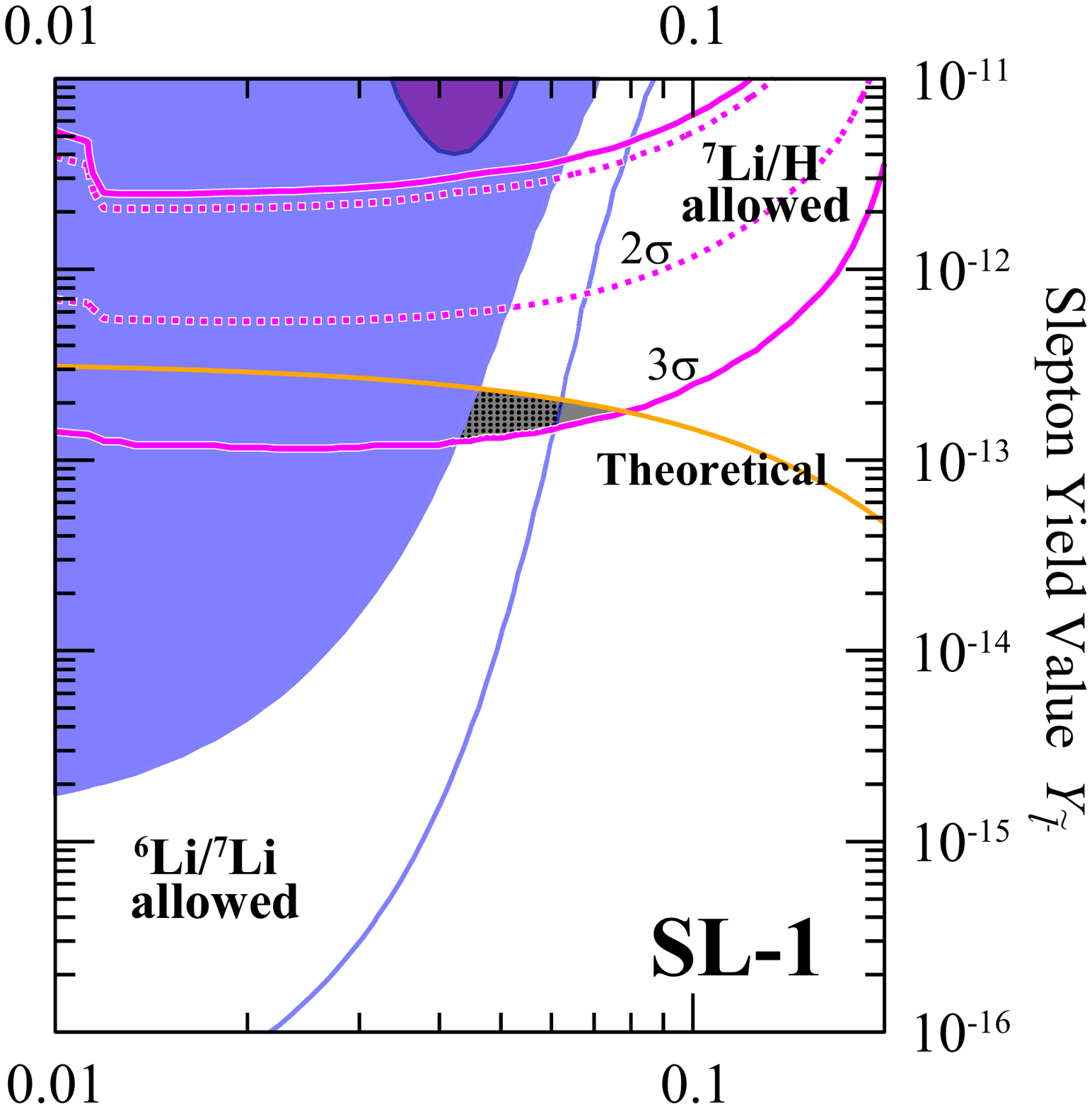}
 \label{fig:}
\end{center}
\end{minipage}
\\[-3mm]
\hspace{-19mm}
\begin{minipage}{79mm}\vspace{-38mm}
\begin{center}
  \includegraphics[width=9cm,clip]{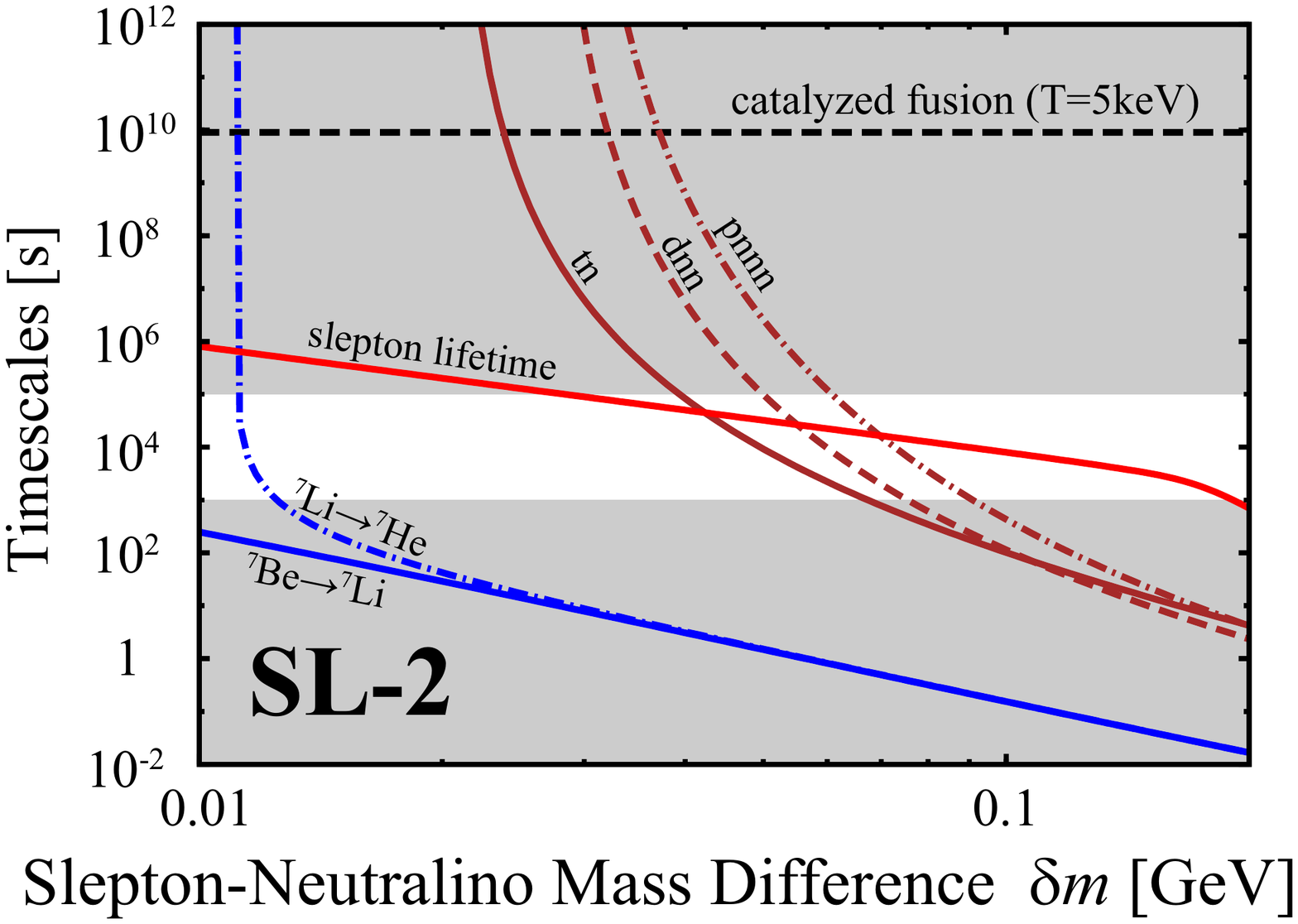}
  \label{fig:}
\end{center}
\end{minipage}
\hspace{-0.5mm}
\begin{minipage}{79.8mm}\vspace{-38mm}
\begin{center}
 \includegraphics[width=9.9cm,clip]{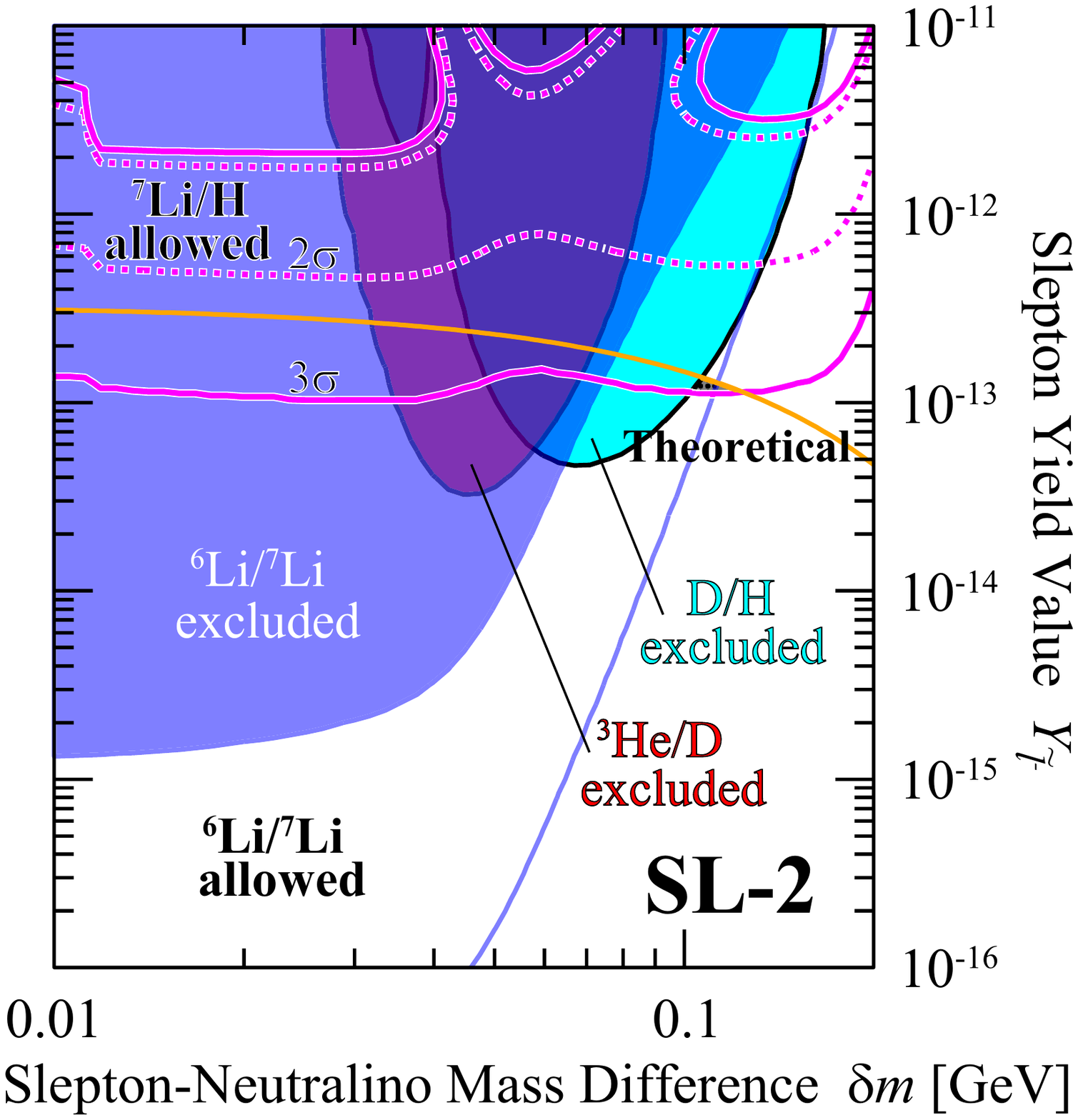}
 \label{fig:}
\end{center}
\end{minipage}
\end{tabular}
\vspace{-28mm}
\caption{
The left panels show 
the slepton lifetime $\tau _{\tilde l}$~(red-solid line; ``slepton lifetime"), 
the timescales of the internal conversion processes \eqref{eq:icbe}~(blue-solid line; ``$^7$Be$\to$$^7$Li"), \eqref{eq:icli}~(blue-dash-dotted line; ``$^7$Li$\to$$^7$He"), 
the $^4$He spallation processes \eqref{eq:tn} (brown-solid line; ``tn"), \eqref{eq:dnn}~(brown-dashed line; ``dnn"), and \eqref{eq:pnnn}~(brown-dash-dotted line; ``pnnn"), 
as a function of the mass difference between the slepton and the neutralino 
at SL-1 (top panel) and SL-2 (bottom panel).
We also show the timescale of the catalyzed fusion \eqref{eq:cf} at the temperature $T=5$~keV ($5\times10^4$~s) when ($^4$He~$\tilde l^-$) is formed as horizontal black-dashed line.
In the shaded regions, Eq.~\eqref{eq:cond-sleptonlife} is not satisfied.
The right panels show the allowed regions from observational light element abundances on $\delta m$-$Y_{\tilde l^-}$ plane at SL-1 (top panel) and SL-2 (bottom panel).
%%%%%%%%%%%%%%
%%%%%%%%%%%%%%
The regions surrounded by magenta-dotted(-solid) lines are allowed by observed 
$^7$Li/H abundance at 2$\sigma$(3$\sigma$).
%%%%%%%%%%%%%%
%%%%%%%%%%%%%%
The regions between the blue-solid line and the blue region are allowed by 
observed $^6$Li/$^7$Li abundance at 2$\sigma$.
%%%%%%%%%%%%%%
%%%%%%%%%%%%%%
The orange-solid lines (``Theoretical") represent the yield value of the slepton 
when the BBN starts as a function of the mass difference.
%%%%%%%%%%%%%%
%%%%%%%%%%%%%%
The colored regions are excluded for 
$^6$Li/$^7$Li (blue region; ``$^6$Li/$^7$Li excluded"), 
$^3$He/D (red region; ``$^3$He/D excluded"), and 
D/H (cyan region; ``D/H excluded"), respectively. 
%%%%%%%%%%%%%%
%%%%%%%%%%%%%%
The shadowed and dotted regions are allowed by only $^7$Li/H (3$\sigma$) 
and both $^7$Li/H (3$\sigma$) and $^6$Li/$^7$Li (2$\sigma$), respectively.
}
\label{fig:large-lk-s_bbn1}
%%%%%%
%%%%%%
\end{figure}
\end{center}
\clearpage
\end{widetext}
%%%%%%%%%%%%%%
%%%%%%%%%%%%%%

\setcounter{figure}{5}

\begin{widetext}
\begin{center}
\begin{figure}[h]
%%%%%%
%%%%%%
\begin{tabular}{l}
\hspace{-23mm}
\begin{minipage}{80mm}\vspace{-20mm}
\begin{center}
 \includegraphics[width=8.6cm,clip]{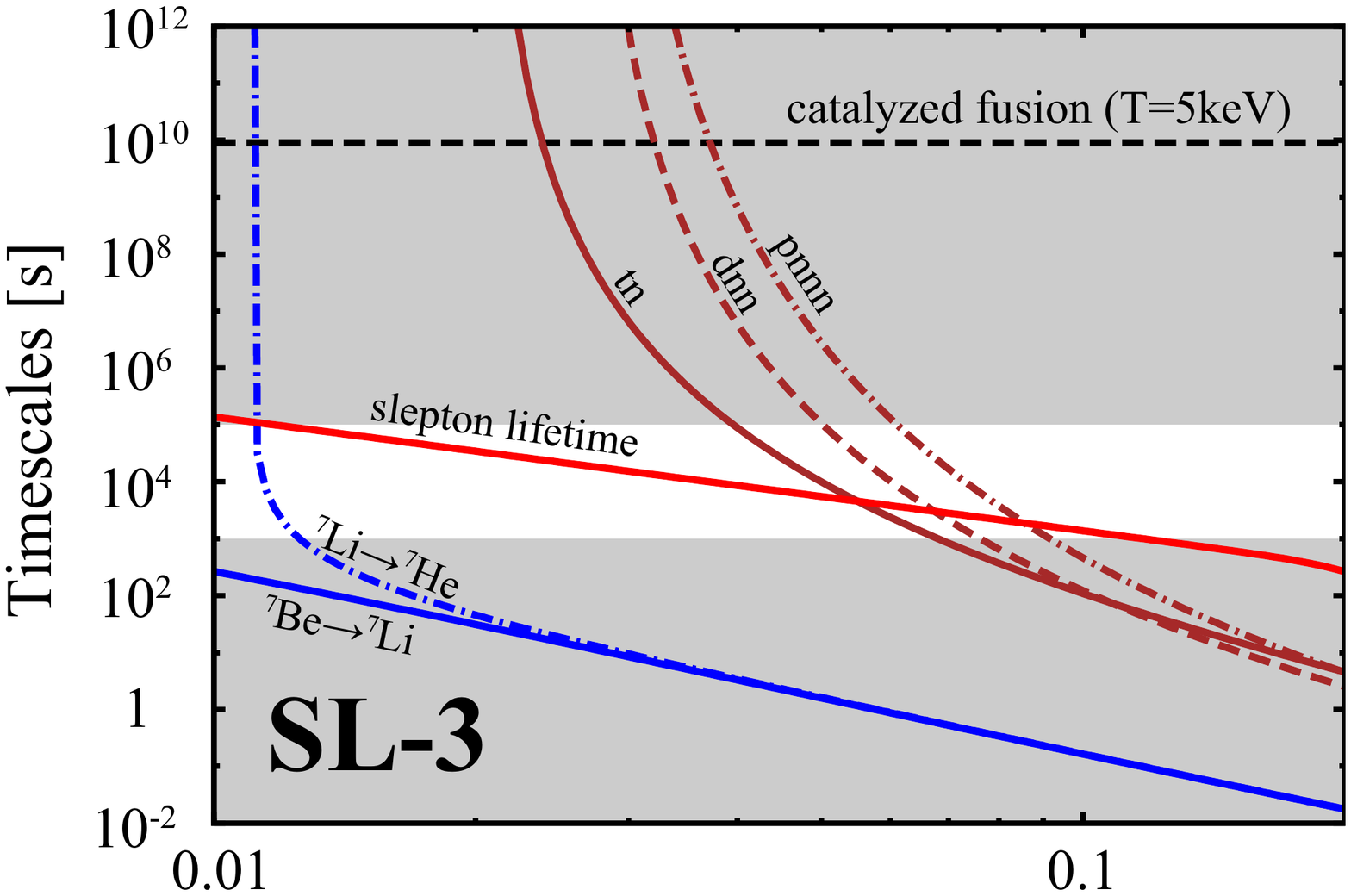}
 \label{fig:}
\end{center}
\end{minipage}
\hspace{-5mm}\vspace{-1.5cm}
\begin{minipage}{80mm}\vspace{-20mm}
\begin{center}
 \includegraphics[width=10.3cm,clip]{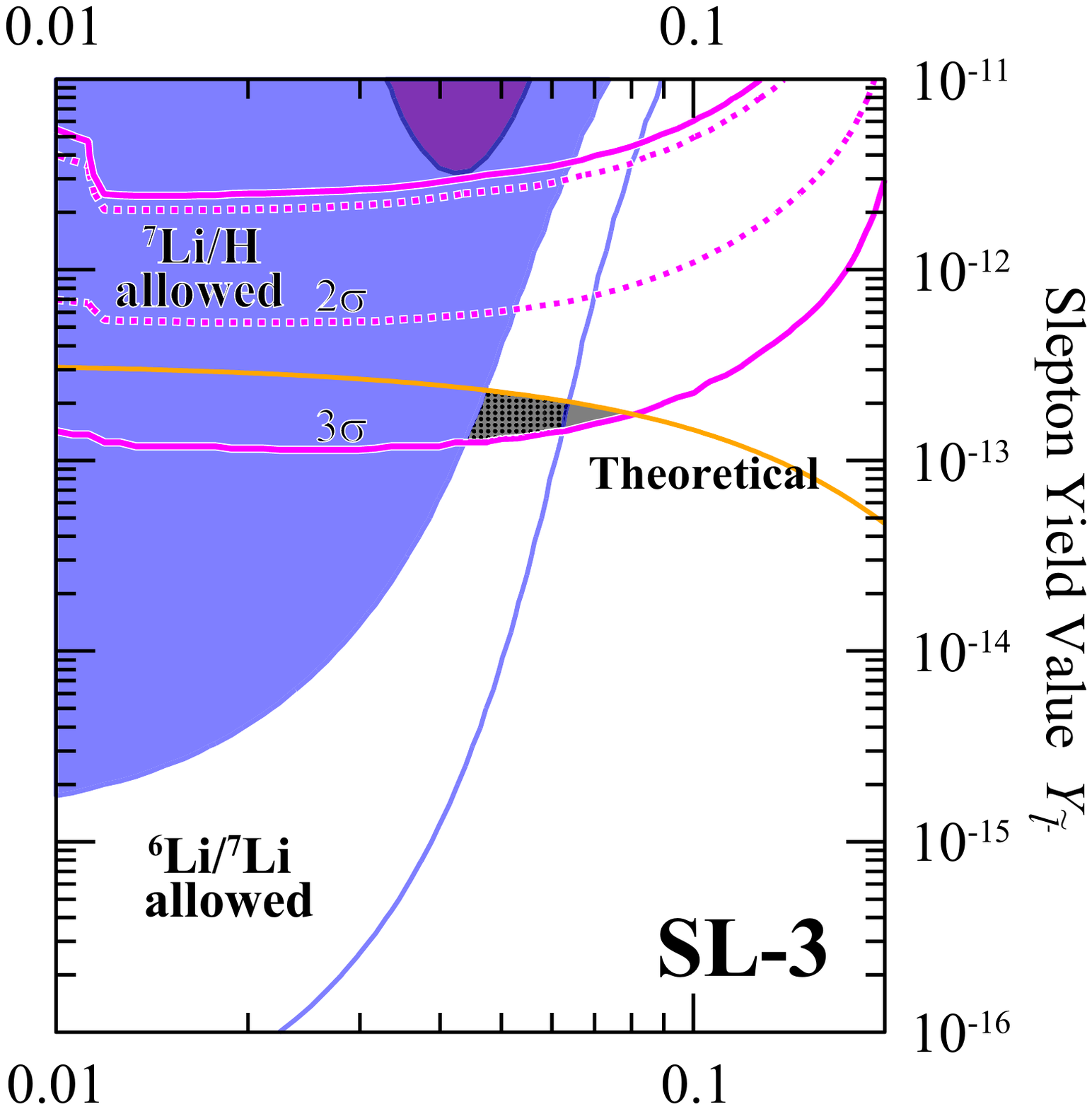}
 \label{fig:}
\end{center}
\end{minipage}
\\[-3mm]
\hspace{-23mm}
\begin{minipage}{79mm}\vspace{-42mm}
\begin{center}
  \includegraphics[width=9.6cm,clip]{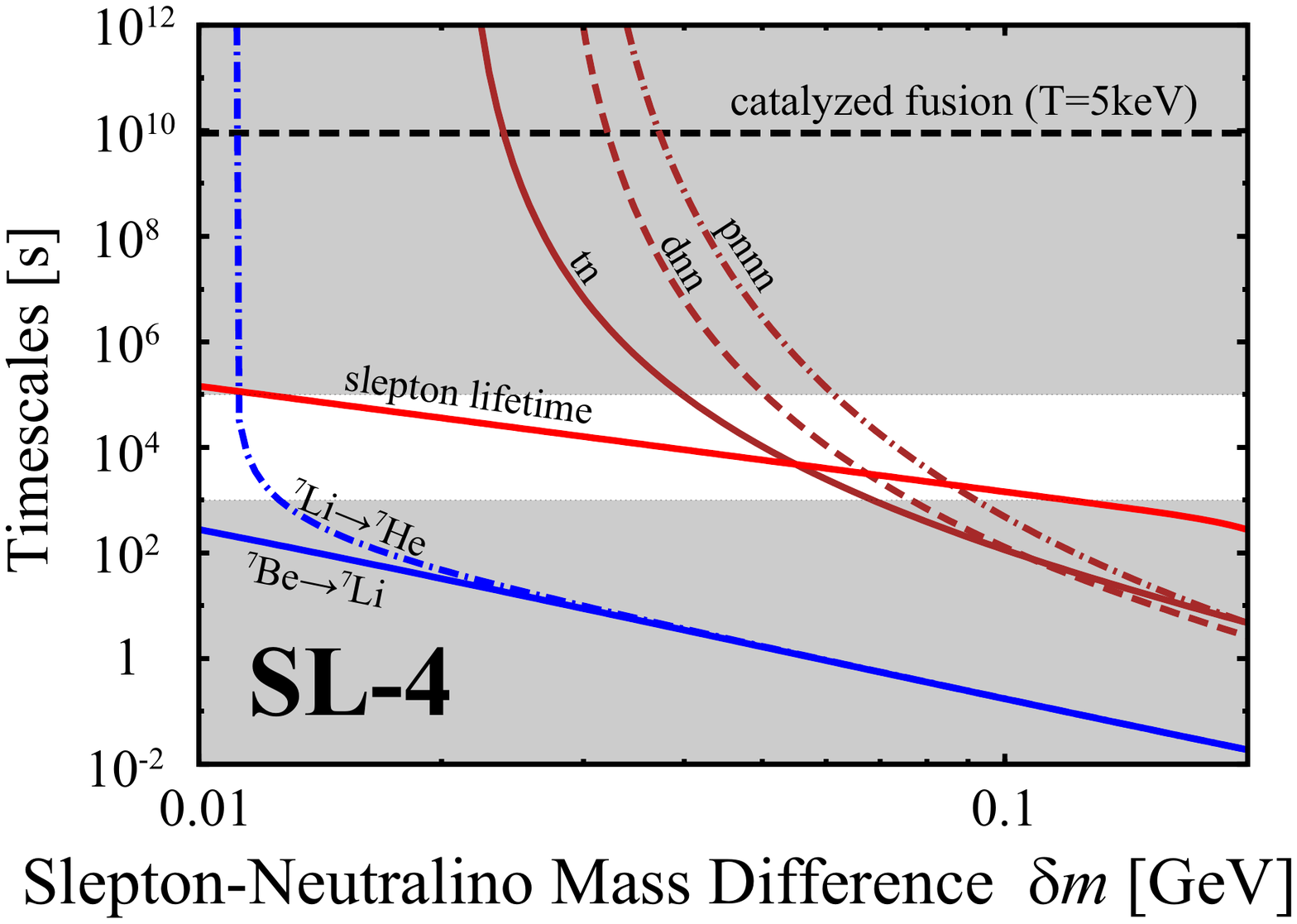}
  \label{fig:}
\end{center}
\end{minipage}
\hspace{-4mm}
\begin{minipage}{79.8mm}\vspace{-42mm}
\begin{center}
 \includegraphics[width=10.3cm,clip]{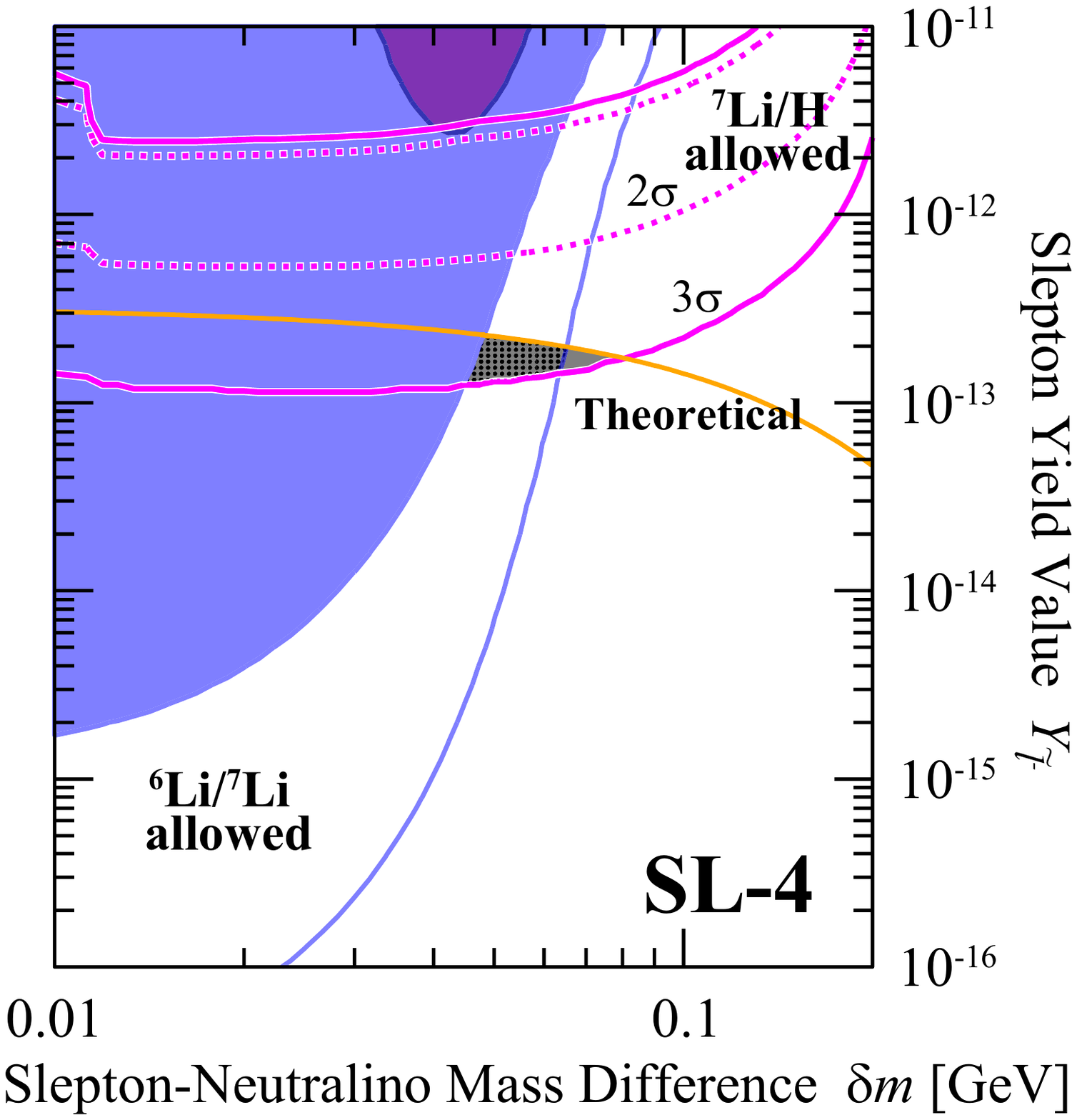}
 \label{fig:}
\end{center}
\end{minipage}
\end{tabular}
\vspace{-28mm}
\caption{
The results at SL-3 (top panels) and SL-4 (bottom panels) are shown.
The meanings of the lines and regions are same as those in the previous page.
}
\label{fig:large-lk-s_bbn2}
%%%%%%
%%%%%%
\end{figure}
\end{center}
\clearpage
\end{widetext}
%%%%%%%%%%%%%%
%%%%%%%%%%%%%%

%%%%%%%%%%%%%%%%%%%%%%%%%%%%%%%%%%%%%%%%%%%%%%%%
%%%%%%%%%%%%%%%%%%%%%%%%%%%%%%%%%%%%%%%%%%%%%%%%
\subsection{Bino-like neutralino LSP; large $\lambda$-$\kappa$ region with small $\tan\beta$} \label{sec:lk-b-large} 
%%%%%%%%%%%%%%%%%%%%%%%%%%%%%%%%%%%%%%%%%%%%%%%%
%%%%%%%%%%%%%%%%%%%%%%%%%%%%%%%%%%%%%%%%%%%%%%%%

%%%%%%%%%%%%%%
%%%%%%%%%%%%%%
We finally show the results in the third case where the neutralino LSP is bino-like with 
relatively large $\lambda$, $\kappa$ and small $\tan\beta$.
%%%%%%%%%%%%%%
%%%%%%%%%%%%%%
In this case, the first term is dominant in the couplings Eqs.~\eqref{eq:gl} and 
\eqref{eq:gr}, and they hardly depend on the parameters, 
$\lambda, \kappa, \tan\beta, \mu _{\rm eff}, M_1$, and $M_2$, since $N_{1\tilde B}^2 \simeq 1$.
%%%%%%%%%%%%%%
%%%%%%%%%%%%%%
Therefore, we do not have to check the dependence on these parameters of the slepton lifetime and relevant timescales of the exotic BBN reactions 
\eqref{eq:ic} and \eqref{eq:spa} since they also do not depend on these parameters.
%%%%%%%%%%%%%%
%%%%%%%%%%%%%%
%The requirements Eqs.~\eqref{eq:cond-sleptonlife}, \eqref{eq:cond-ictime}, 
%and \eqref{eq:cond-b-rate} are satisfied as long as we apply the values of 
%$c_e, m_{\tilde \chi ^0_1}, \delta m$, and $\sin \theta _f$, which give favorable 
%slepton lifetime and relevant timescales of the exotic BBN reactions 
%in the MSSM with bino-neutralino LSP.
%%%%%%%%%%%%%%
%%%%%%%%%%%%%%
%Thus we skip the step to narrow the parameter space.
%%%%%%%%%%%%%%
%%%%%%%%%%%%%%

%%%%%%%%%%%%%%%%%%%%%%%%%%%%%%%%%%%%%%%%%%%%%%%%
\subsubsection{Benchmark points} \label{sec:bmp-b-large} %%%%%%%%%%%%%%%%%
%%%%%%%%%%%%%%%%%%%%%%%%%%%%%%%%%%%%%%%%%%%%%%%%

%%%%%%%%%%%%%%
%%%%%%%%%%%%%%
We take four reference points in the favored region for $\tan\beta=2$ 
as shown in Table~\ref{tab:points3}.
%%%%%%%%%%%%%%
%%%%%%%%%%%%%%
Table~\ref{tab:points-b-large} shows the spectra and observables at these points. 
%%%%%%%%%%%%%%
%%%%%%%%%%%%%%
All the dimensionful values are represented in GeV.
%%%%%%%%%%%%%%
%%%%%%%%%%%%%%
The top rows show input parameters and the middle rows show output spectra.
%%%%%%%%%%%%%%
%%%%%%%%%%%%%%
Every points give the observed Higgs mass.
%%%%%%%%%%%%%%
%%%%%%%%%%%%%%
As in the results of Sec.~\ref{sec:lk-s-large}, observed Higgs mass is obtained 
by virtue of large contribution of the tree terms.
%%%%%%%%%%%%%%
%%%%%%%%%%%%%%
In this respect, the result is completely different from the case with bino-like neutralino 
in the MSSM, where the 1-loop contribution lifts up the Higgs mass.
%%%%%%%%%%%%%%
%%%%%%%%%%%%%%

%%%%%%%%%%%%%%
%%%%%%%%%%%%%%
In the bottom rows, we show relic density of the lightest neutralino, spin-independent cross 
section between the lightest neutralino and nucleon, the SUSY contribution to the muon 
anomalous magnetic moment, and the branching ratios of rare decays 
$B_s \to \mu ^+ \mu ^-$ and $b\to s\gamma$, from top to bottom.
%%%%%%%%%%%%%%
%%%%%%%%%%%%%%
At each point, the dark matter relic density is in range of the measured value~\cite{Ade:2013zuv}.
%%%%%%%%%%%%%%
%%%%%%%%%%%%%%
The spin-independent cross sections are about one order of magnitude smaller than 
those at points we chose in the previous section, and below the present experimental 
bound.
%%%%%%%%%%%%%%
%%%%%%%%%%%%%%
The calculated values of $\delta a_{\mu}$ at the points are below 3$\sigma$ range
which is caused by small $\tan\beta$.
%%%%%%%%%%%%%%
%%%%%%%%%%%%%%
For the branching ratio of $B_s\to \mu^+\mu^-$ and $b\to s\gamma$, we obtained the 
values within 1$\sigma$ and 2$\sigma$, respectively.
%%%%%%%%%%%%%%
%%%%%%%%%%%%%%

%%%%%%%%%%%%%%%%%%%%%%%%%%%%%%%%%%%%%%%%%%%%%%%%
%%%%%%%%%%%%%%%%%%%%%%%%%%%%%%%%%%%%%%%%%%%%%%%%
\subsubsection{BBN results at the benchmark points} \label{sec:bbn-b-large}%%%%%%%%
%%%%%%%%%%%%%%%%%%%%%%%%%%%%%%%%%%%%%%%%%%%%%%%%
%%%%%%%%%%%%%%%%%%%%%%%%%%%%%%%%%%%%%%%%%%%%%%%%

%%%%%%%%%%%%%%
%%%%%%%%%%%%%%
The left panels in Fig.~\ref{fig:large-lk-b_bbn1}
show the slepton lifetime $\tau _{\tilde l}$~(red-solid line; ``slepton lifetime"), 
the timescales of the internal conversion processes
\eqref{eq:icbe}~(blue-solid line; ``$^7$Be$\to$$^7$Li"),
\eqref{eq:icli}~(blue-dash-dotted line; ``$^7$Li$\to$$^7$He"), 
the $^4$He spallation processes
\eqref{eq:tn}~(brown-solid line; ``tn"), 
\eqref{eq:dnn}~(brown-dashed line; ``dnn"), and 
\eqref{eq:pnnn}~(brown-dash-dotted line; ``pnnn"), 
as a function of the mass difference between the slepton NLSP and the neutralino LSP 
at BL-1, BL-2, BL-3, and BL-4 from top to bottom, respectively.
%%%%%%%%%%%%%%
%%%%%%%%%%%%%%
The horizontal black-dashed line represents the timescale of the catalyzed 
fusion process \eqref{eq:cf}~\cite{Hamaguchi:2007mp} at the temperature $T=5$~keV 
($5\times10^4$~s) when ($^4$He~$\tilde l^-$) is formed.
%%%%%%%%%%%%%%
%%%%%%%%%%%%%%
In the right panels
horizontal axis is the mass difference between the slepton NLSP and 
the neutralino LSP, and vertical axis is the yield value of the slepton at the beginning of the BBN.
%$Y_{\tilde l^-}=n_{\tilde l^-}/s$, where $n_{\tilde l^-}$ is the number density
%of the slepton and $s$ is the entropy density.
%%%%%%%%%%%%%%%
%%%%%%%%%%%%%%%

%%%%%%%%%%%%%%
%%%%%%%%%%%%%%
We show the allowed regions in the right panels of Fig.~\ref{fig:large-lk-b_bbn1} 
which we obtain by comparing theoretical values to observational ones for light 
element abundances at BL-1, BL-2, BL-3, and BL-4 from top to bottom, 
respectively.
%%%%%%%%%%%%%%
%%%%%%%%%%%%%%
The lines and regions are the same as those in the previous section.
%%%%%%%%%%%%%%
%%%%%%%%%%%%%%
At BL-1, BL-2, and BL-3, we obtain allowed region only for $^7$Li/H (3$\sigma$) 
while at BL-4, that for $^6$Li/$^7$Li (2$\sigma$) is also obtained simultaneously.
%%%%%%%%%%%%%%
%%%%%%%%%%%%%%

%%%%%%%%%%%%%%
%%%%%%%%%%%%%%
At BL-2 and BL-3, $\kappa$ and $\lambda$ are small compared with those at BL-1, 
respectively.
%%%%%%%%%%%%%%
%%%%%%%%%%%%%%
As was mentioned in the beginning of this section, the slepton lifetime and timescales of the
internal conversion processes \eqref{eq:ic} and the $^4$He spallation 
processes \eqref{eq:spa} hardly depend on the NMSSM parameters.
%%%%%%%%%%%%%%
%%%%%%%%%%%%%%
%Hence the same is applied for the slepton lifetime and timescales of the
%internal conversion processes \eqref{eq:ic} and the $^4$He spallation 
%processes \eqref{eq:spa}.
%%%%%%%%%%%%%%
%%%%%%%%%%%%%%
This is the reason why almost no difference exists among the results at BL-1, 
BL-2, and BL-3.
%%%%%%%%%%%%%%
%%%%%%%%%%%%%%
The couplings $G_{{\rm L}f}$ and $G_{{\rm R}f}$ are large compared with 
those in the singlino-like LSP scenario, and hence 
the timescales of the internal conversion processes \eqref{eq:ic} and 
the $^4$He spallation processes \eqref{eq:spa} are one or two order of 
magnitude shorter than those in the singlino-like LSP scenario.
%%%%%%%%%%%%%%
%%%%%%%%%%%%%%
Therefore, the observed abundance 
of $^7$Li/H is obtained in smaller slepton yield value compared with 
the singlino-like LSP scenarios, Fig.~\ref{fig:small-lk-s_bbn1} and \ref{fig:large-lk-s_bbn1}.
%%%%%%%%%%%%%%
%%%%%%%%%%%%%%
In addition, the excluded regions by $^3$He/D and D/H lie downward compared 
to those of the singlino-LSP scenario (e.g., compare the results at SL-2 and BL-1).
%%%%%%%%%%%%%%
%%%%%%%%%%%%%%
Indeed, the results at BL-1, BL-2, and BL-3 are almost same as the result in the MSSM, 
Fig.~3 in Ref.~\cite{Jittoh:2011ni}, where the same values for 
$c_e, m_{\tilde \chi ^0_1}, \sin \theta _f$ and CP-violating phase are used.
%%%%%%%%%%%%%%
%%%%%%%%%%%%%%

%%%%%%%%%%%%%%
%%%%%%%%%%%%%%
Only at BL-4, tiny flavor mixing exists.
%%%%%%%%%%%%%%　リチウム６、７どちらも説明できる。
%%%%%%%%%%%%%%
This result is similar to that in the MSSM, Fig.~4 (middle panel) in Ref.~\cite{Kohri:2012gc}, 
where same values for $c_e, m_{\tilde \chi ^0_1}, \sin \theta _f$ 
and CP-violating phase are used.
%%%%%%%%%%%%%%
%%%%%%%%%%%%%%

\clearpage
\begin{widetext}

%%%%%%%%%%%%%%
%%%%%%%%%%%%%%
\begin{table}[t]
\begin{center}
\caption{
Benchmark points on $\lambda$-$\kappa$ plane for $\tan\beta =2$. 
}
\begin{tabular}{ccccc} \hline\hline
~~Parameters~~	& ~~~BL-1~~	& ~~BL-2~~	& ~~BL-3~~	& ~~BL-4~~		\\ \hline %\hline
$c_e$			& 0			& 0			& 0			& $5 \times 10^{-11}$	\\ %\hline
$\lambda$			& 0.68		& 0.68		& 0.6			& 0.68	 		\\ %\hline
$\kappa$			& 0.32 		& 0.2			& 0.32		& 0.32			\\ \hline\hline
\end{tabular}
\label{tab:points3}
\end{center}
\end{table}
%%%%%%%%%%%%%%
%%%%%%%%%%%%%%

%%%%%%%%%%%%%%
%%%%%%%%%%%%%%
\begin{table}[htb]
\begin{center}
\caption{
Spectra and observables at each point (see Tab.~\ref{tab:points3}).
All the dimensionful values are shown in GeV.
The top rows show input parameters.
%We assume the relations for gauging masses, $M_1=M_2/2$ and $M_3=3M_2$, 
%and for squarks and sleptons, 
%$m_{\tilde Q_{1,2}} =m_{\tilde Q_{3}}$,
%$m_{\tilde U_{1,2}}=m_{\tilde U_{3}}$,
%$m_{\tilde D_{1,2}}=m_{\tilde D_{3}}$, 
%$m_{\tilde L_{1,2}}=m_{\tilde L_{3}}$,
%and 
%$m_{\tilde E_{1,2}}=m_{\tilde E_{3}}$.
BL-1 and BL-4 give common results since we omit small flavor mixing of the slepton.
The middle rows show output spectra.
The bottom rows show relic density of the lightest neutralino, spin-independent cross 
section between the lightest neutralino and nucleon, the SUSY contribution to the muon 
anomalous magnetic moment, and the branching ratios of rare decays 
$B_s \to \mu ^+ \mu ^-$ and $b\to s\gamma$, and couplings Eqs.~\eqref{eq:gl} and \eqref{eq:gr} from top to bottom.
}
\begin{tabular}{cccc} \hline\hline
~~{\bf Input}~~			& ~~{\bf BL-1, BL-4}~~	&~~{\bf BL-2}~~	& ~~{\bf BL-3}~~	\\ \hline
%$M_1$				& 500.00			& 500.00			& 500.00		\\ \hline
$M_2$				& 713.36			& 713.25			& 712.97		\\ %\hline
%$M_3$				& 3000.0			& 3000.0			& 3000.0		\\ \hline
$A_t$				& $-1500.0$		& $-1500.0$		& $-1500.0$	\\ %\hline
%$A_b$				& $-1500.0$		& $-1500.0$		& $-1500.0$	\\ \hline
%$A_{\tau}$				& $-1500.0$		& $-1500.0$		& $-1500.0$	\\ %\hline
%$A_{\mu}$			& $-1500.0$		& $-1500.0$		& $-1500.0$	\\ \hline
%$m_{\tilde L_1}$		& 359.32			& 359.32			& 359.32		\\ \hline
%$m_{\tilde L_2}$		& 359.32			& 359.32			& 359.32		\\ \hline
$m_{\tilde L_3}$		& 359.32			& 359.32			& 359.32		\\ %\hline
%$m_{\tilde E_1}$		& 354.76			& 354.76			& 354.76		\\ \hline
%$m_{\tilde E_2}$		& 354.76			& 354.76			& 354.76		\\ \hline	
$m_{\tilde E_3}$		& 354.76			& 354.76			& 354.76		\\ %\hline
%$m_{\tilde Q_1}$		& 1000.0	 		& 1000.0			& 1000.0		\\ \hline	
%$m_{\tilde Q_2}$		& 1000.0			& 1000.0			& 1000.0		\\ \hline
$m_{\tilde Q_3}$		& 1000.0			& 1000.0			& 1000.0		\\ %\hline
%$m_{\tilde U_1}$		& 1000.0			& 1000.0			& 1000.0		\\ \hline
%$m_{\tilde U_2}$		& 1000.0			& 1000.0			& 1000.0		\\ \hline
%$m_{\tilde U_3}$		& 1000.0			& 1000.0			& 1000.0		\\ %\hline
%$m_{\tilde D_1}$		& 1000.0			& 1000.0			& 1000.0		\\ \hline
%$m_{\tilde D_2}$		& 1000.0			& 1000.0			& 1000.0		\\ \hline
%$m_{\tilde D_3}$		& 1000.0			& 1000.0			& 1000.0		\\ %\hline
$\lambda$				& 0.6800			& 0.6800			& 0.6000		\\ %\hline
$\kappa$				& 0.3200			& 0.2000			& 0.3200		\\ %\hline
$A_{\lambda}$			& 1000.0			& 1480.0			& 1500.0		\\ %\hline
$A_{\kappa}$			& $-100.00$		& $-100.00$		& $-100.00$	\\ %\hline
$\mu _{\rm eff}$		& 900.00			& 900.00			& 900.00		\\ %\hline 
$\tan \beta $			& 2.0000			& 2.0000			& 2.0000		\\ \hline %\hline 
~{\bf Output}~			&				&				&			\\ \hline
$h^0_1$				& 125.23			& 126.51			&124.35		\\ %\hline
$h^0_2$				& 844.57			& 552.89			& 960.06		\\ %\hline
$h^0_3$				& 1793.5			& 1983.6			& 2115.8		\\ %\hline
$a^0_1$				& 372.60			& 289.22			& 379.68		\\ %\hline
$a^0_2$				& 1792.3			& 1983.7			& 2114.5		\\ %\hline
$H^{\pm}$				& 1783.8			& 1974.3			& 2108.5		\\ %\hline
%$\tilde d_L$			& 993.18			& 993.22			& 993.31		\\ \hline
%$\tilde d_R$			& 992.31			& 992.35			& 992.45		\\ \hline
%$\tilde u_L$			& 991.24			& 991.28			& 991.37		\\ \hline
%$\tilde u_R$			& 991.72			& 991.76			& 991.85		\\ \hline
%$\tilde s_L$			& 993.18			& 993.22			& 993.32		\\ \hline
%$\tilde s_R$			& 992.31			& 992.35			& 992.45		\\ \hline
%$\tilde c_f$			& 991.24			& 991.27			& 991.37		\\ \hline
%$\tilde c_R$			& 991.72			& 991.76			& 991.85		\\ \hline
%$\tilde b_1$			& 988.67			& 988.71			& 988.80		\\ %\hline
%$\tilde b_2$			& 996.82			& 996.85			& 996.95		\\ %\hline
$\tilde t_1$			& 853.37			& 853.41			& 853.51		\\ %\hline
$\tilde t_2$			& 1132.1			& 1132.1			& 1132.2		\\ %\hline
%$\tilde e_L$			& 361.09			& 361.09			& 361.09		\\ \hline
%$\tilde e_R$			& 356.32			& 356.32			& 356.32		\\ \hline
%$\tilde \nu _{eL}$		& 355.99			& 355.99			& 355.98		\\ \hline
%$\tilde \mu_L$		& 361.09			& 361.09			& 361.09		\\ \hline
%$\tilde \mu_R$		& 356.32			& 356.32			& 356.32		\\ \hline
%$\tilde \nu _{\mu L}$	& 355.99			& 355.99			& 355.98		\\ \hline
$\tilde \tau _1$			& 350.10			& 350.10 			& 350.10  		\\ %\hline
$\tilde \tau _2$			& 367.13			& 367.13			& 367.13		\\ %\hline
%$\tilde \nu _{\tau L}$	& 355.99			& 355.00			& 355.98		\\ \hline
%$\tilde g$				& 2068.8			& 2068.5			& 2067.7		\\ %\hline
$\tilde \chi ^0_1$		& 350.00			& 350.00 			& 350.00 		\\ %\hline
$\tilde \chi ^0_2$		& 696.36			& 534.28			& 697.51		\\ %\hline
$\tilde \chi ^0_3$		& 842.69			& 699.36			& 906.34		\\ %\hline
$\tilde \chi ^0_4$		& 908.00			& 909.78			& 914.24		\\ %\hline
$\tilde \chi ^0_5$		& 944.68			& 934.66			& 984.08		\\ \hline
%$\tilde \chi ^{\pm}_1$		& 697.21			& 697.56			& 697.60		\\ %\hline
%$\tilde \chi ^{\pm}_2$		& 929.28			& 929.50			& 929.61		\\ \hline
$\Omega _{\tilde \chi ^0_1} h^2$
					& 0.11956			& 0.11964			& 0.11968  	\\ %\hline
%$N_{1\tilde B}$		& 0.99687			& 0.99660			& 0.99691  	\\ %\hline
%$\sin \theta _{\tau}$	& 0.79992			& 0.79992			& 0.79992  \\ \hline  %\hline
$\sigma _{\rm SI}$[cm$^2$]
					&  $3.6870\times 10^{-46}$
					&  $3.7874\times 10^{-46}$
					&  $3.7116\times 10^{-46}$  \\
$\delta a_{\mu}$
					&  $1.3398\times 10^{-10} (>3\sigma )$
					&  $1.3370\times 10^{-10} (>3\sigma )$
					&  $1.3296\times 10^{-10} (>3\sigma )$ \\ %\hline
Br$(B_s^0\to \mu^+\mu^-)$
					&  $3.5381\times 10^{-9} (1\sigma )$
					&  $3.5382\times 10^{-9} (1\sigma )$
					&  $3.5383\times 10^{-9} (1\sigma )$ \\ %\hline
Br$(b \to s \gamma)$
					&  $3.1907\times 10^{-4} (2\sigma )$
					&  $3.1754\times 10^{-4} (2\sigma )$
					&  $3.1668\times 10^{-4} (2\sigma )$  \\ 
$G_{{\rm L}\tau}/c_{\tau}$	& 0.15080		& 0.15080		& 0.15076	\\
$G_{{\rm R}\tau}/c_{\tau}$	& 0.40310		& 0.40310		& 0.40313	\\
$G_{{\rm L}e}/c_{e} $  & 0.15108		& 0.15108		& 0.15108	\\
$G_{{\rm R}e}/c_{e} $  & 0.40289		& 0.40289		& 0.40289 	\\
\hline\hline
\end{tabular}
\label{tab:points-b-large}
\end{center}
\end{table}

\clearpage
\end{widetext}

%%%%%%%%%%%%%%
%%%%%%%%%%%%%%

\begin{widetext}
\begin{center}
\begin{figure}[h]
%%%%%%
%%%%%%
\begin{tabular}{l}
\hspace{-17mm}
\begin{minipage}{80mm}\vspace{-16mm}
\begin{center}
 \includegraphics[width=9cm,clip]{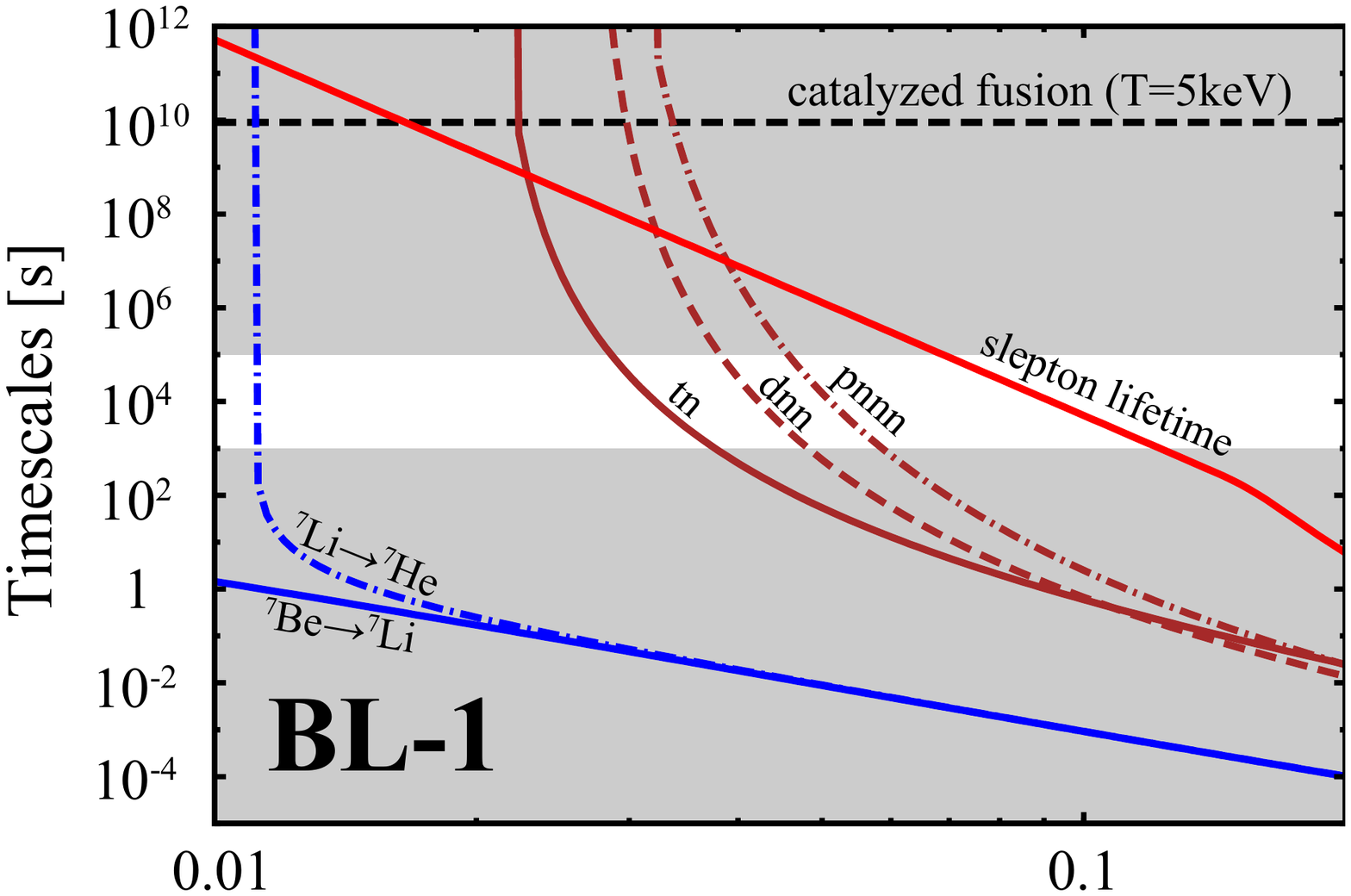}
 \label{fig:}
\end{center}
\end{minipage}
\hspace{8mm}\vspace{-1.5cm}
\begin{minipage}{80mm}\vspace{-16mm}
\begin{center}
 \includegraphics[width=9.3cm,clip]{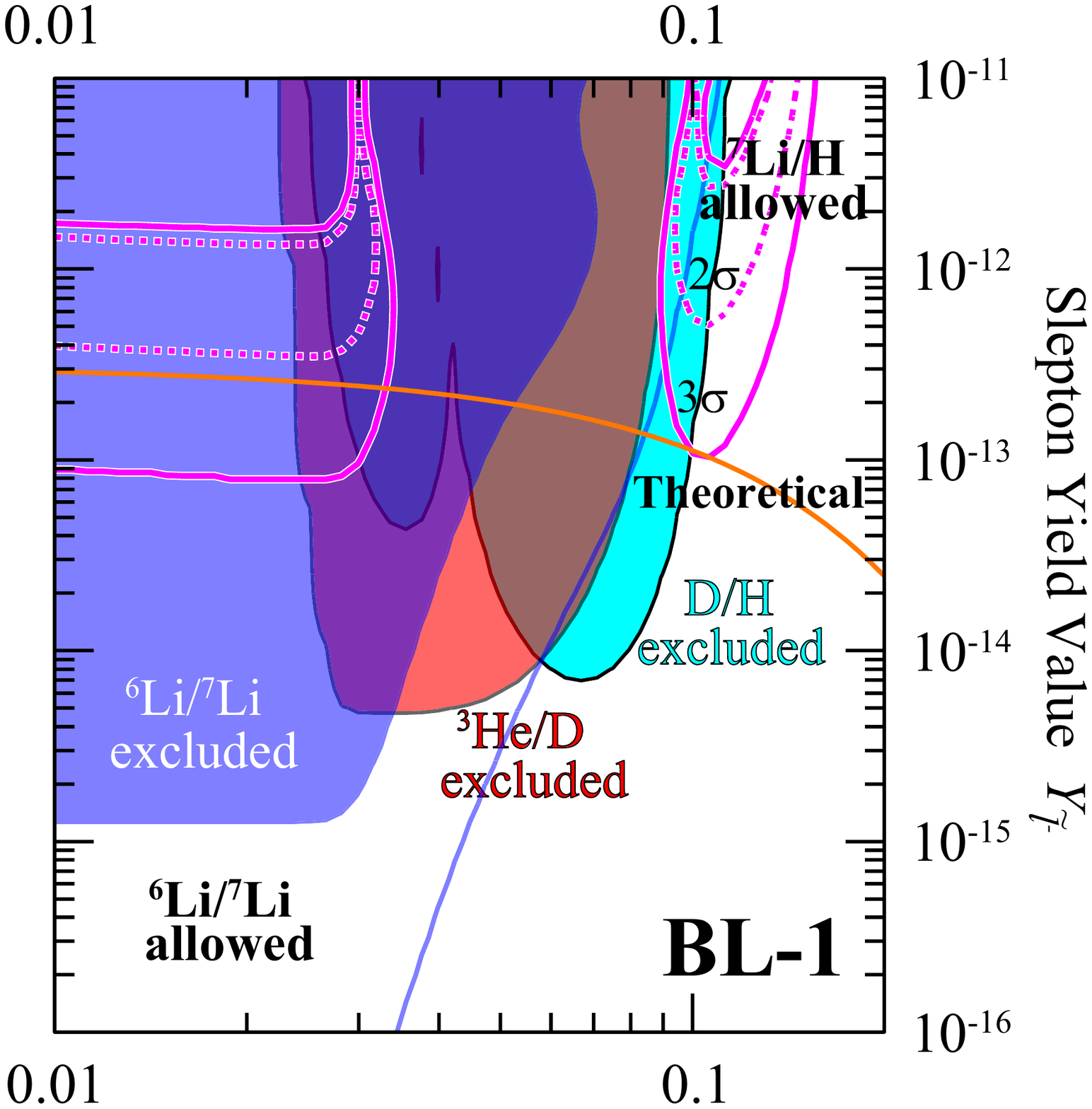}
 \label{fig:}
\end{center}
\end{minipage}
\\[-3mm]
\hspace{-14mm}
\begin{minipage}{79mm}\vspace{-36mm}
\begin{center}
  \includegraphics[width=9cm,clip]{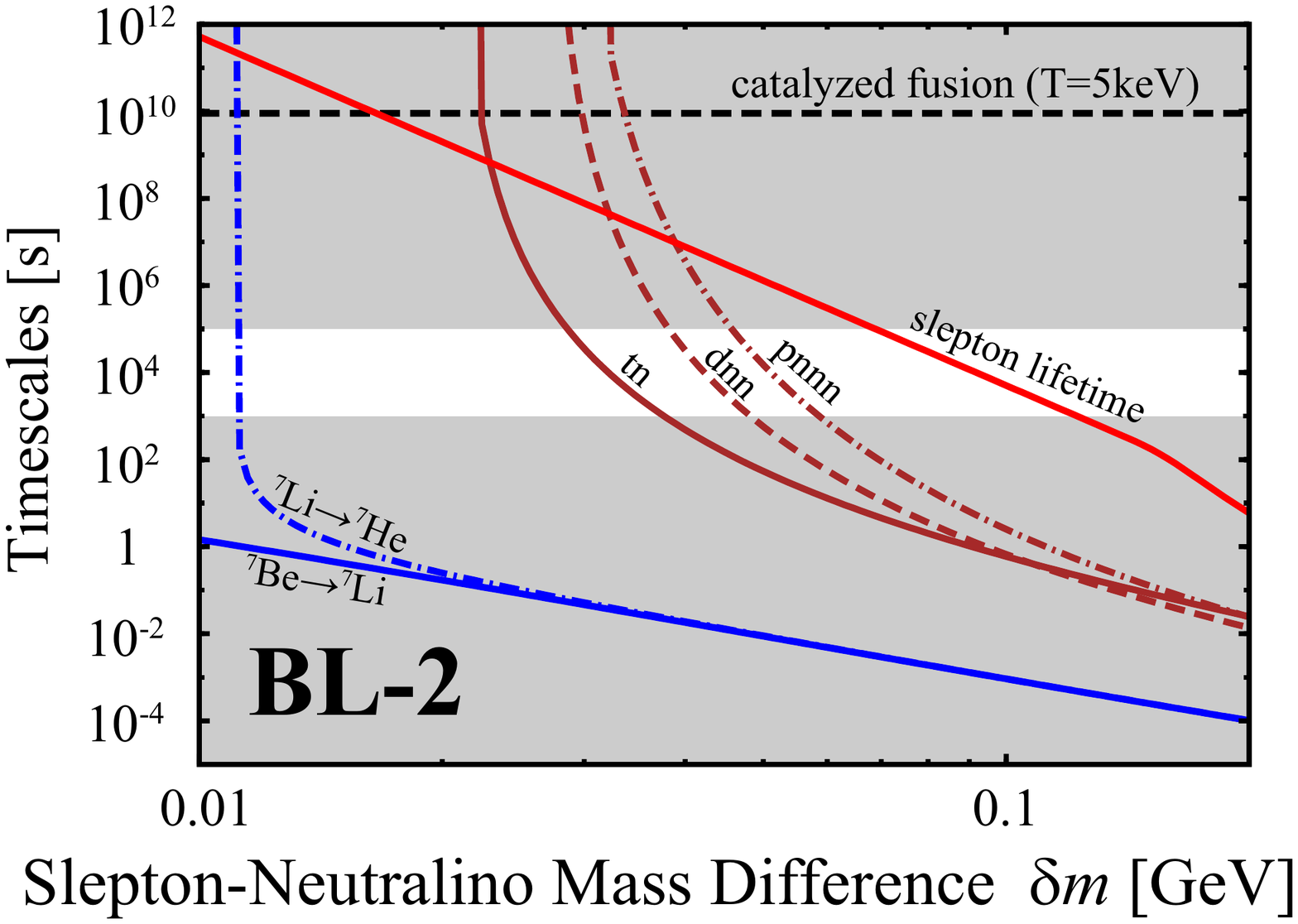}
  \label{fig:}
\end{center}
\end{minipage}
\hspace{1.5mm}
\begin{minipage}{79.8mm}\vspace{-36mm}
\begin{center}
 \includegraphics[width=9.65cm,clip]{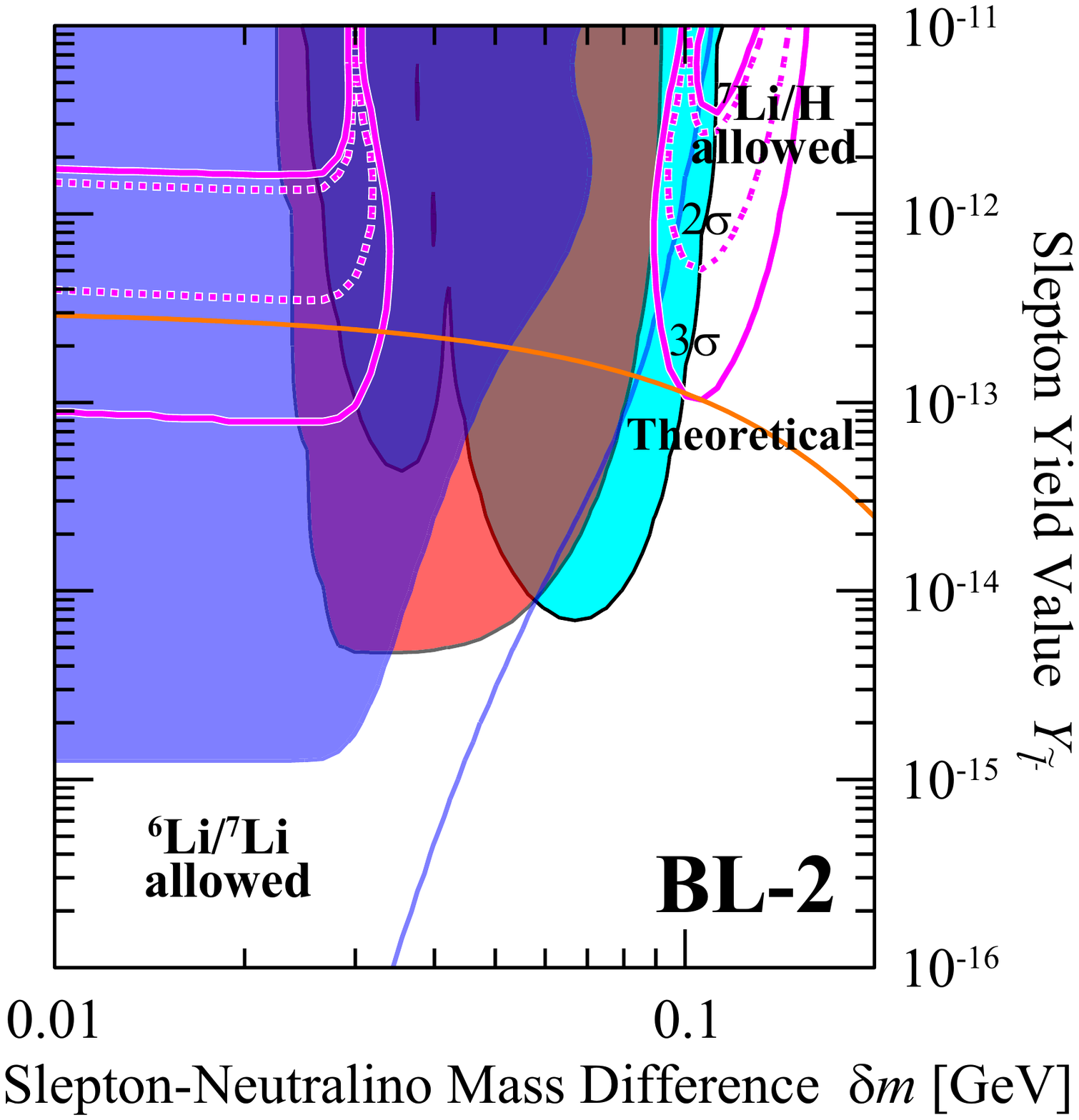}
 \label{fig:}
\end{center}
\end{minipage}
\end{tabular}
\vspace{-24mm}
\caption{
The left panels show 
the slepton lifetime $\tau _{\tilde l}$~(red-solid line; ``slepton lifetime"), 
the timescales of the internal conversion processes, Eqs.~\eqref{eq:icbe}~(blue-solid line; ``$^7$Be$\to$$^7$Li") and \eqref{eq:icli}~(blue-dash-dotted line; ``$^7$Li$\to$$^7$He"), 
the $^4$He spallation processes, Eqs.~\eqref{eq:tn}~(brown-solid line; ``tn"), \eqref{eq:dnn}~(brown-dashed line; ``dnn"), and \eqref{eq:pnnn}~(brown-dash-dotted line; ``pnnn"), 
as a function of the mass difference between the slepton and the neutralino 
at BL-1 (top panel) and BL-2 (bottom panel).
We also show the timescale of the catalyzed fusion \eqref{eq:cf} at the temperature $T=5$~keV ($5\times10^4$~s) when ($^4$He~$\tilde l^-$) is formed as horizontal black-dashed line.
In the shaded regions, Eq.~\eqref{eq:cond-sleptonlife} is not satisfied.
The right panels show the allowed regions from observational light element abundances on $\delta m$-$Y_{\tilde l^-}$ plane at BL-1 (top panel) and BL-2 (bottom panel).
%%%%%%%%%%%%%%
%%%%%%%%%%%%%%
The regions surrounded by magenta-dotted(-solid) lines are allowed by observed 
$^7$Li/H abundance at 2$\sigma$(3$\sigma$).
%%%%%%%%%%%%%%
%%%%%%%%%%%%%%
The regions between the blue-solid line and the blue region are allowed by 
observed $^6$Li/$^7$Li abundance at 2$\sigma$.
%%%%%%%%%%%%%%
%%%%%%%%%%%%%%
The orange-solid lines (``Theoretical") represent the yield value of the slepton 
when the BBN starts as a function of the mass difference.
%%%%%%%%%%%%%%
%%%%%%%%%%%%%%
The colored regions are excluded for 
$^6$Li/$^7$Li (blue region; ``$^6$Li/$^7$Li excluded"), 
$^3$He/D (red region; ``$^3$He/D excluded"), and 
D/H (cyan region; ``D/H excluded"), respectively. 
%%%%%%%%%%%%%%
%%%%%%%%%%%%%%
The shadowed and dotted regions are allowed by only $^7$Li/H (3$\sigma$) 
and both $^7$Li/H (3$\sigma$) and $^6$Li/$^7$Li (2$\sigma$), respectively.
}
\label{fig:large-lk-b_bbn1}
%%%%%%
%%%%%%
\end{figure}
\end{center}
\clearpage
\end{widetext}
%%%%%%%%%%%%%%
%%%%%%%%%%%%%%

\setcounter{figure}{6}

\begin{widetext}
\begin{center}
\begin{figure}[h]
%%%%%%
%%%%%%
\begin{tabular}{l}
\hspace{-14mm}
\begin{minipage}{80mm}\vspace{-16mm}
\begin{center}
 \includegraphics[width=8.7cm,clip]{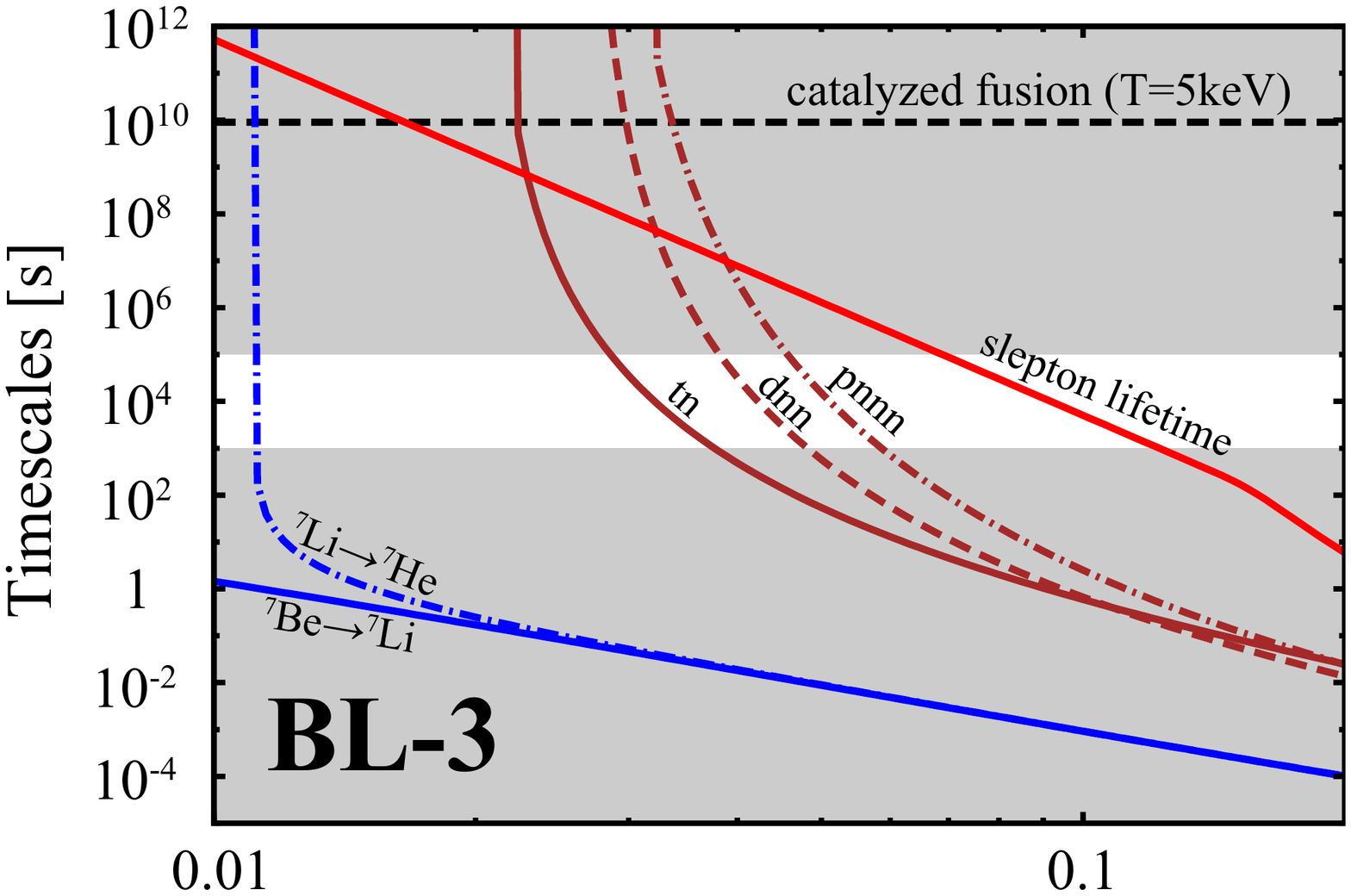}
 \label{fig:}
\end{center}
\end{minipage}
\hspace{2mm}\vspace{-1.5cm}
\begin{minipage}{80mm}\vspace{-16mm}
\begin{center}
 \includegraphics[width=9.5cm,clip]{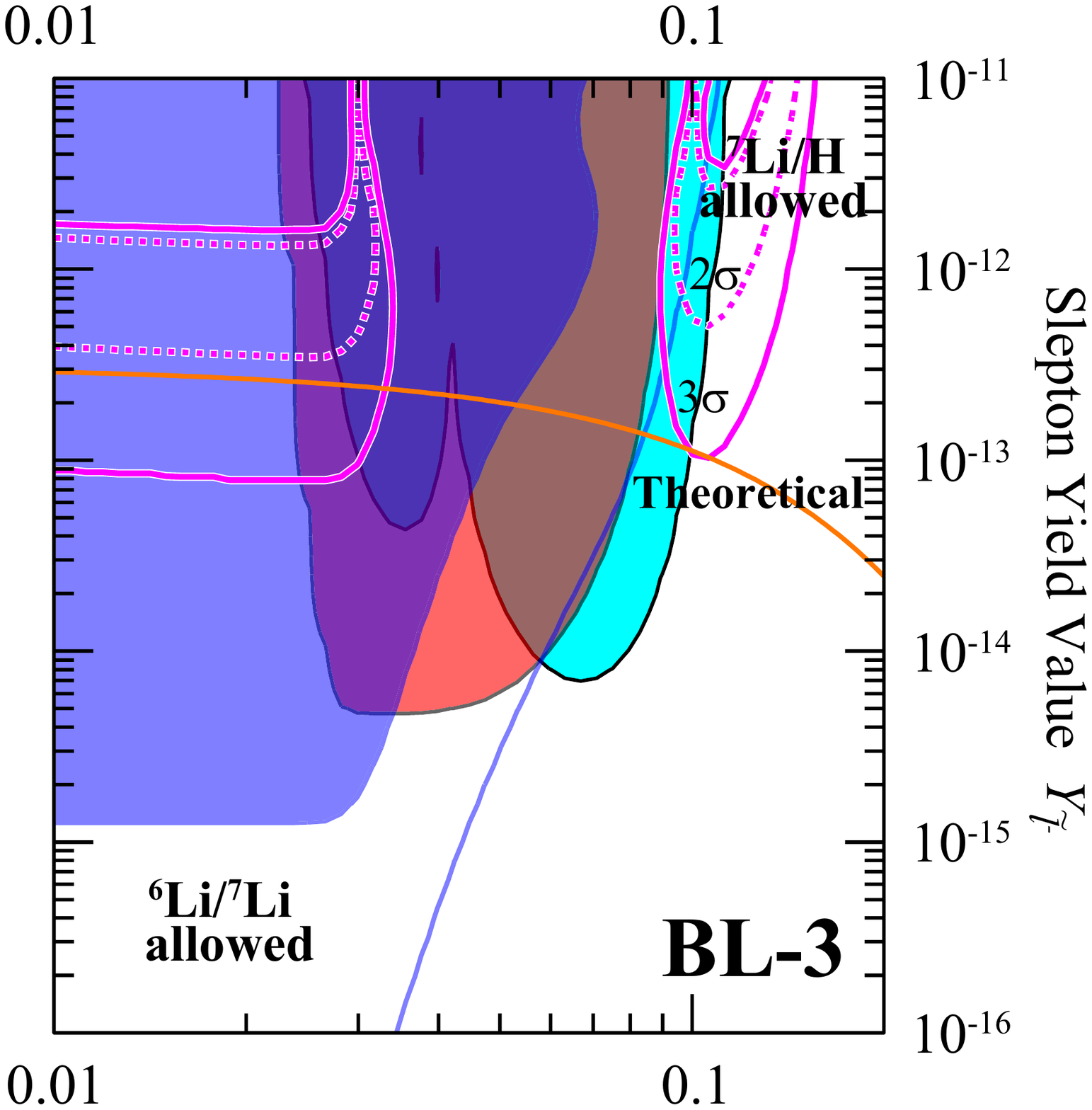}
 \label{fig:}
\end{center}
\end{minipage}
\\[-3mm]
\hspace{-14mm}
\begin{minipage}{79mm}\vspace{-36mm}
\begin{center}
  \includegraphics[width=9cm,clip]{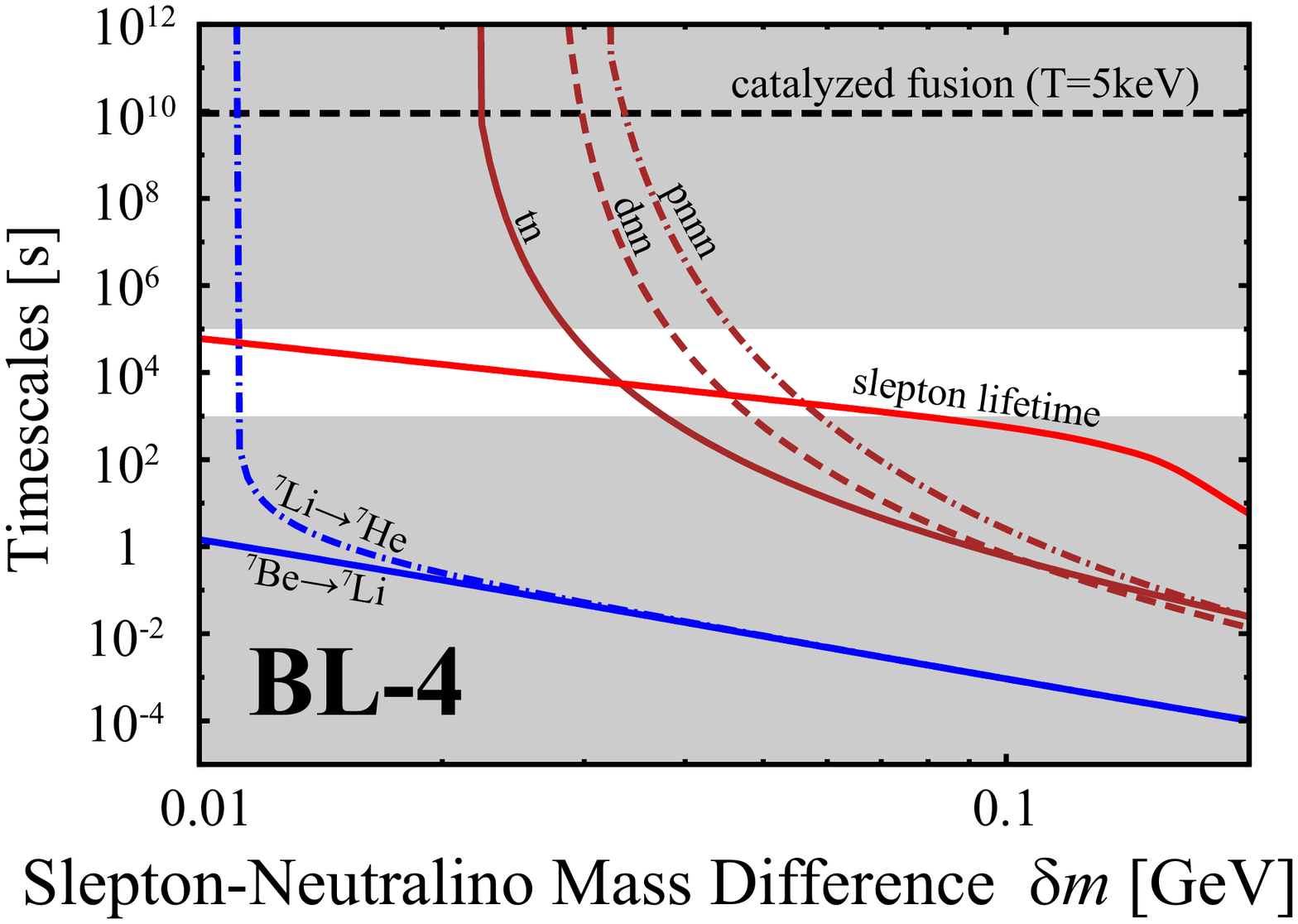}
  \label{fig:}
\end{center}
\end{minipage}
\hspace{2mm}
\begin{minipage}{79.8mm}\vspace{-36mm}
\begin{center}
 \includegraphics[width=9.5cm,clip]{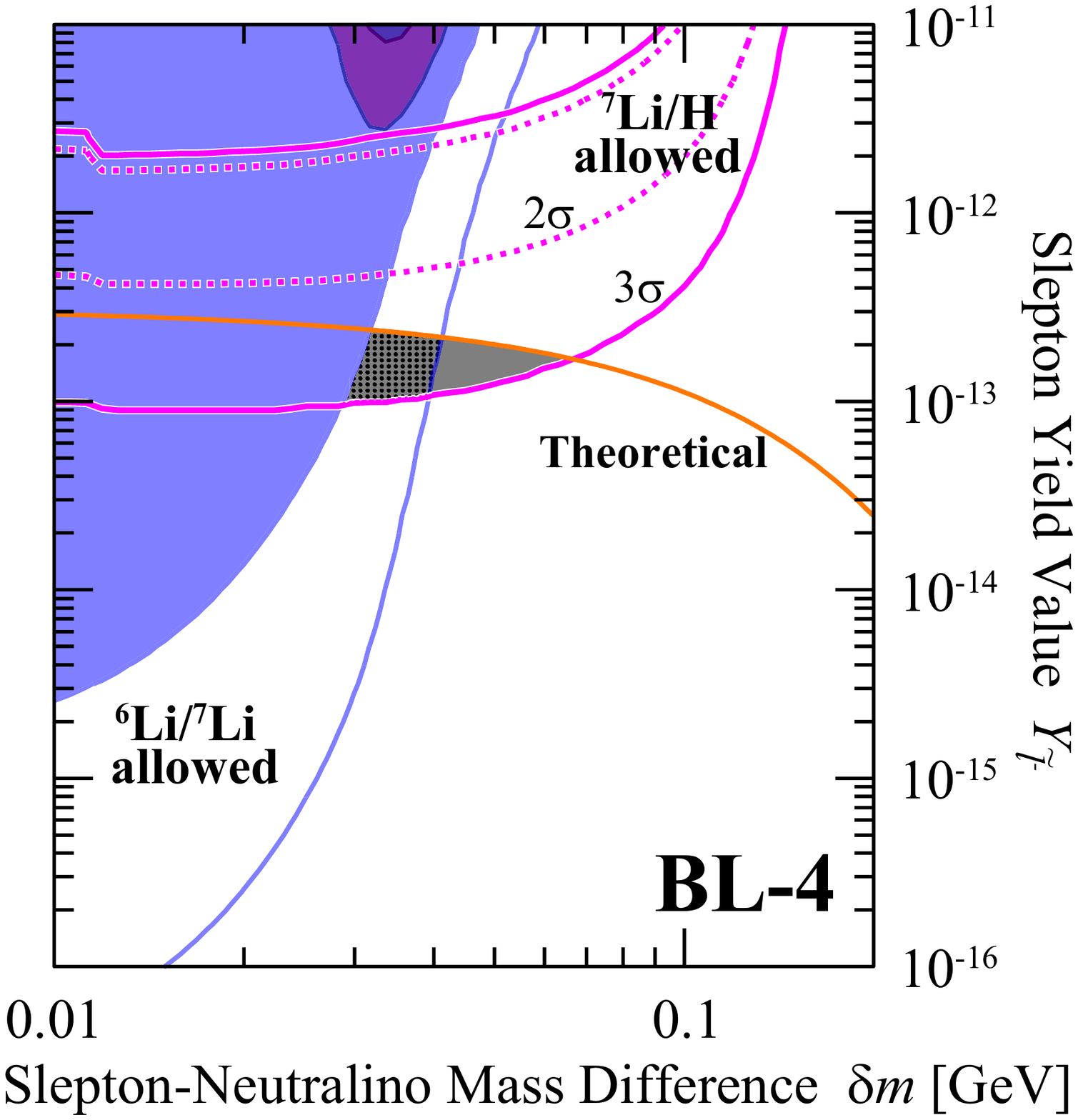}
 \label{fig:}
\end{center}
\end{minipage}
\end{tabular}
\vspace{-24mm}
\caption{
The results at BL-3 (top panels) and BL-4 (bottom panels) are shown.
The meanings of the lines and regions are same as those in the previous page.
}
\label{fig:large-lk-b_bbn2}
%%%%%%
%%%%%%
\end{figure}
\end{center}
\clearpage
\end{widetext}
%%%%%%%%%%%%%%
%%%%%%%%%%%%%%

%%%%%%%%%%%%%%%%%%%%%%%%%%%%%%%%%%%%%%%%%%%%%
%%%%%%%%%%%%%%%%%%%%%%%%%%%%%%%%%%%%%%%%%%%%%
%%%%%%%%%%%%%%%%%%%%%%%%%%%%%%%%%%%%%%%%%%%%%
\section{Summary} \label{sec:summary} %%%%%%%%%%%%%%%
%%%%%%%%%%%%%%%%%%%%%%%%%%%%%%%%%%%%%%%%%%%%%
%%%%%%%%%%%%%%%%%%%%%%%%%%%%%%%%%%%%%%%%%%%%%
%%%%%%%%%%%%%%%%%%%%%%%%%%%%%%%%%%%%%%%%%%%%%

The precise observations of the universe confirmed the presence of dark
matter
and raised a crucial question on its nature.
Another possible problem is the discrepancy of the abundance of $^7$Li
and $^6$Li.
Models are attractive when they can account for these problems and at
the same
time are consistent with the observed mass of Higgs particle.
We demonstrated that the NMSSM can be a candidate for such models.

We specifically considered the case where the neutralino is the
stable LSP and the lightest slepton is the NLSP, and where the mass
difference of the two is so tiny that the slepton becomes long-lived
enough to survive until the time of the nucleosynthesis in the early
universe.
The sleptons interact with the synthesized nuclei and turn into the
LSPs that stay until today as dark matter particles, altering the
relic abundance of the light elements.

First, we searched for the benchmark sets of parameters that can
successfully drive this scenario and simultaneously can reproduce the
mass of Higgs particle within $(125.6 \pm 3.0) \, \mathrm{GeV}$.
Three cases of benchmark parameters are presented: %
(a) Singlino-like neutralino, small $\lambda$-$\kappa$ with large $\tan
\beta$;
(b) Singlino-like neutralino, large $\lambda$-$\kappa$ with small $\tan
\beta$;
and %
(c) Bino-like neutralino, large $\lambda$-$\kappa$ with small $\tan \beta$.
We found the successful benchmark values of $(c_{e}, \lambda, \kappa)$
in all the three cases.
%(Tables~\ref{tab:points1}, \ref{tab:points2}, and \ref{tab:points3}).
%
We confirmed that they lead to the permissible abundance of dark
matter and are consistent with other experimental bounds as presented
in Tables~\ref{tab:points-s-small}, \ref{tab:points-s-large}, and
\ref{tab:points-b-large}.

We then traced the BBN reaction network including the exotic nuclear
reactions.
We employed $Y_{\tilde{l}^-}$ (slepton yield value)-$\delta m$
(LSP-NLSP mass difference) parameter plane to present the regions of
parameters that can account for the observed abundance of light
elements.

The results are illustrated in terms of
timescales for the relevant exotic BBN reactions as follows.
The slepton needs to be long-lived enough to form the bound state with
$\mathrm{^{7}Be}$, while
% Is the term of 'slepton lifetime' right? It's OK.
the couplings ${G_{L,R}}_\tau$ should be large enough so that the
internal conversion processes occur sufficiently.
%Which is internal conversion or internal conversion processes?
The slepton lifetime is thus subject to the lower bound.
Meanwhile, the bound state of $(\mathrm{^{4}He}~\tilde{l}^-$)
accompanies other two relevant processes: one is the
$\mathrm{^{4}He}$ spallation, and the other is the catalyzed fusion.
These processes can readily overproduce the light elements and
should be avoided, but suitable amount of $\mathrm{^{6}Li}$
production is favorable in order to account for its observed
abundance.
Of the two, the $\mathrm{^{4}He}$ spallation generally proceeds more
efficiently and is prone to overproduction, but it can be avoided
when the slepton has not too long lifetime and less of it forms a bound state with 
$\mathrm{^{4}He}$.
In addition, such lifetime can induce appropriate abundance of
$\mathrm{^{6}Li}$ via catalyzed fusion as its
observed abundance is tiny as explained around
Eq.~\eqref{eq:deltaLi6}.
Upper bound on the slepton lifetime is thus brought in.
% This tuning its production puts the upper limit on the slepton lifetime.
%
Combining the above arguments, we obtain an allowed window of
the slepton lifetime.
%

%(a)
In the case of (a) (Fig.~\ref{fig:small-lk-s_bbn1}), the specific relation between $\lambda$ and
$\kappa$ is necessary to make the couplings ${G_{L,R}}_\tau$ large
and thus to reduce %
$\mathrm{^{7}\mathrm{Be}}$ by the internal
conversion processes.
Indeed, we tune the parameters $\lambda$ and $\kappa$ to make
$\mu_{\kappa}^2-\mu_{\rm eff}^2$ small and hence
${G_{L,R}}_\tau$ large at SS-1.
We still need a sizable flavor mixing, $c_e$, in order to
render the slepton lifetime short and
thereby reduce the number of
$(\mathrm{^{4}He}~\tilde{l}^-)$.
Thus we obtain the allowed region at SS-1.
On the contrary, the flavor mixing $c_e$ is small at SS-2 and hence
the slepton lifetime becomes longer.
Allowed region is thus shifted toward larger $\delta m$.
% the allowed region is reduced in the tuned
%parameter region with small flavor mixing $c_e$ at SS-2.
%
The values of $G_{L,R}$ become rapidly small outside the tuned
parameter region of $\lambda$ and $\kappa$ %
so that the allowed region becomes
small (SS-4).
%(b)
In the case of (b) (Fig.~\ref{fig:large-lk-s_bbn1}), the values of $G_{L,R}$ are large due to
large $\lambda$ and $\kappa$, and thus the
internal conversion processes occur efficiently.
Suitable flavor
mixings are also necessary as in the case of (a) to avoid the
excessive amount of $({}^4
\mathrm{He}~\tilde{l}^-)$. 
Otherwise, we may miss the allowed
region (SL-2).
%(c)
In the case of (c) (Fig.~\ref{fig:large-lk-b_bbn1}),
the couplings ${G_{L,R}}_\tau$
hardly depend on $\lambda$ and $\kappa$,
%Since the coupling ${G_{L,R}}_\tau$ for the case of (c)
%does not depend mainly on $\lambda$ and $\kappa$,
%the lifetime of slepton is short
%because the ${G_{L,R}}_\tau$ are determined from the sizable gauge
%couplings.
%
%Therefore, we can explain the observed abundance of ${}^6 \mathrm{Li}$
%even with no
%flavor mixings although the observed abundance of ${}^7 \mathrm{Li}$ is not
%realized due to the spallation process in BL-1-3.
%Taking the small flavor mixings into account,
%we can explain both $\mathrm{Li}$ problems in BL-4
%because these mixings make slepton lifetime shorter
%and suppress the spallation process.
%These results of the case (c) are same as in the case of MSSM
and thus the results are same as in
the case of the MSSM.
%(We found that same results are shown in the case of MSSM)
%

We conclude that our scenario successfully works in the NMSSM and can
simultaneously account for the abundance of dark matter, that of light
elements, and the mass of Higgs particle.
Since all the three cases we considered here are consistent to the
present phenomenological bounds, they should be distinguished through
the characteristic signals of accelerator experiments.
Search for such signals are left for future works.

%%%%%%%%%%%%%%%%%%%%%%%%%%%%%%%%%%%%%%%%%%%%%
%%%%%%%%%%%%%%%%%%%%%%%%%%%%%%%%%%%%%%%%%%%%%
%%%%%%%%%%%%%%%%%%%%%%%%%%%%%%%%%%%%%%%%%%%%%
\section*{Acknowledgments}   %%%%%%%%%%%%%%%%%%%%%%%%%%%%%%
%%%%%%%%%%%%%%%%%%%%%%%%%%%%%%%%%%%%%%%%%%%%%
%%%%%%%%%%%%%%%%%%%%%%%%%%%%%%%%%%%%%%%%%%%%%
%%%%%%%%%%%%%%%%%%%%%%%%%%%%%%%%%%%%%%%%%%%%%
The work of Y.K. was financially supported by the Sasakawa Scientific
Research  Grant from The Japan Science Society.
This swork was supported in part by 
the Grant-in-Aid for the Ministry of Education, Culture, Sports, Science, and Technology, Government of Japan,
Nos.~21111006, 22244030, 23540327~(K.K.), 
No.~24740145~(M.K.),  
Nos.~24340044 and 25105009~(J.S.), 
No. 23740190 (T.S.), 
Nos.~23740208 and 25003345~(M.Y.), 
and by the Center for the Promotion
of Integrated Science (CPIS) of Sokendai
(1HB5804100) (K.K.).

%%%%%%%%%%%%%%%%%%%%%%%%%%%%%%%%%%%%%%%%%%%%%
%%%%%%%%%%%%%%%%%%%%%%%%%%%%%%%%%%%%%%%%%%%%%
%%%%%%%%%%%%%%%%%%%%%%%%%%%%%%%%%%%%%%%%%%%%%
   %%%%%%%%%%%%%%%%%%%%%%%%%%%%%%%%%
%%%%%%%%%%%%%%%%%%%%%%%%%%%%%%%%%%%%%%%%%%%%

%%%%%%%%%%%%%%%%%%%%%%%%%%%%%%%%%%%%%%%%%%%%

\begin{thebibliography}{99}  %%%%%%%%%%%%%%%%%%%%%%%%%%%%%%%
%%%%%%%%%%%%%%%%%%%%%%%%%%%%%%%%%%%%%%%%%%%%%
%%%%%%%%%%%%%%%%%%%%%%%%%%%%%%%%%%%%%%%%%%%%%
%%%%%%%%%%%%%%%%%%%%%%%%%%%%%%%%%%%%%%%%%%%%%


%%%%%%%%%%%%%%%%%%%%%%%%%%%%
% Higgs discovery
%%%%%%%%%%%%%%%%%%%%%%%%%%%%

%\cite{Aad:2012tfa}
\bibitem{Aad:2012tfa} 
  G.~Aad {\it et al.}  [ATLAS Collaboration],
  %``Observation of a new particle in the search for the Standard Model Higgs boson with the ATLAS detector at the LHC,''
  Phys.\ Lett.\ B {\bf 716}, 1 (2012)
  [arXiv:1207.7214 [hep-ex]].
  %%CITATION = ARXIV:1207.7214;%%
  %2287 citations counted in INSPIRE as of 07 Mar 2014


%\cite{Chatrchyan:2012ufa}
\bibitem{Chatrchyan:2012ufa} 
  S.~Chatrchyan {\it et al.}  [CMS Collaboration],
  %``Observation of a new boson at a mass of 125 GeV with the CMS experiment at the LHC,''
  Phys.\ Lett.\ B {\bf 716}, 30 (2012)
  [arXiv:1207.7235 [hep-ex]].
  %%CITATION = ARXIV:1207.7235;%%
  %2269 citations counted in INSPIRE as of 07 Mar 2014


%%%%%%%%%%%%%%%%%%%%%%%%%%%%
% WMAP9
%%%%%%%%%%%%%%%%%%%%%%%%%%%%

%\cite{Bennett:2012zja}
\bibitem{Bennett:2012zja} 
  C.~L.~Bennett {\it et al.}  [WMAP Collaboration],
  %``Nine-Year Wilkinson Microwave Anisotropy Probe (WMAP) Observations: Final Maps and Results,''
  Astrophys.\ J.\ Suppl.\  {\bf 208}, 20 (2013)
  [arXiv:1212.5225 [astro-ph.CO]].
  %%CITATION = ARXIV:1212.5225;%%
  %320 citations counted in INSPIRE as of 07 Mar 2014


%%%%%%%%%%%%%%%%%%%%%%%%%%%%
% Planck
%%%%%%%%%%%%%%%%%%%%%%%%%%%%

\bibitem{Ade:2013zuv}
  P.~A.~R.~Ade {\it et al.}  [Planck Collaboration],
  %``Planck 2013 results. XVI. Cosmological parameters,''
  arXiv:1303.5076 [astro-ph.CO].
  %%CITATION = ARXIV:1303.5076;%%



%%%%%%%%%%%%%%%%%%%%%%%%%%%%
% mu-problem
%%%%%%%%%%%%%%%%%%%%%%%%%%%%

%\cite{Kim:1983dt}
\bibitem{Kim:1983dt} 
  J.~E.~Kim and H.~P.~Nilles,
  %``The mu Problem and the Strong CP Problem,''
  Phys.\ Lett.\ B {\bf 138}, 150 (1984).
  %%CITATION = PHLTA,B138,150;%%
  %597 citations counted in INSPIRE as of 07 Mar 2014






%%%%%%%%%%%%%%%%%%%%%%%%%%%%
% Stau BBN
%%%%%%%%%%%%%%%%%%%%%%%%%%%%
%\cite{Jittoh:2005pq}
\bibitem{Jittoh:2005pq} 
  T.~Jittoh, J.~Sato, T.~Shimomura and M.~Yamanaka,
  %``Long life stau in the minimal supersymmetric standard model,''
  Phys.\ Rev.\ D {\bf 73}, 055009 (2006)
  [Erratum-ibid.\ D {\bf 87}, 019901 (2013)]
  [hep-ph/0512197].
  %%CITATION = HEP-PH/0512197;%%
  %39 citations counted in INSPIRE as of 07 Mar 2014

%\cite{Jittoh:2007fr}
\bibitem{Jittoh:2007fr} 
  T.~Jittoh, K.~Kohri, M.~Koike, J.~Sato, T.~Shimomura and M.~Yamanaka,
  %``Possible solution to the Li-7 problem by the long lived stau,''
  Phys.\ Rev.\ D {\bf 76}, 125023 (2007)
  [arXiv:0704.2914 [hep-ph]].
  %%CITATION = ARXIV:0704.2914;%%
  %61 citations counted in INSPIRE as of 07 Mar 2014

%\cite{Jittoh:2008eq}
\bibitem{Jittoh:2008eq} 
  T.~Jittoh, K.~Kohri, M.~Koike, J.~Sato, T.~Shimomura and M.~Yamanaka,
  %``Big-bang nucleosynthesis and the relic abundance of dark matter in a stau-neutralino coannihilation scenario,''
  Phys.\ Rev.\ D {\bf 78}, 055007 (2008)
  [arXiv:0805.3389 [hep-ph]].
  %%CITATION = ARXIV:0805.3389;%%
  %34 citations counted in INSPIRE as of 07 Mar 2014

%\cite{Jittoh:2010wh}
\bibitem{Jittoh:2010wh} 
  T.~Jittoh, K.~Kohri, M.~Koike, J.~Sato, T.~Shimomura and M.~Yamanaka,
  %``Stau relic density at the Big-Bang nucleosynthesis era consistent with the abundance of the light element nuclei in the coannihilation scenario,''
  Phys.\ Rev.\ D {\bf 82}, 115030 (2010)
  [arXiv:1001.1217 [hep-ph]].
  %%CITATION = ARXIV:1001.1217;%%
  %12 citations counted in INSPIRE as of 07 Mar 2014

%\cite{Jittoh:2011ni}
\bibitem{Jittoh:2011ni} 
  T.~Jittoh, K.~Kohri, M.~Koike, J.~Sato, K.~Sugai, M.~Yamanaka and K.~Yazaki,
  %``Big-bang nucleosynthesis with a long-lived charged massive particle including $^4$He spallation processes,''
  Phys.\ Rev.\ D {\bf 84}, 035008 (2011)
  [arXiv:1105.1431 [hep-ph]].
  %%CITATION = ARXIV:1105.1431;%%
  %8 citations counted in INSPIRE as of 07 Mar 2014

%\cite{Kohri:2012gc}
\bibitem{Kohri:2012gc} 
  K.~Kohri, S.~Ohta, J.~Sato, T.~Shimomura and M.~Yamanaka,
  %``Allowed slepton intergenerational mixing in light of light element abundances,''
  Phys.\ Rev.\ D {\bf 86}, 095024 (2012)
  [arXiv:1208.5533 [hep-ph]].
  %%CITATION = ARXIV:1208.5533;%%
  %4 citations counted in INSPIRE as of 07 Mar 2014
 


%%%%%%%%%%%%%%%%%%
% lithium abundances (obs)
%%%%%%%%%%%%%%%%%%

\bibitem{Melendez:2004ni}
  J.~Melendez and I.~Ramirez,
  %``Reappraising the Spite Lithium Plateau: Extremely Thin and Marginally
  Astrophys.\ J.\  {\bf 615} (2004) L33.
  %[arXiv:astro-ph/0409383].
  %%CITATION = ASTRO-PH/0409383;%% 
 
\bibitem{Asplund:2005yt}
  M.~Asplund, D.~L.~Lambert, P.~E.~Nissen, F.~Primas and V.~V.~Smith,
  %``Lithium isotopic abundances in metal-poor halo stars,''
  Astrophys.\ J.\  {\bf 644} (2006) 229
  [arXiv:astro-ph/0510636].
  %%CITATION = ASTRO-PH/0510636;%% 


%%%%%%%%%%%%%%%%%%
% CMSSM
%%%%%%%%%%%%%%%%%%

\bibitem{Konishi:2013gda}
  Y.~Konishi, S.~Ohta, J.~Sato, T.~Shimomura, K.~Sugai and M.~Yamanaka,
  %``A first evidence of the CMSSM is appearing soon,''
  arXiv:1309.2067 [hep-ph].
  %%CITATION = ARXIV:1309.2067;%%


%%%%%%%%%%%%%%%%%%%%%%%%%%%%
% Coannihilation
%%%%%%%%%%%%%%%%%%%%%%%%%%%%

\bibitem{Griest:1990kh}
  K.~Griest and D.~Seckel,
  %``Three exceptions in the calculation of relic abundances,''
  Phys.\ Rev.\ D {\bf 43} (1991) 3191.
  %%CITATION = PHRVA,D43,3191;%%

%%%%%%%%%%%%%%%%%%%%%%%%%%%%
% long lived charged massive particles
%%%%%%%%%%%%%%%%%%%%%%%%%%%%

\bibitem{Kohri:2006cn} 
  K.~Kohri and F.~Takayama,
  %``Big bang nucleosynthesis with long lived charged massive particles,''
  Phys.\ Rev.\ D {\bf 76}, 063507 (2007)
  [hep-ph/0605243].
  %%CITATION = HEP-PH/0605243;%%
  %112 citations counted in INSPIRE as of 07 Mar 2014


%%%%%%%%%%%%%%%%%%%%%%%%%%%%
% Internal conversion
%%%%%%%%%%%%%%%%%%%%%%%%%%%%

 \bibitem{Bird:2007ge}
  C.~Bird, K.~Koopmans and M.~Pospelov,
  %``Primordial Lithium Abundance in Catalyzed Big Bang Nucleosynthesis,''
  Phys.\ Rev.\  D {\bf 78}, 083010 (2008)
  [arXiv:hep-ph/0703096].
  %%CITATION = HEP-PH/0703096;%%


%%%%%%%%%%%%%%%%%%%%%%%%%%%%
% Catalyzed fusion
%%%%%%%%%%%%%%%%%%%%%%%%%%%%

\bibitem{Pospelov:2006sc}
  M.~Pospelov,
  %``Particle physics catalysis of thermal Big Bang Nucleosynthesis,''
  Phys.\ Rev.\ Lett.\  {\bf 98}, 231301 (2007)
  [arXiv:hep-ph/0605215].
  %%CITATION = HEP-PH/0605215;%%
 
\bibitem{Hamaguchi:2007mp}
  K.~Hamaguchi, T.~Hatsuda, M.~Kamimura, Y.~Kino and T.~T.~Yanagida,
  %``Stau-catalyzed Li-6 production in big-bang nucleosynthesis,''
  Phys.\ Lett.\  B {\bf 650} (2007) 268
  [arXiv:hep-ph/0702274].   
  %%CITATION = HEP-PH/0702274;%%

\bibitem{Kawasaki:2007xb} 
  M.~Kawasaki, K.~Kohri and T.~Moroi,
  %``Big-Bang Nucleosynthesis with Long-Lived Charged Slepton,''
  Phys.\ Lett.\ B {\bf 649}, 436 (2007)
  [hep-ph/0703122].
  %%CITATION = HEP-PH/0703122;%%







%%%%%%%%%%%%%%%%%%%%%%%%%%%%
% NMSSMTools
%%%%%%%%%%%%%%%%%%%%%%%%%%%%
\bibitem{Muhlleitner:2003vg}
  M.~Muhlleitner, A.~Djouadi and Y.~Mambrini,
  %``SDECAY: A Fortran code for the decays of the supersymmetric particles in the the MSSM,''
  Comput.\ Phys.\ Commun.\  {\bf 168} (2005) 46
  [hep-ph/0311167].
  %%CITATION = HEP-PH/0311167;%%

\bibitem{Ellwanger:2004xm}
  U.~Ellwanger, J.~F.~Gunion and C.~Hugonie,
  %``NMHDECAY: A Fortran code for the Higgs masses, couplings and decay widths in the NMSSM,''
  JHEP {\bf 0502} (2005) 066
  [hep-ph/0406215].
  %%CITATION = HEP-PH/0406215;%%

\bibitem{Belanger:2005kh}
  G.~Belanger, F.~Boudjema, C.~Hugonie, A.~Pukhov and A.~Semenov,
  %``Relic density of dark matter in the NMSSM,''
  JCAP {\bf 0509} (2005) 001
  [hep-ph/0505142].
  %%CITATION = HEP-PH/0505142;%%

\bibitem{Ellwanger:2005dv}
  U.~Ellwanger and C.~Hugonie,
  %``NMHDECAY 2.0: An Updated program for sparticle masses, Higgs masses, couplings and decay widths in the NMSSM,''
  Comput.\ Phys.\ Commun.\  {\bf 175} (2006) 290
  [hep-ph/0508022].
  %%CITATION = HEP-PH/0508022;%%
 
\bibitem{Ellwanger:2006rn}
  U.~Ellwanger and C.~Hugonie,
  %``NMSPEC: A Fortran code for the sparticle and Higgs masses in the NMSSM with GUT scale boundary conditions,''
  Comput.\ Phys.\ Commun.\  {\bf 177} (2007) 399
  [hep-ph/0612134].
  %%CITATION = HEP-PH/0612134;%%
 
 





%%%%%%%%%%%%%%%%%%%%%%%%%%%%
% for lambda kappa ranges
%%%%%%%%%%%%%%%%%%%%%%%%%%%%

\bibitem{King:2012is}
  S.~F.~King, M.~Muhlleitner and R.~Nevzorov,
  %``NMSSM Higgs Benchmarks Near 125 GeV,''
  Nucl.\ Phys.\ B {\bf 860} (2012) 207
  [arXiv:1201.2671 [hep-ph]].
  %%CITATION = ARXIV:1201.2671;%%



 

%%%%%%%%%%%%%%%%%%%%%%%%%%%%
% micrOMEGAs
%%%%%%%%%%%%%%%%%%%%%%%%%%%% 
 
\bibitem{Belanger:2013oya}
  G.~Belanger, F.~Boudjema, A.~Pukhov and A.~Semenov,
  %``micrOMEGAs_3: A program for calculating dark matter observables,''
  Comput.\ Phys.\ Commun.\  {\bf 185} (2014) 960
  [arXiv:1305.0237 [hep-ph]].
  %%CITATION = ARXIV:1305.0237;%%


%%%%%%%%%%%%%%%%%%%%%%%%%%%%
% Mhiggs
%%%%%%%%%%%%%%%%%%%%%%%%%%%% 

\bibitem{CMS:yva}
  [CMS Collaboration],
  %``Combination of standard model Higgs boson searches and measurements of the properties of the new boson with a mass near 125 GeV,''
  CMS-PAS-HIG-13-005.
  %%CITATION = CMS-PAS-HIG-13-005;%%
 
\bibitem{ATLAS:2013mma}
  [ATLAS Collaboration],
  %``Combined measurements of the mass and signal strength of the Higgs-like boson with the ATLAS detector using up to 25 fb$^{-1}$ of proton-proton collision data,''
  ATLAS-CONF-2013-014.
  %%CITATION = ATLAS-CONF-2013-014;%%


%%%%%%%%%%%%%%%%%%%%%%
% uncertainty in the Higgs mass calculation
%%%%%%%%%%%%%%%%%%%%%%

\bibitem{Allanach:2001hm}
  B.~C.~Allanach,
  %``Theoretical uncertainties in sparticle mass predictions,''
  eConf C {\bf 010630}, P319 (2001)
  [hep-ph/0110227].
  %%CITATION = HEP-PH/0110227;%%

\bibitem{Djouadi:2002nh}
  A.~Djouadi,
  %``SUSY calculation tools,''
  hep-ph/0211357.
  %%CITATION = HEP-PH/0211357;%%

\bibitem{Allanach:2003jw}
  B.~C.~Allanach, S.~Kraml and W.~Porod,
  %``Theoretical uncertainties in sparticle mass predictions from computational tools,''
  JHEP {\bf 0303}, 016 (2003)
  [hep-ph/0302102].
  %%CITATION = HEP-PH/0302102;%%
 
\bibitem{Allanach:2004rh}
  B.~C.~Allanach, A.~Djouadi, J.~L.~Kneur, W.~Porod and P.~Slavich,
  %``Precise determination of the neutral Higgs masses in the the MSSM,''
  JHEP {\bf 0409}, 044 (2004)
  [hep-ph/0406166].
  %%CITATION = HEP-PH/0406166;%%


%%%%%%%%%%%%%%%%%%%%%%
% deviation in calc. of mHiggs between FeynHiggs and NMSSMTools
%%%%%%%%%%%%%%%%%%%%%%

\bibitem{Cao:2012fz}
  J.~-J.~Cao, Z.~-X.~Heng, J.~M.~Yang, Y.~-M.~Zhang and J.~-Y.~Zhu,
  %``A SM-like Higgs near 125 GeV in low energy SUSY: a comparative study for the MSSM and NMSSM,''
  JHEP {\bf 1203} (2012) 086
  [arXiv:1202.5821 [hep-ph]].
  %%CITATION = ARXIV:1202.5821;%%



%%%%%%%%%%%%%%%%%%%%%%
% observed abundances of light elements
%%%%%%%%%%%%%%%%%%%%%%


 
\bibitem{Pettini:2008mq}
  M.~Pettini, B.~J.~Zych, M.~T.~Murphy, A.~Lewis and C.~C.~Steidel,
  %``Deuterium Abundance in the Most Metal-Poor Damped Lyman alpha
  %System:   Converging on $\Omega_b$,''
  MNRAS, {\bf 391}  (2008) 1499.
  %}[arXiv:0805.0594 [astro-ph]].
  %%CITATION = ARXIV:0805.0594;%%
 
\bibitem{GG03}
  J. Geiss and  G. Gloeckler,
  Space Sience Reviews {\bf 106}, 3 (2003).

\bibitem{Pettini:2012ph}
  M.~Pettini and R.~Cooke,
  %``A new, precise measurement of the primordial abundance of Deuterium,''
  Mon.\ Not.\ Roy.\ Astron.\ Soc.\  {\bf 425}, 2477 (2012)
  [arXiv:1205.3785 [astro-ph.CO]].
  %%CITATION = ARXIV:1205.3785;%%

\bibitem{Cooke:2013cba}
  R.~Cooke, M.~Pettini, R.~A.~Jorgenson, M.~T.~Murphy and C.~C.~Steidel,
  %``Precision measures of the primordial abundance of deuterium,''
  arXiv:1308.3240 [astro-ph.CO].
  %%CITATION = ARXIV:1308.3240;%%
  %8 citations counted in INSPIRE as of 03 Mar 2014




%%%%%%%%%%%%%%%%%%%%%%%%%%%%
% muon g-2
%%%%%%%%%%%%%%%%%%%%%%%%%%%%

\bibitem{Bennett:2006fi}
  G.~W.~Bennett {\it et al.}  [Muon G-2 Collaboration],
  %``Final Report of the Muon E821 Anomalous Magnetic Moment Measurement at BNL,''
  Phys.\ Rev.\ D {\bf 73} (2006) 072003
  [hep-ex/0602035].
  %%CITATION = HEP-EX/0602035;%%

\bibitem{Davier:2010nc}
  M.~Davier, A.~Hoecker, B.~Malaescu and Z.~Zhang,
  %``Reevaluation of the Hadronic Contributions to the Muon g-2 and to alpha(MZ),''
  Eur.\ Phys.\ J.\ C {\bf 71} (2011) 1515
   [Erratum-ibid.\ C {\bf 72} (2012) 1874]
  [arXiv:1010.4180 [hep-ph]].
  %%CITATION = ARXIV:1010.4180;%%

\bibitem{Hagiwara:2011af}
  K.~Hagiwara, R.~Liao, A.~D.~Martin, D.~Nomura and T.~Teubner,
  %``(g-2)_mu and alpha(M_Z^2) re-evaluated using new precise data,''
  J.\ Phys.\ G {\bf 38} (2011) 085003
  [arXiv:1105.3149 [hep-ph]].
  %%CITATION = ARXIV:1105.3149;%%

%%%%%%%%%%%%%%%%%%%%%%%%%%%%
% bsmumu
%%%%%%%%%%%%%%%%%%%%%%%%%%%%

\bibitem{Chatrchyan:2013bka}
  S.~Chatrchyan {\it et al.}  [CMS Collaboration],
  %``Measurement of the B(s) to mu+ mu- branching fraction and search for B0 to mu+ mu- with the CMS Experiment,''
  Phys.\ Rev.\ Lett.\  {\bf 111} (2013) 101804
  [arXiv:1307.5025 [hep-ex]].
  %%CITATION = ARXIV:1307.5025;%%

\bibitem{Aaij:2013aka}
  RAaij {\it et al.}  [LHCb Collaboration],
  %``Measurement of the $B^0_s \to \mu^+ \mu^-$ branching fraction and search for $B^0 \to \mu^+ \mu^-$ decays at the LHCb experiment,''
  Phys.\ Rev.\ Lett.\  {\bf 111} (2013) 101805
  [arXiv:1307.5024 [hep-ex]].
  %%CITATION = ARXIV:1307.5024;%%


%%%%%%%%%%%%%%%%%%%%%%%%%%%%
% bsgamma
%%%%%%%%%%%%%%%%%%%%%%%%%%%%

\bibitem{Amhis:2012bh}
  Y.~Amhis {\it et al.}  [Heavy Flavor Averaging Group Collaboration],
  %``Averages of B-Hadron, C-Hadron, and tau-lepton properties as of early 2012,''
  arXiv:1207.1158 [hep-ex].
  %%CITATION = ARXIV:1207.1158;%%


%%%%%%%%%%%%%%%%%%%%%%%%%%%%
% direct detection (LUS latest)
%%%%%%%%%%%%%%%%%%%%%%%%%%%%

\bibitem{Akerib:2013tjd}
  D.~S.~Akerib {\it et al.}  [LUX Collaboration],
  %``First results from the LUX dark matter experiment at the Sanford Underground Research Facility,''
  arXiv:1310.8214 [astro-ph.CO].
  %%CITATION = ARXIV:1310.8214;%%









%%%%%%%%%%%%%%%%%%%%%%%%%%%%%%%%%%%%%%%%%%%%
\end{thebibliography}
\end{document}